\documentstyle[12pt,epsf,a4wide,rotate]{article}

\newcommand{\nwc}{\newcommand}
\nwc{\cl}  {\clubsuit}
\nwc{\di}  {\diamondsuit}
\nwc{\sps}  {\spadesuit}

\nwc{\fs}  {\footnotesize\em}
\nwc{\ns}  {\normalsize}

\nwc{\hyp} {\hyphenation}
\nwc{\be}  {\begin{equation}}
\nwc{\ee}  {\end{equation}}
\nwc{\ba}  {\begin{array}}
\nwc{\ea}  {\end{array}}
\nwc{\bdm} {\begin{displaymath}}
\nwc{\edm} {\end{displaymath}}
\nwc{\bea} {\be\ba{rcl}}
\nwc{\eea} {\ea\ee}
\nwc{\ben} {\begin{eqnarray}}
\nwc{\een} {\end{eqnarray}}
\nwc{\bda} {\bdm\ba{lcl}}
\nwc{\eda} {\ea\edm}
\nwc{\bc}  {\begin{center}}
\nwc{\ec}  {\end{center}}
\nwc{\ds}  {\displaystyle}
\nwc{\bmat}{\left(\ba}
\nwc{\emat}{\ea\right)}
\nwc{\non} {\nonumber}
\nwc{\bib} {\bibitem}
\nwc{\lra} {\longrightarrow}
\nwc{\Llra}{\Longleftrightarrow}
\nwc{\ra}  {\rightarrow}
\nwc{\Ra}  {\Rightarrow}
\nwc{\lmt} {\longmapsto}
\nwc{\prl} {\partial}
\nwc{\iy}  {\infty}
\nwc{\ol}  {\overline}
\nwc{\ul}  {\underline}
\nwc{\hm}  {\hspace{3mm}}
\nwc{\lf}  {\left}
\nwc{\ri}  {\right}
\nwc{\lm}  {\limits}
\nwc{\lb}  {\lbrack}
\nwc{\rb}  {\rbrack}
\nwc{\ov}  {\over}
\nwc{\pr}  {\prime}
\nwc{\nnn} {\nonumber \vspace{.2cm} \\ }
\nwc{\Sc}  {{\cal S}}
\nwc{\Lc}  {{\cal L}}
\nwc{\Rc}  {{\cal R}}
\nwc{\Dc}  {{\cal D}}
\nwc{\Oc}  {{\cal O}}
\nwc{\Cc}  {{\cal C}}
\nwc{\Pc}  {{\cal P}}
\nwc{\Mc}  {{\cal M}}
\nwc{\Ec}  {{\cal E}}
\nwc{\Fc}  {{\cal F}}
\nwc{\Hc}  {{\cal H}}
\nwc{\Kc}  {{\cal K}}
\nwc{\Xc}  {{\cal X}}
\nwc{\Gc}  {{\cal G}}
\nwc{\Zc}  {{\cal Z}}
\nwc{\Nc}  {{\cal N}}
\nwc{\fca} {{\cal f}}
\nwc{\xc}  {{\cal x}}
\nwc{\Ac}  {{\cal A}}
\nwc{\Bc}  {{\cal B}}
\nwc{\Uc}  {{\cal U}}
\nwc{\Vc}  {{\cal V}}
\nwc{\Th} {\Theta}
\nwc{\vth} {\vartheta}
\nwc{\eps}{\epsilon}
\nwc{\si} {\sigma}
\nwc{\Gm} {\Gamma}
\nwc{\gm} {\gamma}
\nwc{\bt} {\beta}
\nwc{\La} {\Lambda}
\nwc{\la} {\lambda}
\nwc{\om} {\omega}
\nwc{\Om} {\Omega}
\nwc{\dt} {\delta}
\nwc{\Si} {\Sigma}
\nwc{\Dt} {\Delta}
\nwc{\al} {\alpha}
\nwc{\vph}{\varphi}
\nwc{\zt} {\zeta}
\def\tr{\mathop{\rm tr}}
\def\Tr{\mathop{\rm Tr}}

\def\VEV#1{\left\langle #1\right\rangle}

\def\abs#1{\left| #1\right|}
\def\pr#1{#1^\prime}
\def\ltap{\raisebox{-.4ex}{\rlap{$\sim$}} \raisebox{.4ex}{$<$}}

\nwc{\Id}  {{\bf 1}}
\nwc{\sgn}  {{\rm sgn}}
\nwc{\diag} {{\rm diag}}
\nwc{\inv}  {{\rm inv}}
\nwc{\mod}  {{\rm mod}}
\nwc{\hal} {\frac{1}{2}}
\nwc{\tpi}  {2\pi i}

\def\slash#1{#1\!\!\!/\!\,\,}

\def\pr#1{Phys. Rev. {\bf #1}}

\def\APP#1{Acta Phys.~Pol.~{\bf #1}}

\def\CMP#1{Comm. Math. Phys.~{\bf #1}}
\def\CNPP#1{Comm. Nucl. Part. Phys.~{\bf #1}}
\def\HPA#1{Helv. Phys. Acta~{\bf #1}}
\def\IJMP#1{Int. J. Mod. Phys.~{\bf #1}}

\def\MPL#1{Mod. Phys. Lett.~{\bf #1}}
\def\NP#1{Nucl. Phys.~{\bf #1}}
\def\NPPS#1{Nucl. Phys. Proc. Suppl.~{\bf #1}}
\def\NC#1{Nuovo Cim.~{\bf #1}}

\def\PL#1{Phys. Lett.~{\bf #1}}
\def\PR#1{Phys. Rev.~{\bf #1}}
\def\PRP#1{Phys. Rep.~{\bf #1}}
\def\PRL#1{Phys. Rev. Lett.~{\bf #1}}
\def\PNAS#1{Proc. Nat. Acad. Sc.~{\bf #1}}
\def\PTP#1{Progr. Theor. Phys.~{\bf #1}}
\def\RMP#1{Rev. Mod. Phys.~{\bf #1}}

\def\ZP#1{Z. Phys.~{\bf #1}}

\def\MeV {\,{\rm  MeV}}
\def\GeV {\,{\rm  GeV}}

\def \lta {\mathrel{\vcenter
     {\hbox{$<$}\nointerlineskip\hbox{$\sim$}}}}
\def \gta {\mathrel{\vcenter
     {\hbox{$>$}\nointerlineskip\hbox{$\sim$}}}} 
\newsavebox{\nnin} \sbox{\nnin}{$\hspace{1mm}\in\kern -.8em /
                   \hspace{1mm}$}

\newcommand{\sub}{\subset}
\newsavebox{\nnsub} \sbox{\nnsub}{$\hspace{1mm}\sub\kern -.9em /
            \hspace{1mm}$}

\def\KK{{\rm I\kern -.2em  K}}
\def\NN{{\rm I\kern -.16em N}}
\def\RR{{\rm I\kern -.2em  R}}
\def\ZZ{Z \kern -.43em Z}
\def\QQ{{\rm \kern .25em
             \vrule height1.4ex depth-.12ex width.06em\kern-.31em Q}}
\def\CC{{\rm \kern .25em
             \vrule height1.4ex depth-.12ex width.06em\kern-.31em C}}
\def\ZZZ{Z\kern -0.31em Z}

\nwc{\site}[1]{\refnote{\cite{#1}}}
\nwc{\Ref}[1]{Eq.(\ref{#1})}
\nwc{\iddq}  {\int\frac{d^dq}{(2\pi)^d}}

\nwc{\cst}     {SU_L(3)\times SU_R(3)}
\nwc{\csN}     {SU_L(N)\times SU_R(N)}
\nwc{\rmcl}    {{\rm cl}}
\nwc{{\rmeff}}   {{\rm eff}}

\newcounter{app}
\def\app{\par
 \addtocounter{app}{1}
 \def\thesection{\Alph{app}}
 \def\ksection{\Alph{app}}}

\def\appendix#1{\app\sect{#1}}

\renewcommand{\theequation}{\thesection.\arabic{equation}}
\newcommand{\sect}[1]{ \section{#1} \setcounter{equation}{0} }

\begin{document}

\title{Nonperturbative Flow Equations,\\ Low--Energy QCD and\\ the
  Chiral Phase Transition\thanks{Lectures given at the {\em NATO
      Advanced Study Institute: Confinement, Duality and
      Non--perturbative Aspects of QCD}, Newton Institute, Cambridge,
    UK, 23 June -- 4 July, 1997.}}

\author{{\sc D.--U.~Jungnickel\thanks{Email: 
D.Jungnickel@thphys.uni-heidelberg.de}} \\
 \\ and \\ \\
{\sc C.~Wetterich\thanks{Email: C.Wetterich@thphys.uni-heidelberg.de}} 
\\ \\ \\
{\em Institut f\"ur Theoretische Physik} \\
{\em Universit\"at Heidelberg} \\
{\em Philosophenweg 16} \\
{\em 69120 Heidelberg, Germany}}

\date{}

\thispagestyle{empty}

\maketitle

\begin{picture}(5,2.5)(-350,-500)
\put(12,-115){HD--THEP--97--53}
\put(12,-138){October, 1997}
\end{picture}

\begin{abstract}
  We review the formalism of the effective average action in quantum
  field theory which corresponds to a coarse grained free energy in
  statistical mechanics. The associated exact renormalization group
  equation and possible nonperturbative approximations for its
  solution are discussed. This is applied to QCD where one observes
  the consecutive emergence of mesonic bound states and spontaneous
  chiral symmetry breaking as the coarse graining scale is lowered. We
  finally present a study of the chiral phase transition in two flavor
  QCD. A precision estimate of the universal critical equation of
  state for the three--dimensional $O(4)$ Heisenberg model is
  presented.  We explicitly connect the $O(4)$ universal behavior near
  the critical temperature and zero quark mass with the physics at
  zero temperature and a realistic pion mass.  For realistic quark
  masses the pion correlation length near $T_c$ turns out to be
  smaller than its zero temperature value.
\end{abstract}

\sect{Effective average action}
\label{EffectiveAverageAction}

Quantum chromodynamics (QCD) describes qualitatively different physics
at different length scales. At short distances the relevant degrees of
freedom are quarks and gluons which can be treated perturbatively. At
long distances we observe hadrons, and an essential part of the
dynamics can be encoded in the masses and interactions of mesons. Any
attempt to deal with this situation analytically and to predict the
meson properties from the short distance physics (as functions of the
strong gauge coupling $\alpha_s$ and the current quark masses $m_q$)
has to bridge the gap between two qualitatively different effective
descriptions. Two basic problems have to be mastered for an
extrapolation from short distance QCD to mesonic length scales:
\begin{itemize}
\item The effective couplings change with scale. This does not only
  concern the running gauge coupling, but also the coefficients of
  non--renormalizable operators as, for example, four quark operators.
  Typically, these non--renormalizable terms become important in the
  momentum range where $\alpha_s$ is strong and deviate substantially
  from their perturbative values. Consider the four--point function
  which results after integrating out the gluons. For heavy quarks it
  contains the information about the shape of the heavy quark
  potential whereas for light quarks the complicated spectrum of light
  mesons and chiral symmetry breaking are encoded in it. At distance
  scales around $1{\rm fm}$ one expects that the effective action
  resembles very little the form of the classical QCD action which is
  relevant at short distances.
\item Not only the couplings, but even the relevant variables or
  degrees of freedom are different for long distance and short
  distance QCD. It seems forbiddingly difficult to describe the
  low--energy scattering of two mesons in a language of quarks and
  gluons only. An appropriate analytical field theoretical method
  should be capable of introducing field variables for composite
  objects such as mesons.
\end{itemize}
A conceptually very appealing idea for our task is the block--spin
action~\cite{Kad66-1,Wil71-1}. It realizes that physics with a given
characteristic length scale $l$ is conveniently described by a
functional integral with an ultraviolet (UV) cutoff $\Lambda$ for the
momenta. Here $\Lambda$ should be larger than $l^{-1}$ but not
necessarily by a large factor. The Wilsonian effective action
$S_\Lambda^{\rm W}$ replaces then the classical action in the
functional integral. It is obtained by integrating out the
fluctuations with momenta $q^2\gta\Lambda^2$. An exact renormalization
group equation~\cite{Wil71-1,WH73-1,Wei76-1,Pol84-1,Has86-1} describes
how $S_\Lambda^{\rm W}$ changes with the UV cutoff $\Lambda$.

We will use here the somewhat different but related concept of the
effective average action~\cite{Wet91-1} $\Gamma_k$ which, in the
language of statistical physics, is a coarse grained free energy with
coarse graining scale $k$. The effective average action is based on
the quantum field theoretical concept of the effective
action~\cite{Sch51-1} $\Gamma$ which is obtained by integrating out all
quantum fluctuations.  The effective action contains all information
about masses, couplings, form factors and so on, since it is the
generating functional of the $1PI$ Green functions. The field
equations derived from $\Gamma$ are exact including all quantum
effects. For a field theoretical description of thermal equilibrium
this concept is easily generalized to a temperature dependent
effective action which includes now also the thermal fluctuations. In
statistical physics $\Gamma$ describes the free energy as a functional
of some convenient (space dependent) order parameter, for instance the
magnetization. In particular, the behavior of $\Gamma$ for a constant
order parameter (the effective potential) specifies the equation of
state. The effective average action $\Gamma_k$ is a simple
generalization of the effective action, with the distinction that only
quantum fluctuations with momenta $q^2\gta\ k^2$ are included. This
can be achieved by introducing an explicit infrared cutoff $\sim k$ in
the functional integral defining the partition function (or the
generating functional for the $n$--point functions).  Typically, this
IR--cutoff is quadratic in the fields and modifies the inverse
propagator, for example by adding a mass--like term $\sim k^2$. The
effective average action can then be defined in complete analogy to
the effective action (via a Legendre transformation of the logarithm
of the partition function). The mass--like term in the propagator
suppresses the contributions from the small momentum modes with
$q^2\lta k^2$ and $\Gamma_k$ accounts effectively only for the
fluctuations with $q^2\gta k^2$.

Following the behavior of $\Gamma_k$ for different $k$ is like looking
at the world through a microscope with variable resolution: For large
$k$ one has a very precise resolution $\sim k^{-1}$ but one also
studies effectively only a small volume $\sim k^d$. Taking in QCD the
coarse graining scale $k$ much larger than the confinement scale
guarantees that the complicated nonperturbative physics does not play
a role yet.  In this case, $\Gamma_k$ will look similar to the
classical action, typically with a running gauge coupling evaluated at
the scale $k$.  (This does not hold for Green functions with much
larger momenta $p^2\gg k^2$ where the relevant IR cutoff is $p$, and
the effective coupling is $\alpha_s(p)$.)  For lower $k$ the
resolution is smeared out and the detailed information of the short
distance physics can be lost.  (Again, this does not concern Green
functions at high momenta.)  On the other hand, the ``observable
volume'' is increased and long distance aspects such as collective
phenomena become visible. In a theory with a physical UV cutoff
$\Lambda$ we may associate $\Gamma_\Lambda$ with the classical action
$S$ since no fluctuations are effectively included. By definition, the
effective average action equals the effective action for $k=0$,
$\Gamma_{0}=\Gamma$, since the infrared cutoff is absent. Thus
$\Gamma_k$ interpolates between the classical action $S$ and the
effective action $\Gamma$ as $k$ is lowered from $\Lambda$ to zero.
The ability to follow the evolution $k\ra0$ is equivalent to the
ability to solve the quantum field theory.

For a formal description we will consider in the first two sections a
model with real scalar fields $\chi^a$ (the index $a$ labeling
internal degrees of freedom) in $d$ Euclidean dimensions with
classical action $S[\chi]$. We define the generating functional for
the connected Green functions by
\begin{equation}
  W_k[J]=\ln\int\Dc\chi\exp\left\{-S_k[\chi]+\int d^d x
  J_a(x)\chi^a(x)\right\}
\end{equation} 
where we have added to the classical action an IR cutoff $\Delta_kS$
\begin{equation}
  S_k[\chi] = S[\chi]+\Dt S_k[\chi]
\end{equation}
which is quadratic in the fields and conveniently formulated in
momentum space
\begin{equation}
  \Dt S_k[\chi] = \hal\iddq
  R_k(q^2)\chi_a(-q)\chi^a(q)\; .  
\end{equation}
Here $J_a$ are the usual scalar sources introduced to define
generating functionals and $R_k(q^2)$ denotes an appropriately chosen
(see below) IR cutoff function. We require that $R_k(q^2)$ vanishes
rapidly for $q^2\gg k^2$ whereas for $q^2\ll k^2$ it behaves as
$R_k(q^2)\simeq k^2$. This means that all Fourier components of
$\chi^a$ with momenta smaller than the IR cutoff $k$ acquire an
effective mass $m_\rmeff\simeq k$ and therefore decouple while the
high momentum components of $\chi^a$ should not be affected by $R_k$.
The classical fields
\begin{equation}
  \Phi^a\equiv\VEV{\chi^a}=\frac{\dt W_k[J]}{\dt J_a} 
\end{equation} 
now depend on $k$. In terms of $W_k$ the effective average action is
defined via a Legendre transformation
\begin{equation}
  \label{AAA60}
  \Gm_k[\Phi]=-W_k[J]+\int d^dx J_a(x)\Phi^a(x)- 
  \Delta S_k[\Phi] \; .
\end{equation}
In order to define a reasonable coarse grained free energy we have
subtracted in \Ref{AAA60} the infrared cutoff piece. This guarantees
that the only difference between $\Gamma_k$ and $\Gamma$ is the
effective IR cutoff in the fluctuations. Furthermore, this has the
consequence that $\Gamma_k$ does not need to be convex whereas a pure
Legendre transform is always convex by definition. (The coarse grained
free energy becomes convex~\cite{RW90-1} only for $k\ra0$.) This is
very important for the description of phase transitions, in particular
first order ones. One notes
\begin{equation}\begin{array}{lcl}
 \ds{\lim_{k\ra0}R_k(q^2)=0} &\Ra& 
 \ds{\lim_{k\ra0}\Gm_k[\Phi]=\Gm[\Phi]}\nnn
 \ds{\lim_{k\ra\La}R_k(q^2)=\infty} &\Ra&
 \ds{\lim_{k\ra\La}\Gm_k[\Phi]=S[\Phi]}
 \label{ConditionsForRk}
\end{array}\end{equation}
for a convenient choice of $R_k$ like
\begin{equation}
 R_k(q^2)=Z_{\Phi,k} q^2
 \frac{1+e^{-q^2/k^2}-e^{-q^2/\La^2}}
 {e^{-q^2/\La^2}-e^{-q^2/k^2}}\; .
 \label{Rk}
\end{equation}
Here $Z_{\Phi,k}$ denotes the wave function renormalization to be
specified below and we will often use for $R_k$ the limit
$\Lambda\ra\infty$
\begin{equation}
  \label{AAA61}
  R_k(q^2)=\frac{Z_{\Phi,k} q^2}
  {e^{q^2/k^2}-1}\; .
\end{equation}
We note that the property $\Gamma_\Lambda=S$ is not essential since
the short distance laws may be parameterized by $\Gamma_\Lambda$ as
well as by $S$. In addition, for momentum scales much smaller than
$\Lambda$ universality implies that the precise form of
$\Gamma_\Lambda$ is irrelevant, up to the values of a few relevant
renormalized couplings.

A few properties of the effective average action are worth mentioning:
\begin{enumerate}
\item All symmetries of the model which are respected by the IR cutoff
  $\Delta_k S$ are automatically symmetries of $\Gamma_k$. In
  particular, this concerns translation and rotation invariance and
  one is not plagued by many of the problems encountered by a
  formulation of the block--spin action on a lattice.
\item In consequence, $\Gamma_k$ can be expanded in terms of
  invariants with respect to these symmetries with couplings depending
  on $k$. For the example of a scalar theory one may use a derivative
  expansion ($\rho=\Phi^a\Phi_a/2$)
  \begin{equation}
    \label{AAA63}
    \Gamma_k=\int d^d x\left\{
    U_k(\rho)+\frac{1}{2}Z_{\Phi,k}(\rho)
    \prl^\mu\Phi_a\prl_\mu\Phi^a+\ldots\right\}
  \end{equation}
  and expand further in powers of $\rho$
  \begin{eqnarray}
    \label{AAA64}
    \ds{U_k(\rho)} &=& \ds{
    \frac{1}{2}\ol{\la}_k\left(
    \rho-\rho_0(k)\right)^2+
    \frac{1}{6}\ol{\gamma}_k\left(
    \rho-\rho_0(k)\right)^3+\ldots}\nnn
    \ds{Z_{\Phi,k}(\rho)} &=& \ds{
    Z_{\Phi,k}(\rho_0)+Z_{\Phi,k}^\prime(\rho_0)
    \left(\rho-\rho_0\right)+\ldots}\; 
  \end{eqnarray}
  where $\rho_0$ denotes the ($k$--dependent) minimum of the effective
  average potential $U_k(\rho)$.  We see that $\Gamma_k$ describes
  infinitely many running couplings.  ($Z_{\Phi,k}$ in \Ref{Rk} can
  be identified with $Z_{\Phi,k}(\rho_0)$.)
\item There is no problem incorporating chiral fermions since a
  chirally invariant cutoff $R_k$ can be
  formulated~\cite{Wet90-1,CKM97-1}.
\item Gauge theories can be formulated along similar
  lines~\cite{RW93-1,Bec96-1,BAM94-1,EHW94-1,Wet95-2,EHW96-1} even
  though $\Delta_k S$ may not be gauge invariant. In this case the
  usual Ward identities receive corrections for which one can derive
  closed expressions~\cite{EHW94-1}. These corrections vanish for
  $k\ra0$.
\item The high momentum modes are very effectively integrated out
  because of the exponential decay of $R_k$ for $q^2\gg k^2$.
  Nevertheless, it is sometimes technically easier to use a cutoff
  without this fast decay property (e.g.~$R_k\sim k^2$ or $R_k\sim
  k^4/q^2$). In the latter case one has to be careful with possible
  remnants of an incomplete integration of the short distance modes.
  Also our cutoff does not introduce any non--analytical behavior as
  would be the case for a sharp cutoff~\cite{Wet91-1}.
\item Despite a similar spirit and many analogies there remains also a
  conceptual difference to the Wilsonian effective action
  $S_\Lambda^{\rm W}$. The Wilsonian effective action describes a set
  of different actions (parameterized by $\Lambda$) for one and the
  same theory --- the $n$--point functions are independent of
  $\Lambda$ and have to be computed from $S_\Lambda^{\rm W}$ by
  further functional integration. In contrast, $\Gamma_k$ describes
  the effective action for different theories --- for any value of $k$
  the effective average action is related to the generating functional
  of a theory with a different action $S_k=S+\Delta_k S$. The
  $n$--point functions depend on $k$. The Wilsonian effective action
  does not generate the $1PI$ Green functions~\cite{KKS92-1}.
\item Because of the incorporation of an infrared cutoff, $\Gamma_k$
  is closely related to an effective action for averages of
  fields~\cite{Wet91-1}, where the average is taken over a volume
  $\sim k^d$.
\end{enumerate}

\sect{Exact renormalization group equation}
\label{AnExactRGE}

The dependence of $\Gamma_k$ on the coarse graining scale $k$ is
governed by an exact renormalization group equation
(ERGE)~\cite{Wet93-1}
\begin{equation}
  \prl_t\Gm_k[\Phi] =  \hal\Tr\left\{\left[
  \Gm_k^{(2)}[\Phi]+R_k\right]^{-1}\prl_t R_k\right\} \; .
  \label{ERGE}
\end{equation}
Here $t=\ln(k/\La)$ with some arbitrary momentum scale $\La$, and the
trace includes a momentum integration as well as a summation over
internal indices, $\Tr=\int\frac{d^d q}{(2\pi)^d}\sum_a$. The second
functional derivative $\Gm_k^{(2)}$ denotes the {\em exact} inverse
propagator
\begin{equation}
 \label{AAA69}
 \left[\Gm_k^{(2)}\right]_{ab}(q,q^\prime)=
 \frac{\dt^2\Gm_k}{\dt\Phi^a(-q)\dt\Phi^b(q^\prime)}\; .
\end{equation}
The flow equation \Ref{ERGE} can be derived from \Ref{AAA60} in a
straightforward way using
\begin{eqnarray}
  \label{AAA70}
  \ds{\prl_t \left.\Gamma_k\right|_\Phi} &=& \ds{
  -\prl_t \left.W_k\right|_J-
  \prl_t\Delta_kS[\Phi]}\nnn
  &=& \ds{
  \frac{1}{2}\Tr\left\{\prl_t R_k\left(
  \VEV{\Phi\Phi}-\VEV{\Phi}\VEV{\Phi}\right)
  \right\}}\nnn
  &=& \ds{
  \frac{1}{2}\Tr\left\{
  \prl_t R_k W^{(2)}_k\right\} }
\end{eqnarray}
and
\begin{eqnarray}
  \label{AAA71}
  \ds{W_{k,ab}^{(2)}(q,q^\prime)} &=& \ds{
  \frac{\delta^2 W_k}
  {\delta J^a(-q)\delta J^b(q^\prime)} }\nnn
  \ds{\frac{\delta^2 W_k}
  {\delta J_a(-q)\delta J_b(q^\prime)}
  \frac{\delta^2\left(\Gamma_k+\Delta_k S\right)}
  {\delta\Phi_b(-q^{\prime})\delta\Phi_c
  (q^{\prime\prime})}  } &=& \ds{
  \delta_{ac}\delta_{q q^{\prime\prime}} }\; .
\end{eqnarray}
It has the form of a renormalization group improved one--loop
expression~\cite{Wet91-1}. Indeed, the one--loop formula for
$\Gamma_k$ reads
\begin{equation}
  \label{AAA72}
  \Gamma_k[\Phi]=S[\Phi]+
  \frac{1}{2}\Tr\ln\left(
  S^{(2)}[\Phi]+R_k\right)
\end{equation}
with $S^{(2)}$ the second functional derivative of the {\em classical}
action, similar to \Ref{AAA69}.  (Remember that $S^{(2)}$ is the field
dependent classical inverse propagator. Its first and second
derivative with respect to the fields describe the classical three--
and four--point vertices, respectively.) Taking a $t$--derivative of
\Ref{AAA72} gives a one--loop flow equation very similar to
\Ref{ERGE} with $\Gamma_k^{(2)}$ replaced by $S^{(2)}$. It may seem
surprising, but it is nevertheless true, that the renormalization
group improvement $S^{(2)}\ra\Gamma_k^{(2)}$ promotes the one--loop
flow equation to an exact nonperturbative flow equation which includes
the effects from all loops as well as all contributions which are
non--analytical in the couplings like instantons, etc.! For practical
computations it is actually often quite convenient to write the flow
equation \Ref{ERGE} as a formal derivative of a renormalization
group improved one--loop expression
\begin{equation}
  \label{AAA73}
  \prl_t\Gamma_k=
  \frac{1}{2}\Tr\tilde{\prl}_t\ln\left(
  \Gamma_k^{(2)}+R_k\right)
\end{equation}
with $\tilde{\prl}_t$ acting only on $R_k$ and not on
$\Gamma_k^{(2)}$, i.e. $\tilde{\prl}_t=\left(\prl R_k/\prl
t\right)\left(\prl/\prl R_k\right)$. Flow equations for $n$--point
functions follow from appropriate functional derivatives of \Ref{ERGE}
or \Ref{AAA73} with respect to the fields. For their derivation it is
sufficient to evaluate the corresponding one--loop expressions (with
the vertices and propagators derived from $\Gamma_k$) and then to take
a formal $\tilde{\prl}_t$--derivative.  (If the one--loop expression
is finite or properly regularized the $\tilde{\prl}_t$--derivative can
be taken after the evaluation of the trace.) This permits the use of
(one--loop) Feynman diagrams and standard perturbative techniques in
many circumstances. Most importantly, it establishes a very direct
connection between the solution of flow--equations and perturbation
theory. If one uses on the right hand side of \Ref{ERGE} a
truncation for which the propagator and vertices appearing in
$\Gamma_k^{(2)}$ are replaced by the ones derived from the classical
action, but with running $k$--dependent couplings, and then expands
the result to lowest non--trivial order in the coupling constants one
recovers standard renormalization group improved one--loop
perturbation theory. The formal solution of the flow equation can also
be employed for the development of a systematically resummed
perturbation theory~\cite{Wet96-1}.

For a choice of the cutoff function similar to \Ref{AAA61} the
momentum integral contained in the trace on the right hand side of the
flow equation is both infrared and ultraviolet finite. Infrared
finiteness arises through the presence of the infrared regulator $\sim
R_k$. We note that all eigenvalues of the matrix $\Gamma_k^{(2)}+R_k$
must be positive semi--definite. The proof follows from the
observation that the functional $\Gamma_k+\Delta_k S$ is convex since
it is obtained from $W_k$ by a Legendre transform. On the other hand,
ultraviolet finiteness is related to the fast decay of $\prl_t R_k$
for $q^2\gg k^2$. This expresses the fact that only a narrow range of
fluctuations with $q^2\simeq k^2$ contributes effectively if the
infrared cutoff $k$ is lowered by a small amount. If for some other
choice of $R_k$ the right hand side of the flow equation would not
remain UV finite this would indicate that the high momentum modes have
not yet been integrated out completely in the computation of
$\Gamma_k$.  Since the flow equation is manifestly finite this can be
used to define a regularization scheme. The ``ERGE--scheme'' is
specified by the flow equation, the choice of $R_k$ and the ``initial
condition'' $\Gamma_\Lambda$. This is particularly important for gauge
theories where other regularizations in four dimensions and in the
presence of chiral fermions are difficult to construct. For gauge
theories $\Gamma_\Lambda$ has to obey appropriately modified Ward
identities. In the context of perturbation theory a first proposal how
to regularize gauge theories by use of flow equations can be found
in~\cite{Bec96-1,BAM94-1}. We note that in contrast to previous
versions of exact renormalization group equations there is no need in
the present formulation to construct an ultraviolet momentum cutoff
--- a task known to be very difficult in non--Abelian gauge theories.

Despite the conceptual differences between the Wilsonian effective
action $S_\Lambda^{\rm W}$ and the effective average action
$\Gamma_k$ the exact flow equations describing the $\Lambda$--dependence
of $S_\Lambda^{\rm W}$ and the $k$--dependence of $\Gamma_k$ are
simply related. Polchinski's continuum version of the Wilsonian flow
equation~\cite{Pol84-1} can be transformed into \Ref{ERGE} by means
of a Legendre transform and a suitable variable
redefinition~\cite{BAM93-1}.

Even though intuitively simple, the replacement of the (RG--improved)
classical propagator by the full propagator turns the solution of the
flow equation \Ref{ERGE} into a difficult mathematical problem: The
evolution equation is a functional differential equation. Once
$\Gamma_k$ is expanded in terms of invariants (e.g.~Eqs.(\ref{AAA63}),
(\ref{AAA64})) this is equivalent to a coupled system of non--linear
partial differential equations for infinitely many couplings. General
methods for the solution of functional differential equations are not
developed very far. They are mainly restricted to iterative procedures
that can be applied once some small expansion parameter is identified.
This covers usual perturbation theory in the case of a small coupling,
the $1/N$--expansion or expansions in the dimensionality $4-d$ or
$2-d$. It may also be extended to less familiar expansions like a
derivative expansion which is related in critical three dimensional
scalar theories to a small anomalous dimension. In the absence of a
clearly identified small parameter one nevertheless needs to truncate
the most general form of $\Gamma_k$ in order to reduce the infinite
system of coupled differential equations to a (numerically) manageable
size. This truncation is crucial. It is at this level that
approximations have to be made and, as for all nonperturbative
analytical methods, they are often not easy to control. The challenge
for nonperturbative systems like low momentum QCD is to find flow
equations which (a) incorporate all the relevant dynamics such that
neglected effects make only small changes, and (b) remain of
manageable size. The difficulty with the first task is a reliable
estimate of the error. For the second task the main limitation is a
practical restriction for numerical solutions of differential
equations to functions depending only on a small number of variables.
The existence of an exact functional differential flow equation is a
very useful starting point and guide for this task. At this point the
precise form of the exact flow equation is quite important.
Furthermore, it can be used for systematic expansions through
enlargement of the truncation and for an error estimate in this way.
Nevertheless, this is not all. Usually, physical insight into a model
is necessary to device a useful nonperturbative truncation!

So far, two complementary approaches to nonperturbative truncations
have been explored: an expansion of the effective Lagrangian in powers
of derivatives ($\rho\equiv\frac{1}{2}\Phi_a\Phi^a$)
\begin{equation}
\label{DerExp}
  \Gamma_k[\Phi]=\int d^d x\left\{
  U_k(\rho)+\frac{1}{2}Z_{\Phi,k}(\Phi)
  \prl_\mu\Phi^a\prl^\mu\Phi_a+
  \frac{1}{4}Y_{\Phi,k}(\rho)\prl_\mu\rho
  \prl^\mu\rho+
  \Oc(\prl^4)\right\}
\end{equation}
or one in powers of the fields
\begin{equation}
\label{FieldExp}
  \Gamma_k[\Phi]=
  \sum_{n=0}^\infty\frac{1}{n!}\int
  \left(\prod_{j=0}^n d^d x_j
  \left[\Phi(x_j)-\Phi_0\right]\right)
  \Gamma_k^{(n)}(x_1,\ldots,x_n)\; .
\end{equation}
If one chooses $\Phi_0$ as the $k$--dependent VEV of $\Phi$, the
series \Ref{FieldExp} starts effectively at $n=2$. The flow
equations for the $1PI$ $n$--point functions $\Gamma_k^{(n)}$ are
obtained by functional differentiation of \Ref{ERGE}. Such flow
equations have been discussed earlier from a somewhat different
viewpoint~\cite{Wei76-1}. They can also be interpreted as a
differential form of Schwinger--Dyson equations~\cite{DS49-1}.

The formation of mesonic bound states, which typically appear as poles
in the (Min\-kowskian) four--quark Green function, is most efficiently
described by expansions like \Ref{FieldExp}.  This is also the form
needed to compute the nonperturbative momentum dependence of the gluon
propagator and the heavy quark
potential~\cite{Wet95-2,EHW96-1,BBW97-1}.  On the other hand, a
parameterization of $\Gamma_k$ as in \Ref{DerExp} seems particularly
suited for the study of phase transitions. The evolution equation for
the average potential $U_k$ follows by evaluating \Ref{DerExp} for
constant $\Phi$. In the limit where the $\Phi$--dependence of
$Z_{\Phi,k}$ is neglected and $Y_{\Phi,k}=0$ one finds~\cite{Wet91-1}
for the $O(N)$--symmetric scalar model
\begin{equation}
  \label{AAA80}
  \prl_t U_k(\rho)=\frac{1}{2}
  \int\frac{d^d q}{(2\pi)^d}
  \frac{\prl R_k}{\prl t}\left(
  \frac{N-1}{Z_{\Phi,k}q^2+R_k+U_k^\prime}+
  \frac{1}{Z_{\Phi,k}q^2+R_k(q)+U_k^\prime+
  2\rho U_k^{\prime\prime}}\right)
\end{equation}
with $U_k^\prime\equiv\frac{\prl U_k}{\prl\rho}$, etc. One observes
the appearance of $\rho$--dependent mass terms in the effective
propagators of the right hand side of \Ref{AAA80}. Once
$\eta_\Phi\equiv-\prl_t\ln Z_{\Phi,k}$ is determined~\cite{Wet91-1} in
terms of the couplings parameterizing $U_k$ this is a partial
differential equation for a function $U_k$ depending on two variables
$k$ and $\rho$ which can be solved
numerically~\cite{ABB95-1,BTW96-1,Tet96-1,BW96-1}. (The Wilson--Fisher
fixed point relevant for a second order phase transition ($d=3$)
corresponds to a scaling solution~\cite{TW94-1,Mor94-1} where $\prl_t
U_k=0$.) A suitable truncation of a flow equation of the type
\Ref{DerExp} will play a central role in the description of chiral
symmetry breaking below.

It should be mentioned at this point that the weakest point in the
ERGE approach seems to be a reliable estimate of the truncation error
in a nonperturbative context. This problem is common to all known
analytical approaches to nonperturbative phenomena and appears often
even within systematic (perturbative) expansions. One may hope that
the existence of an exact flow equation could also be of some help for
error estimates. An obvious possibility to test a given truncation is
its enlargement to include more variables --- for example, going one
step higher in a derivative expansion. This is similar to computing
higher orders in perturbation theory and limited by practical
considerations. As an alternative, one may employ different
truncations of comparable size --- for instance, by using different
definitions of retained couplings. A comparison of the results can
give a reasonable picture of the uncertainty if the used set of
truncations is wide enough. In this context we should also note the
dependence of the results on the choice of the cutoff function
$R_k(q)$.  Of course, for $k\ra0$ the physics should not depend on a
particular choice of $R_k$ and, in fact, it does not for full
solutions of \Ref{ERGE}. Different choices of $R_k$ just correspond
to different trajectories in the space of effective average actions
along which the unique IR limit $\Gamma[\Phi]$ is reached. Once
approximations are used to solve the ERGE \Ref{ERGE}, however, not
only the trajectory but also its end point will depend on the precise
definition of the function $R_k$. This is very similar to the
renormalization scheme dependence usually encountered in perturbative
computations of Green functions. One may use this scheme dependence as
a tool to study the robustness of a given approximation scheme.

Before applying a new nonperturbative method to a complicated theory
like QCD it should be tested for simpler models. A good criterion for
the capability of the ERGE to deal with nonperturbative phenomena
concerns the critical behavior in three dimensional scalar theories.
In a first step the well known results of other methods for the
critical exponents have been reproduced within a few percent
accuracy~\cite{TW94-1}. The ability of the method to produce new
results has been demonstrated by the computation of the critical
equation of state for Ising and Heisenberg models~\cite{BTW96-1} which
has been verified by lattice simulations~\cite{Tsy94-1}. This has been
extended to first order transitions in matrix models~\cite{BW96-1} or
for the Abelian Higgs model relevant for
superconductors~\cite{BLL95-1,Tet96-1}.  Analytical investigations of
the high temperature phase transitions in $d=4$ scalar theories
($O(N)$--models) have correctly described the second order nature of
the transition~\cite{TW93-1}, in contrast to earlier attempts within high
temperature perturbation theory.

For an extension of the flow equations to Abelian and non--Abelian
gauge theories we refer the reader
to~\cite{RW93-1,Bec96-1,BAM94-1,EHW94-1,Wet95-2,EHW96-1,BBW97-1}. The
other necessary piece for a description of low--energy QCD, namely the
transition from fundamental (quark and gluon) degrees of freedom to
composite (meson) fields within the framework of the ERGE can be found
in~\cite{EW94-1}. We will describe the most important aspects of this
formalism for mesons below.

\sect{Chiral symmetry breaking in QCD}
\label{ChiralSymmetryBreaking}

The strong interaction dynamics of quarks and gluons at short
distances or high energies is successfully described by quantum
chromodynamics (QCD). One of its most striking features is asymptotic
freedom~\cite{GW73-1} which makes perturbative calculations reliable
in the high energy regime. On the other hand, at scales around a few
hundred $\MeV$ confinement sets in. As a consequence, the low--energy
degrees of freedom in strong interaction physics are mesons, baryons
and glueballs rather than quarks and gluons. When constructing
effective models for these IR degrees of freedom one usually relies on
the symmetries of QCD as a guiding principle, since a direct
derivation of such models from QCD is still missing. The most
important symmetry of QCD, its local color $SU(3)$ invariance, is of
not much help here, since the IR spectrum appears to be color neutral.
When dealing with bound states involving heavy quarks the so called
``heavy quark symmetry'' may be invoked to obtain approximate symmetry
relations between IR observables~\cite{Neu94-1}. We will rather focus
here on the light scalar and pseudoscalar meson spectrum and therefore
consider QCD with only the light quark flavors $u$, $d$ and $s$. To a
good approximation the masses of these three flavors can be considered
as small in comparison with other typical strong interaction scales.
One may therefore consider the chiral limit of QCD (vanishing current
quark masses) in which the classical QCD Lagrangian does not couple
left-- and right--handed quarks. It therefore exhibits a global chiral
invariance under $U_L(N)\times U_R(N)=SU_L(N)\times SU_R(N)\times
U_V(1)\times U_A(1)$ where $N$ denotes the number of massless quarks
($N=2$ or $3$) which transform as
\begin{eqnarray}
  \ds{\psi_R\equiv\frac{1-\gm_5}{2}\psi} &\longrightarrow&
  \ds{\Uc_R \psi_R\; ;\;\;\;\Uc_R\in U_R(N)}\nnn
  \ds{\psi_L\equiv\frac{1+\gm_5}{2}\psi} &\longrightarrow& \ds{\Uc_L
  \psi_L\; ;\;\;\;\Uc_L\in U_L(N)}\; .
 \label{ChiralTransoformation}
\end{eqnarray}
Even for vanishing quark masses only the diagonal $SU_V(N)$ vector--like
subgroup can be observed in the hadron spectrum (``eightfold
way''). The symmetry $SU_L(N)\times SU_R(N)$ must therefore be
spontaneously broken to $SU_V(N)$
\begin{equation}
  SU_L(N)\times SU_R(N)\longrightarrow
  SU_{L+R}(N)\equiv SU_V(N)\; .
  \label{CSBPattern}
\end{equation}
Chiral symmetry breaking is one of the most prominent features of
strong inter\-action dynamics and phenomenologically well
established~\cite{Leu95-1}, though a rigorous derivation of this
phenomenon starting from first principles is still missing. In
particular, the chiral symmetry breaking \Ref{CSBPattern} predicts
for $N=3$ the existence of eight light parity--odd (pseudo--)Goldstone
bosons: $\pi^0$, $\pi^\pm$, $K^0$, $\ol{K}^0$, $K^\pm$ and $\eta$.
Their comparably small masses are a consequence of the explicit chiral
symmetry breaking due to small but non--vanishing current quark
masses.  The axial Abelian subgroup $U_A(1)=U_{L-R}(1)$ is broken in
the quantum theory by an anomaly of the axial--vector current. This
breaking proceeds without the occurrence of a Goldstone
boson~\cite{Hoo86-1}.  Finally, the $U_V(1)=U_{L+R}(1)$ subgroup
corresponds to baryon number conservation.

The light pseudoscalar and scalar mesons are thought of as color
neutral quark--anti\-quark bound states $\Phi^{ab}\sim \ol{\psi}_L^b
\psi_R^a$, $a,b=1,\ldots,N$, which therefore transform under chiral
rotations \Ref{ChiralTransoformation} as
\begin{equation} 
  \Phi\longrightarrow
  \Uc_R\Phi\Uc_L^\dagger\; .  
\end{equation} 
Hence, the chiral symmetry breaking pattern \Ref{CSBPattern} is
realized if the meson potential develops a VEV
\begin{equation}
  \label{AAA50}
  \VEV{\Phi^{ab}}=\ol{\si}_0\dt^{ab}\;
  ;\;\;\; \ol{\si}_0\neq0\; .  
\end{equation} 
One of the most crucial and yet unsolved problems of strong
interaction dynamics is to derive an effective field theory for the
mesonic degrees of freedom directly from QCD which exhibits this
behavior.

\sect{A semi--quantitative picture}
\label{ASemiQuantitativePicture}

Before turning to a quantitative description of chiral symmetry
breaking using flow equations it is perhaps helpful to give a brief
overview of the relevant scales which appear in relation to this
phenomenon and the physical degrees of freedom associated to them.
Some of this will be explained in more detail in the remainder of
these lectures whereas other parts are rather well established
features of strong interaction physics.

At scales above approximately $1.5\GeV$, the relevant degrees of
freedom of strong interactions are quarks and gluons and their
dynamics appears to be well described by perturbative QCD. At somewhat
lower energies this changes dramatically. Quark and gluon bound states
form and confinement sets in. Concentrating on the physics of scalar
and pseudoscalar mesons~\footnote{One may assume that all other bound
  states are integrated out. We will comment on this issue below.}
there are three important momentum scales which appear to be rather
well separated:
\begin{itemize}
\item The compositeness scale $k_\Phi$ at which mesonic
  $\ol{\psi}\psi$ bound states form because of the increasing strength
  of the strong interaction. It will turn out to be somewhere in the
  range $(600-700)\MeV$.
\item The chiral symmetry breaking scale $k_{\chi SB}$ at which the
  chiral condensate $\VEV{\ol{\psi}^b\psi^a}$ or $\VEV{\Phi^{ab}}$
  assumes a non--vanishing value, therefore breaking chiral symmetry
  according to \Ref{CSBPattern}.  This scale is found to be around
  $(400-500)\MeV$. For $k$ below $k_{\chi SB}$ the quarks acquire
  constituent masses $M_q\simeq350\MeV$ due to their Yukawa coupling
  to the chiral condensate \Ref{AAA50}.
\item The confinement scale $\Lambda_{\rm QCD}\simeq200\MeV$ which
  corresponds to the Landau pole in the perturbative evolution of the
  strong coupling constant $\alpha_s$. In our context, this is the
  scale where possible deviations of the effective quark propagator
  from its classical form and multi--quark interactions not included
  in the meson physics may become very important.
\end{itemize}
For scales $k$ in the range $k_{\chi SB}\lta k\lta k_\Phi$ the most
relevant degrees of freedom are mesons and quarks. Typically, the
dynamics in this range is dominated by the strong Yukawa coupling $h$
between quarks and mesons: $h^2/(4\pi)\gg\alpha_s$.  One may therefore
assume that the dominant QCD effects are included in the meson physics
and consider a simple model of quarks and mesons only.  As one evolves
to scales below $k_{\chi SB}$ the Yukawa coupling decreases whereas
$\alpha_s$ increases. Of course, getting closer to $\Lambda_{\rm QCD}$
it is no longer justified to neglect the QCD effects which go beyond
the dynamics of effective meson degrees of freedom. On the other hand,
the final IR value of the Yukawa coupling $h$ is fixed by the typical
values of constituent quark masses $M_q\simeq350\MeV$ to be
$h^2/(4\pi)\simeq4.5$. One may therefore speculate that the domination
of the Yukawa interaction persists down to scales $k\simeq M_q$ at
which the quarks decouple from the evolution of the mesonic degrees of
freedom altogether due to their mass. Of course, details of the
gluonic interactions are expected to be crucial for an understanding
of quark and gluon confinement. Strong interaction effects may
dramatically change the momentum dependence of the quark $n$--point
functions for $k$ around $\Lambda_{\rm QCD}$.  Yet, as long as one is
only interested in the dynamics of the mesons one is led to expect
that these effects are quantitatively no too important. Because of the
effective decoupling of the quarks and therefore the whole colored
sector the details of confinement have only little influence on the
mesonic flow equations for $k\lta\Lambda_{\rm QCD}$. We conclude that
there are good prospects that the meson physics can be described by an
effective action for mesons and quarks for $k<k_\Phi$. The main part
of the work presented here is concerned with this effective quark
meson model.

In order to obtain this effective action at the compositeness scale
$k_\Phi$ from short distance QCD two steps have to be carried out. In
a first step one computes at the scale $k_p\simeq1.5\GeV$ an effective
action involving only quarks. This step integrates out the gluon
degrees of freedom in a ``quenched approximation''. More precisely,
one solves a truncated flow equation for QCD with quark and gluon
degrees of freedom in presence of an effective infrared cutoff
$k_p\simeq1.5\GeV$ in the quark propagators. The exact flow equation
to be used for this purpose is obtained by lowering the infrared
cutoff $R_k$ for the gluons to zero while keeping the one for the
quarks fixed. Subsequently, the gluons are eliminated by solving the
field equations for the gluon fields as functionals of the quarks.
This will result in a non--trivial momentum dependence of the quark
propagator and effective non--local four and higher quark
interactions. Because of the infrared cutoff $k_p$ the resulting
effective action for the quarks resembles closely the one for heavy
quarks (at least for Euclidean momenta). The dominant effect is the
appearance of an effective quark potential (similar to the one for the
charm quark) which describes the effective four--quark interactions.
For the effective quark action at $k_p$ we only retain this
four--quark interaction in addition to the two--point
function, while neglecting $n$--point functions involving six and more
quarks. 

\begin{figure}
\unitlength1.0cm
\begin{picture}(13.,8.)
\put(0.8,7.5){\bf $\ds{q^2V(q^2)}$}
\put(6.7,0.0){\bf $\sqrt{q^2}/\GeV$}
\put (5.5,7.5) {2-loop}
\put (2.75,3.3) {\rm Rich.}
\put (3.2,3.55) {\vector(1,1){1.05}}
\put (9.75,2.65) {\rm Rich.}
\put (9.7,2.6) {\vector(-1,-1){1.0}}
\put(1.5,0.5){
\epsfysize=8.cm
\epsffile{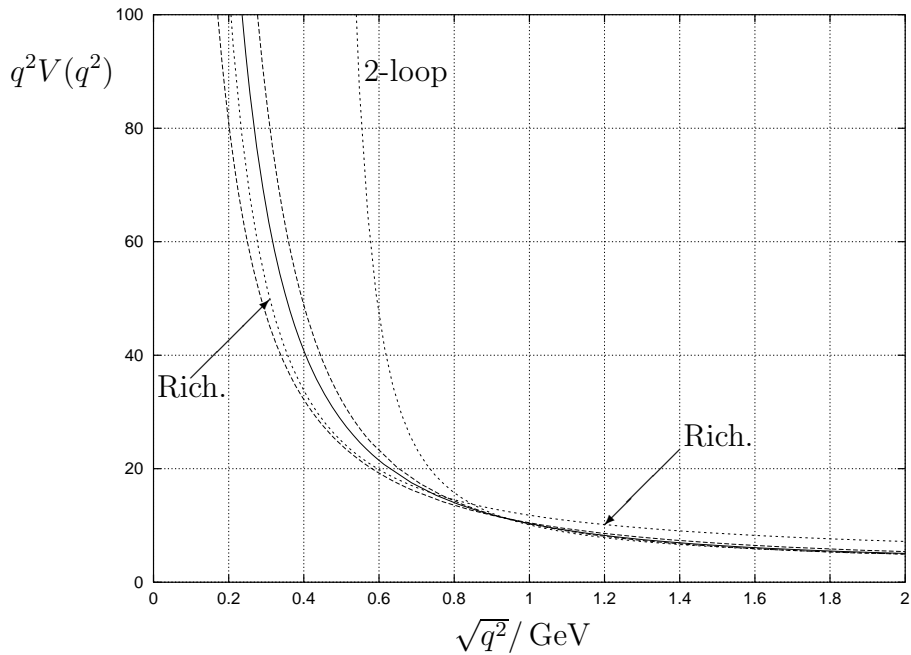}
}
\end{picture}
\caption{Heavy
  quark potential for different values of $\alpha_\star$.  The solid
  line corresponds to $\alpha_\star = 1.5$ whereas the dashed lines
  are for $\alpha_\star = 1$ (lower dashed line) and $\alpha_\star =
  2$ (upper dashed line).  Also shown are the two loop potential and
  the Richardson fit as dotted lines.}
\label{HQP} 
\end{figure}

For typical momenta larger than $\Lambda_{\rm QCD}$ a reliable
computation of the effective quark action should be possible by using
in the quark--gluon flow equation a relatively simple truncation. The
result~\cite{BBW97-1} for the Fourier transform of the potential
$V(q^2)$ is shown in figure~\ref{HQP}.
(See ref~\cite{BBW97-1} for the relation between the four--quark
interaction and the heavy quark potential which involves a rescaling
in dependence on $k_p$.)  A measure for the truncation uncertainties
is the parameter $\alpha_*$ which corresponds to the value to which an
appropriately defined running strong coupling $\alpha_s$ evolves for
$k\ra0$. We see that these uncertainties affect principally the
low--$q^2$ region. For high values of $q^2$ the potential is very
close to the perturbative two--loop potential whereas for intermediate
$q^2$ relevant for quarkonium spectra it is quite close to
phenomenologically acceptable potentials (e.g.~the Richardson
potential~\cite{Ric79-1}). The inverse quark propagator is found in
this computation to remain very well approximated by the simple
classical momentum dependence $\slash{q}$.

In the second step one has to lower the infrared cutoff in the
effective non--local quark model in order to extrapolate from $k_p$ to
$k_\Phi$.  This task can be carried out by means of the exact flow
equation for quarks only, starting at $k_p$ with an initial value
$\Gamma_{k_p}[\psi]$ as obtained after integrating out the gluons. For
fermions the trace in \Ref{ERGE} has to be replaced by a supertrace in
order to account for the minus sign related to Grassmann
variables~\cite{Wet90-1}.  A first investigation in this
direction~\cite{EW94-1} has used a truncation with a chirally invariant
four quark interaction whose most general momentum dependence was
retained
\begin{eqnarray}
 \ds{\Gm_k} &=& \ds{\int\frac{d^4 p}{(2\pi)^4}
 \ol{\psi}_a^i(p)Z_{\psi,k}(p)\left[
 \slash{p}\delta^{ab}+m^{ab}(p)\gamma_5+
 i\tilde{m}^{ab}(p)\right]\psi_{ib}(p)}\nnn 
  &+& \ds{
 \frac{1}{2}\int\left(\prod_{l=1}^4
 \frac{d^4 p_l}{(2\pi)^4}\right)
 \left(2\pi\right)^4\delta(p_1+p_2-p_3-p_4)}\nnn
 &\times& \ds{
 \la_k^{(\psi)}(p_1,p_2,p_3,p_4)
 \left\{
 \left[\ol{\psi}_a^i(-p_1)\psi_i^b(p_2)\right]
 \left[\ol{\psi}_b^j(p_4)\psi_j^a(-p_3)\right]
 \right. }\nnn
 && \ds{\left.\hspace{2.95cm}
 -\left[\ol{\psi}_a^i(-p_1)\gamma_5\psi_i^b(p_2)\right]
 \left[\ol{\psi}_b^j(p_4)\gamma_5\psi_j^a(-p_3)\right]
 \right\} }\; .
 \label{QCDFourFermi}
\end{eqnarray}
Here $i,j$ run from one to $N_c$ which is the number of quark colors.
The indices $a,b$ label the different light quark flavors and run from
$1$ to $N$. The matrices $m$ and $\tilde{m}$ are hermitian and
$m+i\tilde{m}\gamma_5$ forms therefore the most general quark mass
matrix. (Our chiral conventions~\cite{Wet90-1} where the hermitean part
of the mass matrix is multiplied by $\gamma_5$ may be somewhat unusual
but they are quite convenient for Euclidean calculations.) The ansatz
\Ref{QCDFourFermi} does not correspond to the most general chirally
invariant four--quark interaction. It neglects similar interactions in
the $\rho$--meson and pomeron channels which are also obtained from a
Fierz transformation of the heavy quark potential~\cite{Wet95-2}. With
$V(q^2)$ the heavy quark potential in a Fourier representation, the
initial value at $k_p=1.5\GeV$ was taken as
($\hat{Z}_{\psi,k}=Z_{\psi,k}(p^2=-k_p^2)$)
\begin{equation}
  \label{FFCBC}
  \la_{k_p}^{(\psi)}(p_1,p_2,p_3,p_4)
  \hat{Z}_{\psi,k_p}^{-2}=
  \frac{1}{2}V((p_1-p_3)^2)=
  \frac{2\pi\alpha_s}{(p_1-p_3)^2}+
  \frac{8\pi\la}{\left((p_1-p_3)^2\right)^2}\; .
\end{equation}
This corresponds to an approximation by a one gluon exchange term
$\sim\alpha_s(k_p)$ and a string tension $\la\simeq0.18\GeV^2$ and is
in reasonable agreement with the form computed recently~\cite{BBW97-1}
from the solution of flow equations (see fig\ref{HQP}).  In the
simplified ansatz \Ref{FFCBC} the string tension introduces a second
scale in addition to $k_p$ and it becomes clear that the incorporation
of gluon fluctuations is a crucial ingredient for the emergence of
mesonic bound states. For a more precise treatment~\cite{BBW97-1} of
the four--quark interaction at the scale $k_\Phi$ this second scale is
set by the running of $\alpha_s$ or $\Lambda_{\rm QCD}$.

The evolution equation for the function $\la_k^{(\psi)}$ for $k<k_p$
can be derived from the fermionic version of \Ref{ERGE} and the
truncation \Ref{QCDFourFermi}. Since $\la_k^{(\psi)}$ depends on six
independent momentum invariants it is a partial differential equation
for a function depending on seven variables and has to be solved
numerically~\cite{EW94-1}.  The ``initial value'' \Ref{FFCBC}
corresponds to the $t$--channel exchange of a ``dressed'' colored
gluonic state and it is by far not clear that the evolution of
$\la_k^{(\psi)}$ will lead at lower scales to a momentum dependence
representing the exchange of colorless mesonic bound states. Yet, at
the compositeness scale
\begin{equation}
\label{kphi}
  k_\Phi\simeq630\MeV
\end{equation}
one finds an approximate factorization
\begin{equation}
\label{BSFact}
  \la_{k_\Phi}^{(\psi)}(p_1,p_2,p_3,p_4)=
  g(p_1,p_2)\tilde{G}(s)g(p_3,p_4)+\ldots
\end{equation}
which indicates the formation of mesonic bound states.  Here
$g(p_1,p_2)$ denotes the amputated Bethe--Salpeter wave function and
$\tilde{G}(s)$ is the mesonic bound state propagator displaying a
pole--like structure in the $s$--channel if it is continued to
negative $s=(p_1+p_2)^2$. The dots indicate the part of
$\la_k^{(\psi)}$ which does not factorize and which will be
neglected in the following. In the limit where the momentum dependence
of $g$ and $\tilde{G}$ is neglected we recover the four--quark
interaction of the Nambu--Jona-Lasinio model~\cite{NJL61-1,Bij95-1}.
It is therefore not surprising that our description of the dynamics
for $k<k_\Phi$ will parallel certain aspects of other investigations
of this model, even though we are not bound to the approximations used
typically in such studies (large--$N_c$ expansion, perturbative
renormalization group, etc.).

It is clear that for scales $k\lta k_\Phi$ a description of strong
interaction physics in terms of quark fields alone would be rather
inefficient. Finding physically reasonable truncations of the
effective average action should be much easier once composite fields
for the mesons are introduced.  The exact renormalization group
equation can indeed be supplemented by an exact formalism for the
introduction of composite field variables or, more generally, a change
of variables~\cite{EW94-1}. For our purpose, this amounts in practice
to inserting at the scale $k_\Phi$ the identities
\begin{eqnarray}
 1 &=& \displaystyle{
 {\rm const}\;
 \int{\cal D}\si_A}\nnn
 && \ds{\hspace{-1.2cm}\times
 \exp\left\{ -\tr
 \left(\si_A^\dagger -
 K_A^\dagger \tilde{G}-m_A^\dagger-
 {\cal O}^\dagger \tilde{G}\right)
 \frac{1}{2\tilde{G}}
 \left(\si_A -\tilde{G}K_A -m_A-
 \tilde{G}{\cal O} \right)\right\} }\nnn
 \label{identity}
  1 &=& \displaystyle{
 {\rm const}\;
 \int{\cal D}\si_H}\\[2mm]
 && \displaystyle{\hspace{-1.2cm}\times
 \exp\left\{-\tr
 \left(\si_H^\dagger -
 K_H^\dagger \tilde{G}-m_H^\dagger-
 {\cal O}^{(5)\dagger} \tilde{G}\right)
 \frac{1}{2\tilde{G}}\left(\si_H -\tilde{G}K_H -m_H-
 \tilde{G}{\cal O}^{(5)} \right)\right\} }\nonumber
\end{eqnarray}
into the functional integral which formally defines the quark
effective average action. Here we have used the shorthand notation
$A^\dagger G B\equiv\int\frac{d^d q}{(2\pi)^d}A_a^*(q)G^{ab}(q)
B_b(q)$, and $K_{A,H}$ are sources for the collective fields
$\si_{A,H}$ which correspond in turn to the anti-hermitian and
hermitian parts of the meson field $\Phi$. They are associated to the
fermion bilinear operators ${\cal O} [\psi]$, ${\cal O}^{(5)}[\psi ]$
whose Fourier components read
\begin{eqnarray}
 \ds{{\cal O}_{\;\; b}^a (q)} &=& \ds{ 
 -i\int\frac{d^4 p}{(2\pi)^4} g(-p,p+q)
 \ol{\psi}^a (p)\psi_b (p+q) }\nnn
 \ds{{\cal O}_{\;\;\;\;\;\; b}^{(5)a} (q)} &=& \ds{ 
 -\int\frac{d^4 p}{(2\pi)^4} g(-p,p+q)
 \ol{\psi}^a (p)\gamma_5\psi_b (p+q) }\; .
\end{eqnarray}
The choice of $g(-p,p+q)$ as the bound state wave function
renormalization and of $\tilde{G}(q)$ as its propagator guarantees
that the four--quark interaction contained in \Ref{identity} cancels
the dominant factorizing part of the QCD--induced non--local
four--quark interaction Eqs.(\ref{QCDFourFermi}), (\ref{BSFact}). In
addition, one may choose
\begin{eqnarray}
 \ds{m_{Hab}^T} &=& \ds{
 m_{ab}(0)g^{-1}(0,0)Z_{\psi,k_\Phi}(0)}\nnn
 \ds{m_{Aab}^T} &=& \ds{
 \tilde{m}_{ab}(0)g^{-1}(0,0)Z_{\psi,k_\Phi}(0)}
\end{eqnarray}
such that the explicit quark mass term cancels out for $q=0$. The
remaining quark bilinear is $\sim m(q)-m(0)Z_{\psi,k_\Phi}(0)g(-q,q)/
[Z_{\psi,k_\Phi}(q)g(0,0)]$. It vanishes for zero momentum and will be
neglected in the following. Without loss of generality we can take $m$
real and diagonal and $\tilde{m}=0$. 

In consequence, we have replaced at the scale $k_\Phi$ the effective
quark action \Ref{QCDFourFermi} with \Ref{BSFact} by an effective
quark meson action given by
\begin{eqnarray}
 \ds{\hat{\Gamma}_k} &=& \ds{
 \Gamma_k-\frac{1}{2}\int d^4 x\tr
 \left(\Phi^\dagger\jmath+\jmath^\dagger\Phi\right)}\nnn
 \ds{\Gamma_{k}} &=& \displaystyle{ \int d^4 x
 U_k(\Phi,\Phi^\dagger)
  \label{EffActAnsatz}
 }\\[2mm]
 &+& \displaystyle{ 
 \int\frac{d^4 q}{(2\pi)^d}\Bigg\{
 Z_{\Phi,k}(q) q^2 \tr\left[
 \Phi^\dagger (q)\Phi (q)\right] +
 Z_{\psi,k}(q)\ol{\psi}_a(q)
 \gamma^\mu q_\mu \psi^a (q)
 }\nonumber\vspace{.2cm}\\
 &+& \displaystyle{
 \int\frac{d^4 p}{(2\pi)^d}\ol{h} _k (-q,q-p)}\nnn
 &\times& \ds{
 \ol{\psi}^a(q) \left(
 \frac{1+\gamma_5}{2}\Phi _{ab}(p)-
 \frac{1-\gamma_5}{2}\Phi_{ab}^\dagger (-p) \right)
 \psi^b (q-p) \Bigg\}\nonumber \; .}
\end{eqnarray}
At the scale $k_\Phi$ the inverse scalar propagator is related to
$\tilde{G}(q)$ in \Ref{BSFact} by
\begin{equation}
 \tilde{G}^{-1}(q^2) = 2\ol{m}^2_{k_\Phi} +
 2Z_{\Phi,k_\Phi}(q)q^2\; .
\end{equation}
This fixes the term in $U_{k_\Phi}$ which is quadratic in $\Phi$ to be
positive, $U_{k_\Phi}=\ol{m}^2_{k_\Phi}\tr\Phi^\dagger\Phi+\ldots$.
The higher order terms in $U_{k_\Phi}$ cannot be determined in the
approximation \Ref{QCDFourFermi} since they correspond to terms
involving six or more quark fields.  The initial value of the Yukawa
coupling corresponds to the ``quark wave function in the meson'' in
\Ref{BSFact}, i.e.
\begin{equation}
 \ol{h} _{k_\Phi}(-q,q-p) = g(-q,q-p)
\end{equation}
which can be normalized with $\ol{h}_{k_\Phi}(0,0)=g(0,0)=1$.
We observe that the explicit chiral symmetry breaking from
non--vanishing current quark masses appears now in the form of a meson
source term with
\begin{equation}
  \label{AAA22}
  \jmath=2\ol{m}^2_{k_\Phi} Z_{\psi,k_\Phi}(0)
  g^{-1}(0,0)\left(
  m_{ab}+i\tilde{m}_{ab}\right)=
  2Z_{\psi,k_\Phi}\ol{m}^2_{k_\Phi}
  {\rm diag}(m_u,m_d,m_s)\; .
\end{equation}
This induces a non--vanishing $\VEV{\Phi}$ and an effective quark mass
$M_q$ through the Yukawa coupling. We note that the current quark mass
$m_q$ and the constituent quark mass $M_q\sim\ol{h}_k\VEV{\Phi}$ are
identical at the scale $k_\Phi$. (By solving the field equation for
$\Phi$ as a functional of $\ol{\psi}$, $\psi$ (with
$U_k=\ol{m}^2_k\tr\Phi^\dagger\Phi$) one recovers from
\Ref{EffActAnsatz} the effective quark action \Ref{QCDFourFermi}.  For
a generalization beyond the approximation of a four--quark interaction
or a quadratic potential see ref~\cite{BJW97-1}.)  Spontaneous chiral
symmetry breaking can be described in this language by a
non--vanishing $\VEV{\Phi}$ in the limit $\jmath\ra0$. Because of
spontaneous chiral symmetry breaking the constituent quark mass $M_q$
can differ from zero even for $m_q=0$. It is a nice feature of our
formalism that it provides for a unified description of the concepts
of the current and the constituent quark masses. As long as the
effective average potential $U_k$ has a unique minimum at $\Phi=0$
there is simply no difference between the two. The running of the
current quark mass in the pure quark model should be equivalent in the
quark meson language to the running of $\ol{h}_k\VEV{\Phi}(k)$.  (A
verification of this property would actually provide a good check for
the truncation errors.) Nevertheless, the formalism is now adapted to
account for the quark mass contribution from chiral symmetry breaking
since the absolute minimum of $U_k$ may be far from the perturbative
one for small $k$. We also note that $\Gamma_k$ in \Ref{EffActAnsatz}
is chirally invariant.  The explicit chiral symmetry breaking in
$\hat{\Gamma}_k$ appears only in the form of a linear source term
which is independent of $k$ and does not affect the flow of
$\Gamma_k$.  The flow equation for $\Gamma_k$ therefore respects
chiral symmetry even in the presence of quark masses. This leads to a
considerable simplification.

At the scale $k_\Phi$ the propagator $\tilde{G}$ and the wave function
$g(-q,q-p)$ should be optimized for a most complete elimination of
terms quartic in the quark fields. In the present context we will,
however, neglect the momentum dependence of $Z_{\psi,k}$, $Z_{\Phi,k}$
and $\ol{h}_k$.  The mass $\ol{m}_{k_\Phi}$ was found in~\cite{EW94-1}
for the simple truncation \Ref{QCDFourFermi} with $Z_{\psi,k_\Phi}=1$,
$m=\tilde{m}=0$ to be $\ol{m}_{k_\Phi}\simeq120\MeV$.  In view of the
possible large truncation errors we will take this only as an order of
magnitude estimate. Below we will consider the range
$\ol{m}_{k_\Phi}=(45-120)\MeV$ for which chiral symmetry breaking can
be obtained in a two flavor model.  Furthermore, we will assume, as
usually done in large--$N_c$ computations within the NJL--model, that
$Z_{\Phi,k_\Phi}\equiv Z_{\Phi,k_\Phi}(q=0)\ll1$.  The quark wave
function renormalization $Z_{\psi,k}\equiv Z_{\psi,k}(q=0)$ is set to
one at the scale $k_\Phi$ for convenience.  For $k<k_\Phi$ we will
therefore study an effective action for quarks and mesons in the
truncation
\begin{eqnarray}
  \ds{\Gm_k} &=& \ds{
    \int d^4x\Bigg\{
    Z_{\psi,k}\ol{\psi}_a i\slash{\prl}\psi^a+
    Z_{\Phi,k}\tr\left[\prl_\mu\Phi^\dagger\prl^\mu\Phi\right]+
    U_k(\Phi,\Phi^\dagger)
    }\nnn
  &+& \ds{
    \ol{h}_k\ol{\psi}^a\left(\frac{1+\gm_5}{2}\Phi_{ab}-
    \frac{1-\gm_5}{2}(\Phi^\dagger)_{ab}\right)\psi^b
    \Bigg\} }
  \label{GammaEffective}
\end{eqnarray}
with compositeness conditions
\begin{eqnarray}
 \ds{U_{k_\Phi}(\Phi,\Phi^\dagger)} &=& \ds{
 \ol{m}^2_{k_\Phi}\tr\Phi^\dagger\Phi-
 \frac{1}{2}\ol{\nu}_{k_\Phi}\left(
 \det\Phi+\det\Phi^\dagger\right)}\nnn
 &+& \ds{
 \frac{1}{2}\ol{\la}_{1,k_\Phi}\left(\tr\Phi^\dagger\Phi\right)^2+
 \frac{N-1}{4}\ol{\la}_{2,k_\Phi}\tr
 \left(\Phi^\dagger\Phi-\frac{1}{N}\tr\Phi^\dagger\Phi\right)^2+
 \ldots}\nnn
 \ds{\ol{m}^2_{k_\Phi}} &\equiv& \ds{\frac{1}{2\tilde{G}(0)}\simeq
 (45\MeV)^2-(120\MeV)^2}\nnn
 \ds{\ol{h}_{k_\Phi}} &=&
 Z_{\psi,k_\Phi}=1\nnn
 \ds{Z_{\Phi,k_\Phi}} &\ll& 1\; .
 \label{CompositenessConditions}
\end{eqnarray}
As a consequence, the initial value of the renormalized Yukawa
coupling which is given by
$h_{k_\Phi}=\ol{h}_{k_\Phi}Z_{\psi,k_\Phi}^{-1}Z_{\Phi,k_\Phi}^{-1/2}$
is large!  Note that we have included in the potential an explicit
$U_A(1)$ breaking term~\footnote{The anomaly term in the fermionic
  effective average action has been computed in~\cite{Paw96-1}.}
$\sim\ol{\nu}_k$ which mimics the effect of the chiral anomaly of QCD
to leading order in an expansion of the effective potential in powers
of $\Phi$. Because of the infrared stability of the evolution of
$\Gamma_k$ which will be discussed below the precise form of the
potential, i.e.~the values of the quartic couplings
$\ol{\la}_{i,k_\Phi}$ and so on, will turn out to be unimportant.

We have refrained here for simplicity from considering four quark
operators with vector and pseudo--vector spin structure.  Their
inclusion is straightforward and would lead to vector and
pseudo--vector mesons in the effective action \Ref{GammaEffective}.
We will concentrate first on two flavors and consider only the two
limiting cases $\ol{\nu}_k=0$ and $\ol{\nu}_k\ra\infty$.  We also omit
first the explicit quark masses and study the chiral limit $\jmath=0$.
Because of the positive mass term $\ol{m}^2_{k_\Phi}$ one has at the
scale $k_\Phi$ a vanishing expectation value $\VEV{\Phi}=0$ (for
$\jmath=0$). There is no spontaneous chiral symmetry breaking at the
compositeness scale.  This means that the mesonic bound states at
$k_\Phi$ and somewhat below are not directly connected to chiral
symmetry breaking.

The question remains how chiral symmetry is broken. We will try to
answer it by following the evolution of the effective potential $U_k$
from $k_\Phi$ to lower scales using the exact renormalization group
method outlined above with the compositeness conditions
\Ref{CompositenessConditions} defining the initial values.  In this
context it is important that the formalism for composite
fields~\cite{EW94-1} also induces an infrared cutoff in the meson
propagator. The flow equations are therefore exactly of the form
\Ref{ERGE} (except for the supertrace), with quarks and mesons
treated on an equal footing.  In fact, one would expect that the large
renormalized Yukawa coupling will rapidly drive the scalar mass term
to negative values as the IR cutoff $k$ is lowered~\cite{Wet90-1}.
This will then finally lead to a potential minimum away from the
origin at some scale $k_{\rm \chi SB}<k_\Phi$ such that
$\VEV{\Phi}\neq0$. The ultimate goal of such a procedure, besides from
establishing the onset of chiral symmetry breaking, would be to
extract phenomenological quantities, like $f_\pi$ or meson masses,
which can be computed in a straightforward manner from $\Gamma_k$ in
the IR limit $k\ra0$.

At first sight, a reliable computation of $\Gamma_{k\ra0}$ seems a
very difficult task. Without a truncation $\Gamma_k$ is described by
an infinite number of parameters (couplings, wave function
renormalizations, etc.) as can be seen if $\Gamma_k$ is expanded in
powers of fields and derivatives. For instance, the pseudoscalar and
scalar meson masses are obtained as the poles of the exact propagator,
$\lim_{k\ra0}\Gamma_k^{(2)}(q)|_{\Phi=\VEV{\Phi}}$, which receives
formally contributions from terms in $\Gamma_k$ with arbitrarily high
powers of derivatives and the expectation value $\si_0$.  Realistic
nonperturbative truncations of $\Gamma_k$ which reduce the problem to
a manageable size are crucial.  We will argue in the following that
there may be a twofold solution to this problem:
\begin{itemize}
\item Due to an IR fixed point structure of the flow equations in the
  symmetric regime, i.e. for $k_{\chi SB}<k<k_\Phi$, the values of
  many parameters of $\Gamma_k$ for $k\ra0$ will be approximately
  independent of their initial values at the compositeness scale
  $k_{\Phi}$. For small enough $Z_{\Phi,k_\Phi}$ only a few relevant
  parameters ($\ol{m}^2_{k_\Phi}$, $\ol{\nu}_{k_\Phi}$) need to be
  computed accurately from QCD. They can alternatively be determined
  from phenomenology.
\item One can show that physical observables like meson masses, decay
  constants, etc., can be expanded in powers of the quark masses
  within the linear meson model~\cite{JW96-1,JW96-3,JW97-1}. This is
  similar to the way it is usually done in chiral perturbation
  theory~\cite{Leu95-1}. To a given finite order of this expansion only
  a finite number of terms of a simultaneous expansion of $\Gamma_k$
  in powers of derivatives and $\Phi$ are required if the expansion
  point is chosen properly.
\end{itemize}
In combination, these two results open the possibility for a perhaps
unexpected degree of predictive power within the linear meson model.

We wish to stress, though, that a perturbative treatment of the model
at hand, e.g., using perturbative RG techniques, cannot be expected to
yield reliable results. The renormalized Yukawa coupling is expected
to be large at the scale $k_\Phi$. Even the IR value of $h$ is still
relatively big
\begin{equation} 
  \label{IRh}
  \lim_{k\ra0}h=\frac{2M_q}{f_\pi}\simeq7.5
\end{equation} 
and $h$ increases with increasing $k$. The dynamics of the linear
meson model is therefore clearly nonperturbative for all scales $k\leq
k_\Phi$.

\sect{Flow equations for the linear quark meson model}

We will next turn to the ERGE analysis of the linear meson model which
was introduced in the last section\footnote{For a study of chiral
  symmetry breaking in QED using related exact renormalization group
  techniques see ref~\cite{AMST97-1}.}.  In a first approach we will
attack the problem at hand by truncating $\Gm_k$ in such a way that it
contains all perturbatively relevant and marginal operators, i.e.
those with canonical dimensions. This has the advantage that the flow
of a small number of couplings permits quantitative insight into the
relevant mechanisms, e.g., of chiral symmetry breaking. More
quantitative precision will be obtained once we generalize our
truncation (see below) for the $O(4)$--model by allowing for the most
general effective average potential $U_k$.  The effective potential
$U_k$ is a function of only four $\csN$ invariants for $N=3$:
\begin{eqnarray}
  \label{Invariants}
  \ds{\rho} &=&
  \ds{\tr\Phi^\dagger\Phi}\nnn \ds{\tau_2} &=& \ds{
  \frac{N}{N-1}\tr\left(\Phi^\dagger\Phi- \frac{1}{N}\rho\right)^2}\nnn
  \ds{\tau_3} &=& \ds{ \tr\left(\Phi^\dagger\Phi-
  \frac{1}{N}\rho\right)^3}\nnn \ds{\xi} &=&
  \ds{\det\Phi+\det\Phi^\dagger}\; .  
\end{eqnarray}
The invariant $\tau_3$ is only independent for $N\ge3$. For $N=2$ it
can be eliminated by a suitable combination of $\tau_2$ and $\rho$.
The additional $U_A(1)$ breaking invariant
$\omega=i(\det\Phi-\det\Phi^\dagger)$ is $\Cc\Pc$ violating and may
therefore appear only quadratically in $U_k$. It is straightforward to
see that $\omega^2$ is expressible in terms of the invariants
\Ref{Invariants}.  We may expand $U_k$ as a function of these
invariants around its minimum, i.e. $\rho=\rho_0\equiv N\ol{\si}_0^2$,
$\xi=\xi_0=2\ol{\si}_0^N$ and $\tau_2=\tau_3=0$ where
\begin{equation}
  \label{Sigma_0}
  \ol{\si}_0\equiv\frac{1}{N}\tr\VEV{\Phi}\; .
\end{equation}
The expansion coefficients are the $k$--dependent couplings of the
model. In our first version we only keep couplings of canonical
dimension $d_c\leq4$. This yields in the chirally symmetric regime,
i.e., for $k_{\rm\chi SB}\leq k\leq k_\Phi$ where $\ol{\si}_0=0$
\begin{equation}
  \label{SymmetricPotential}
  U_k=\ol{m}^2_k\rho+\hal\ol{\la}_{1,k}\rho^2+
  \frac{N-1}{4}\ol{\la}_{2,k}\tau_2-
  \frac{1}{2}\ol{\nu}_k\xi
\end{equation}
whereas in the SSB regime for $k\leq k_{\rm\chi SB}$ we have
\begin{equation}
 U_k=\hal\ol{\la}_{1,k}\left[\rho-N\ol{\si}_{0.k}^2\right]^2+
 \frac{N-1}{4}\ol{\la}_{2,k}\tau_2+
 \hal\ol{\nu}_k\left[\ol{\si}_{0,k}^{N-2}\rho-\xi\right]\, .
 \label{EffPotentialSSB}
\end{equation}

Before continuing to compute the nonperturbative beta functions for
these couplings it is worthwhile to pause here and emphasize that
naively (perturbatively) irrelevant operators can by no means always
be neglected. The most prominent example for this is QCD itself. It is
the very assumption of our treatment of chiral symmetry breaking
(substantiated by the results of~\cite{EW94-1}) that the momentum
dependence of the coupling constants of some six--dimensional quark
operators $(\ol{\psi}\psi)^2$ develop poles in the $s$--channel
indicating the formation of mesonic bound states. On the other hand,
it is quite natural to assume that $\Phi^6$ or $\Phi^8$ operators are
not really necessary to understand the properties of the potential in
a neighborhood around its minimum.  Yet, truncating higher dimensional
operators does not imply the assumption that the corresponding
coupling constants are small. In fact, this could only be expected as
long as the relevant and marginal couplings are small as well. What is
required, though, is that their {\em influence} on the evolution of
those couplings kept in the truncation, for instance, the set of
equations \Ref{FlowEquationsSSB} below, is small.  A comparison with
the results of the later sections for the full potential $U_k$ (i.e.,
including arbitrarily many couplings in the formal expansion) will
provide a good check for the validity of this assumption.  In this
context it is perhaps also interesting to note that the truncation
\Ref{GammaEffective} includes the known one--loop beta functions of a
small coupling expansion as well as the leading order result of the
large--$N_c$ expansion of the $U_L(N)\times U_R(N)$
model~\cite{BHJ94-1}. This should provide at least some minimal control
over this truncation, even though we believe that our results are
significantly more accurate.

Inserting the truncation Eqs.(\ref{GammaEffective}),
(\ref{SymmetricPotential}), (\ref{EffPotentialSSB}) into \Ref{ERGE}
reduces this functional differential equation for infinitely many
variables to a finite set of ordinary differential equations. This
yields, in particular, the beta functions for the couplings
$\ol{\la}_{1,k}$, $\ol{\la}_{2,k}$, $\ol{\nu}_k$ and $\ol{m}^2_k$ or
$\ol{\si}_{0,k}$. Details of the calculation can be found
in~\cite{JW95-1}. We will refrain here from presenting the full set of
flow equations but rather illustrate the main results with a few
examples. Defining dimensionless renormalized VEV and coupling
constants
\begin{eqnarray}
  \ds{\kappa} &=& \ds{Z_{\Phi,k}
    N\ol{\si}_{0,k}^2 k^{-2}}\nnn 
  \ds{h^2} &=& \ds{
    \ol{h}_k^2 Z_{\Phi,k}^{-1}Z_{\psi,k}^{-2}}\nnn 
  \ds{\la_i} &=& \ds{
    \ol{\la}_{i,k} Z_{\Phi,k}^{-2}\; ;\;\;\; i=1,2}\nnn 
  \ds{\nu} &=& \ds{
    \ol{\nu}_k Z_{\Phi,k}^{-\frac{N}{2}}k^{N-4}}
  \label{DimensionlessCouplings}
\end{eqnarray}
one finds, e.g., for the spontaneous symmetry breaking (SSB) regime
and $\ol{\nu}_k=0$ 
\begin{eqnarray}
  \ds{ \frac{d}{d t}\kappa } &=& \ds{
    -(2+\eta_\Phi )\kappa + \frac{1}{16\pi^2} \Bigg\{ N^2l_1^4(0;\eta_\Phi)
    +3l_1^4 (2\la_1 \kappa;\eta_\Phi) }\nnn 
  &+& \ds{ 
    (N^2-1)\left[
    1+\frac{\la_2}{\la_1}\right] l_1^4 (\la_2\kappa;\eta_\Phi)-4N_c
    \frac{h^2}{\la_1} 
    l_{1}^{(F)4} (\frac{1}{N}h^2 \kappa;\eta_\psi) \Bigg\} }\nnn
  \ds{\frac{d}{d t}\la_1 } &=& \ds{ 
    2\eta_\Phi \la_1
    +\frac{1}{16\pi^2} \Bigg\{ N^2\la_1^2 
    l_2^4(0;\eta_\Phi) +9\la_1^2 l_2^4
    (2\la_1\kappa;\eta_\Phi) }\nnn 
  &+& \ds{ 
    (N^2-1)\left[\la_1 +\la_2\right]^2
    l_2^4 (\la_2\kappa;\eta_\Phi)-
    4\frac{N_c}{N}h^4 l_{2}^{(F)4}
    (\frac{1}{N}h^2\kappa;\eta_\psi) 
    \Bigg\} \label{FlowEquationsSSB} }\\[2mm]
  \ds{\frac{d}{d t}\la_2 } &=& \ds{ 
    2\eta_\Phi \la_2
    +\frac{1}{16\pi^2} \Bigg\{ \frac{N^2}{4}\la_2^2 l_2^4(0;\eta_\Phi) +
    \frac{9}{4}(N^2-4)\la_2^2 l_2^4 (\la_2\kappa;\eta_\Phi) }\nnn 
  &-& \ds{ 
    \hal N^2 \la_2^2 l_{1,1}^4(0,\la_2\kappa;\eta_\Phi) + 
    3[\la_2+4\la_1]\la_2
    l_{1,1}^4(2\la_1\kappa,\la_2\kappa;\eta_\Phi) }\nnn 
  &-& \ds{ 8\frac{N_c}{N}h^4
    l_{2}^{(F)4} (\frac{1}{N}h^2\kappa;\eta_\psi)\Bigg\} }\nnn 
  \ds{\frac{d}{d t} h^2} &=& \ds{ 
    \left[ d-4+2\eta_\psi +\eta_\Phi\right] h^2 -
      \frac{4}{N}v_d h^4 \left\{
      N^2 l_{1,1}^{(FB)d} (\frac{1}{N}\kappa h^2,0;
      \eta_\psi,\eta_\Phi) \right. }\nonumber\vspace{.2cm}\\ 
  &-& \ds{ 
    (N^2-1)l_{1,1}^{(FB)d} (\frac{1}{N}\kappa
    h^2,\kappa\la_2; \eta_\psi,\eta_\Phi)}\nnn
  &-& \ds{\left.
    l_{1,1}^{(FB)d} (\frac{1}{N}\kappa h^2,2\kappa\la_1;
    \eta_\psi,\eta_\Phi) \right\}\nonumber } 
\end{eqnarray} 
Here $\eta_\Phi=-\frac{d}{d t}\ln Z_{\Phi,k}$, $\eta_\psi=-\frac{d}{d
  t}\ln Z_{\psi,k}$ are the meson and quark anomalous dimensions,
respectively~\cite{JW95-1}.  The symbols $l_n^d$, $l_{n_1,n_2}^d$ and
$l_n^{(F)d}$ denote bosonic and fermionic mass threshold functions,
respectively, which are defined in~\cite{JW95-1}.  They describe the
decoupling of massive modes and provide an important nonperturbative
ingredient. For instance, the bosonic threshold functions
\begin{equation}
 \label{AAA85n}
 l_n^d(w;\eta_\Phi)=\frac{n+\delta_{n,0}}{4}v_d^{-1}
 k^{2n-d}\int\frac{d^d q}{(2\pi)^d}
 \frac{1}{Z_{\Phi,k}}\frac{\prl R_k}{\prl t}
 \frac{1}{\left[ P(q^2)+k^2w\right]^{n+1}}
\end{equation}
involve the inverse average propagator
$P(q^2)=q^2+Z_{\Phi,k}^{-1}R_k(q^2)$ where the infrared cutoff is
manifest. These functions decrease $\sim w^{-(n+1)}$ for $w\gg1$.
Since typically $w=M^2/k^2$ with $M$ a mass of the model, the main
effect of the threshold functions is to cut off fluctuations of
particles with masses $M^2\gg k^2$. Once the scale $k$ is changed
below a certain mass threshold, the corresponding particle no longer
contributes to the evolution and decouples smoothly.

Within our truncation the beta functions \Ref{FlowEquationsSSB} for
the dimensionless couplings look almost the same as in one--loop
perturbation theory. There are, however, two major new ingredients
which are crucial for our approach: First, there is a new equation for
the running of the mass term in the symmetric regime or for the
running of the potential minimum in the regime with spontaneous
symmetry breaking. This equation is related to the quadratic
divergence of the mass term in perturbation theory and does not appear
in the Callan--Symanzik~\cite{CS70-1} or Coleman--Weinberg~\cite{CW73-1}
treatment of the renormalization group. Obviously, these equations are
the key for a study of the onset of spontaneous chiral symmetry
breaking as $k$ is lowered from $k_\Phi$ to zero.  Second, the most
important nonperturbative ingredient in the flow equations for the
dimensionless Yukawa and scalar couplings is the appearance of
effective mass threshold functions like \Ref{AAA85n} which account
for the decoupling of modes with masses larger than $k$.  Their form
is different for the symmetric regime (massless fermions, massive
scalars) or the regime with spontaneous symmetry breaking (massive
fermions, massless Goldstone bosons).  Without the inclusion of the
threshold effects the running of the couplings would never stop and no
sensible limit $k\rightarrow0$ could be obtained because of unphysical
infrared divergences. The threshold functions are not arbitrary but
have to be computed carefully. The mass terms appearing in these
functions involve the dimensionless couplings. Expanding the threshold
functions in powers of the mass terms (or the dimensionless couplings)
makes their non--perturbative content immediately visible.  It is
these threshold functions which will make it possible below to use
one--loop type formulae for the necessarily nonperturbative
computation of the critical behavior of the (effectively)
three--dimensional $O(4)$--symmetric scalar model.

\sect{The chiral anomaly and the  $O(4)$--model}
\label{ChiralAnomaly}

We have seen how the mass threshold functions in the flow equations
describe the decoupling of heavy modes from the evolution of
$\Gamma_k$ as the IR cutoff $k$ is lowered. In the chiral limit with
two massless quark flavors ($N=2$) the pions are the massless
Goldstone bosons.  Below, once we will consider the flow of the full
effective average potential without resorting to a quartic truncation,
we will also consider the more general case of non--vanishing current
quark masses. For the time being, the effect of the physical pion mass
of $m_\pi\simeq140\MeV$, or equivalently of the two small but
non--vanishing current quark masses, can easily be mimicked by
stopping the flow of $\Gamma_k$ at $k=m_\pi$ by hand.  This situation
changes significantly once the strange quark is included. Now the
$\eta$ and the four $K$ mesons appear as additional massless Goldstone
modes in the spectrum. They would artificially drive the running of
$\Gamma_k$ at scales $m_\pi\lta k\lta500\MeV$ where they should
already be decoupled because of their physical masses. It is therefore
advisable to focus on the two flavor case $N=2$ as long as the chiral
limit of vanishing current quark masses is considered.

It is straightforward to obtain an estimate of the (renormalized)
coupling $\nu$ which parameterizes the explicit $U_A(1)$ breaking due
to the chiral anomaly. From \Ref{EffPotentialSSB} we find
\begin{equation}
  \label{AAA01}
  m_{\eta^\prime}^2=\frac{N}{2}\nu\si_0^{N-2}\simeq1\GeV
\end{equation}
which translates for $N=2$ for $k\ra0$ into
\begin{equation}
  \label{AAA02}
  \nu(k\ra0)\simeq1\GeV\; .
\end{equation}
This suggests that $\nu\ra\infty$ can be considered as a realistic
limit.  An important simplification occurs for $N=2$ and
$\nu\ra\infty$, related to the fact that for $N=2$ the chiral group
$SU_L(2)\times SU_R(2)$ is (locally) isomorphic to $O(4)$. Thus, the
complex $(\bf 2,\bf 2)$ representation $\Phi$ of $SU_L(2)\times
SU_R(2)$ may be decomposed into two vector representations,
$(\si,\pi^k)$ and $(\eta^\prime,a^k)$ of $O(4)$:
\begin{equation}
 \Phi=\hal\left(\si-i\eta^\prime\right)+
 \hal\left( a^k+i\pi^k\right)\tau_k \; .
\end{equation}
For $\nu\ra\infty$ the masses of the $\eta^\prime$ and the $a^k$
diverge and these particles decouple. We are then left with the
original $O(4)$ symmetric linear $\si$--model of Gell--Mann and
Levy~\cite{GML60-1} coupled to quarks. The flow equations of this model
have been derived previously~\cite{Wet90-1,Wet91-1} for the truncation
of the effective action used here. For mere comparison we also
consider the opposite limit $\nu\ra0$. Here the $\eta^\prime$ meson
becomes an additional Goldstone boson in the chiral limit which
suffers from the same problem as the $K$ and the $\eta$ in the case
$N=3$.  Hence, we may compare the results for two different
approximate limits of the effects of the chiral anomaly:
\begin{itemize}
\item the $O(4)$ model corresponding to $N=2$ and $\nu\ra\infty$
\item the $U_L(2)\times U_R(2)$ model corresponding to $N=2$ and
  $\nu=0$.
\end{itemize}
For the reasons given above we expect the first situation to be closer
to reality. In this case we may imagine that the fluctuations of the
kaons, $\eta$, $\eta^\prime$ and the scalar mesons (as well as vector
and pseudovector mesons) have been integrated out in order to obtain
the initial values of $\Gamma_{k_\Phi}$ --- in close analogy to the
integration of the gluons for the effective quark action
$\Gamma_{k_p}[\psi]$ discussed above. We will keep the initial values
of the couplings $\ol{m}^2_{k_\Phi}$, $\ol{\la}_{1,k_\Phi}$ and
$Z_{\Phi,k_\Phi}$ as free parameters. Our results should be
quantitatively accurate to the extent to which the local polynomial
truncation is a good approximation.

\sect{Infrared stability}
\label{InfraredStability}

\begin{figure}
\unitlength1.0cm
\begin{picture}(13.,8.)
\put(-0.8,5.){\bf $\ds{\frac{m,\si_0}{\MeV}}$}
\put(6.3,0.5){\bf $k/\MeV$}
\put(4.3,3.5){\bf $\si_0$}
\put(9.8,5.){\bf $m$}
\put(-2.0,-11.5){
\epsfysize=22.cm
\epsffile{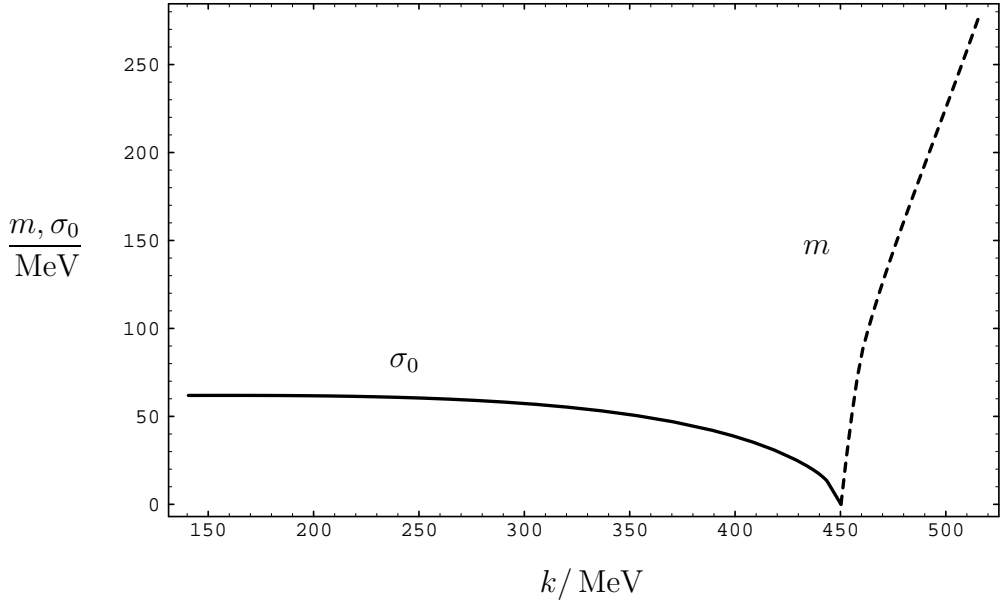}
}
\end{picture}
\caption{Evolution of the renormalized mass $m$ in the symmetric
  regime (dashed line) and the vacuum expectation value
  $\si_0=Z_{\Phi,k}^{1/2}\ol{\si}_{0,k}$ of the scalar field in the
  SSB regime (solid line) as functions of $k$ for the $U_L(2)\times
  U_R(2)$ model. Initial values at $k=k_\Phi$ are $\la_{i,I}=0$,
  $h^2_I=300$ and $\ol{m}_{k_\Phi}=63\MeV$.}
\label{Fig1} 
\end{figure}

\Ref{FlowEquationsSSB} and the corresponding set of flow
equations for the symmetric regime constitute a coupled system of
ordinary differential equations which can be integrated numerically.
Similar equations can be computed for the $O(4)$ model where $N=2$ and
the coupling $\la_2$ is absent.  We have neglected in a first step the
dependence of all threshold functions appearing in the flow equations
on the anomalous dimensions. This dependence will be taken into
account below once we abandon the quartic potential approximation.  The
most important result is that chiral symmetry breaking indeed occurs
for a wide range of initial values of the parameters including the
presumably realistic case of large renormalized Yukawa coupling and a
bare mass $\ol{m}_{k_\Phi}$ of order $100\MeV$. A typical evolution of
the renormalized meson mass $m$ with $k$ is plotted in fig.~\ref{Fig1}.
Driven by the strong Yukawa coupling, $m$ decreases rapidly and goes
through zero at a scale $k_{\chi{\rm SB}}$ not far below $k_\Phi$.
Here the system enters the SSB regime and a non--vanishing
(renormalized) VEV $\si_0$ for the meson field $\Phi$ develops. The
evolution of $\si_0$ with $k$ turns out to be reasonably stable
already before scales $k\simeq m_\pi$ where the evolution is stopped.
We take this result as an indication that our truncation of the
effective action $\Gm_k$ leads at least qualitatively to a
satisfactory description of chiral symmetry breaking. The reason for
the relative stability of the IR behavior of the VEV (and all other
couplings) is that the quarks acquire a constituent mass
$M_q=h\si_0\simeq350\MeV$ in the SSB regime. As a consequence they
decouple once $k$ becomes smaller than $M_q$ and the evolution is then
dominantly driven by the massless Goldstone bosons.  This is also
important in view of potential confinement effects expected to become
important for the quark $n$--point functions for $k$ around $\La_{\rm
  QCD}\simeq200\MeV$.  Since confinement is not explicitly included in
our truncation of $\Gamma_k$, one might be worried that such effects
could spoil our results completely. Yet, as discussed in some more
detail above, only the colored quarks should feel confinement and they
are no longer important for the evolution of the meson couplings for
$k$ around $200\MeV$.  One might therefore hope that a precise
treatment of confinement is not crucial for this approach to chiral
symmetry breaking.

\begin{figure}
\unitlength1.0cm
\begin{picture}(13.,9.0)
\put(-0.5,5.){\bf $\ds{\frac{f_\pi}{\MeV}}$}
\put(6.5,0.5){\bf $\ol{m}_{k_\Phi}^2/k_\Phi^2$}
\put(9.0,5.5){\bf $O(4)$}
\put(3.8,2.5){\bf $U_L(2)\times U_R(2)$}
\put(-2.0,-11.5){
\epsfysize=22.cm
\epsffile{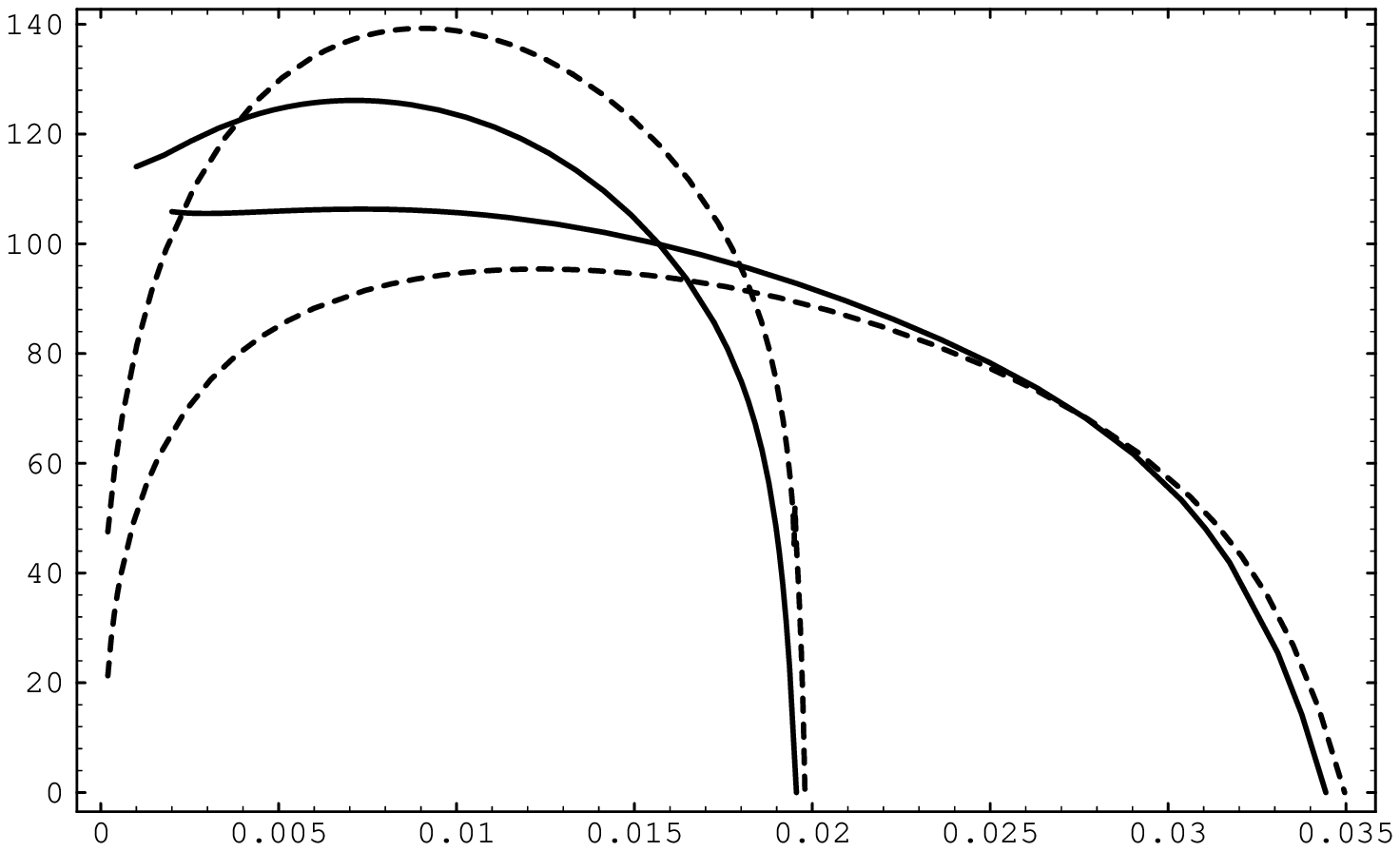}
}
\end{picture}
\caption{The pion decay constant $f_\pi$ as a function of 
  $\ol{m}_{k_\Phi}^2/k_\Phi^2$ for $k_\Phi=630\MeV$ and initial values
  $\la_{i,I}=0$ and $h^2_I=300$ (solid line) as well as $h^2_I=10^4$
  (dashed line).}
\label{Fig8}
\end{figure}

Most importantly, one finds that the system of flow equations exhibits
a partial IR fixed point in the symmetric phase. As already pointed
out one expects $Z_{\Phi,k_\Phi}$ to be rather small.  In turn, one
may assume that, at least for the initial range of running in the
symmetric regime the mass parameter $m^2\sim Z_{\Phi,k}^{-1}$ is
large. This means, in particular, that all threshold functions with
arguments $\sim m^2$ may be neglected in this regime.  As a
consequence, the flow equations simplify considerably.  We find, for
instance, for the $U_L(2)\times U_R(2)$ model
\begin{eqnarray}
  \label{AAA08}
  \ds{\frac{d}{d t}\tilde{\la}_1} &\equiv& \ds{
    \frac{d}{d t}\frac{\la_1}{h^2}\simeq
    \frac{N_c}{4\pi^2}h^2\left[\hal\tilde{\la}_1-\frac{1}{N}\right]}\nnn
  \ds{\frac{d}{d t}\tilde{\la}_2} &\equiv& \ds{
    \frac{d}{d t}\frac{\la_2}{h^2}\simeq
    \frac{N_c}{4\pi^2}h^2\left[\hal\tilde{\la}_2-\frac{2}{N}\right]}\\[2mm]
  \ds{\frac{d}{d t}h^2} &\simeq& \ds{
    \frac{N_c}{8\pi^2}h^4}\nnn
  \ds{\eta_\Phi} &\simeq& \ds{ 
    \frac{N_c}{8\pi^2}h^2\; ,\;\;\;
    \eta_\psi\simeq0}\nonumber\; .
\end{eqnarray} 
This system possesses an attractive IR fixed point for the
quartic scalar self interactions 
\begin{equation}
  \label{AAA10}
  \tilde{\la}_{1*}=\hal\tilde{\la}_{2*}=\frac{2}{N}\; .  
\end{equation}
Furthermore it is exactly soluble~\cite{JW95-1}.  Because of the
strong Yukawa coupling the quartic couplings $\tilde{\la}_1$ and
$\tilde{\la}_2$ generally approach their fixed point values rapidly,
long before the systems enters the broken phase ($m\ra0$) and the
approximation of large $m$ breaks down.  In addition, for large
initial values $h^2_I$ the Yukawa coupling at the scale $k_{\chi SB}$
where $m$ vanishes (or becomes small) only depends on the initial
value $\ol{m}_{k_\Phi}$.  Hence, the system is approximately
independent in the IR upon the initial values of $\la_1$, $\la_2$ and
$h^2$, the only ``relevant'' parameter being $\ol{m}_{k_\Phi}$. (Once
quark masses and a proper treatment of the chiral anomaly are included
for $N=3$ one expects that $m_q$ and $\nu$ are additional relevant
parameters. Their values may be fixed by using the masses of $\pi$,
$K$ and $\eta^\prime$ as phenomenological input.) In other words, the
effective action looses almost all its ``memory'' in the far IR of
where in the UV it came from. This feature of the flow equations leads
to a perhaps surprising degree of predictive power.  In addition, also
the dependence of $f_\pi=2\si_0$ on $\ol{m}_{k_\Phi}$ is not very
strong for a large range in $\ol{m}_{k_\Phi}$, as shown in
fig.~\ref{Fig8}.
The relevant parameter $\ol{m}_{k_\Phi}$ can be fixed by using the
constituent quark mass $M_q\equiv h\si_0\simeq350\MeV$ as a
phenomenological input.  One obtains for the $O(4)$--model
\begin{equation}
 \ds{\frac{\ol{m}_{k_\Phi}^2}{k_\Phi^2}} \simeq \ds{0.02}\; .
\end{equation}
The resulting value for the decay constant is
\begin{equation}
  \label{AAA52}
  f_\pi=(91-100)\MeV
\end{equation}
for $h^2_I=10^4-300$. It is striking that this comes close to the real
value $f_\pi=92.4\MeV$ but we expect that the uncertainty in the
determination of the compositeness scale $k_\Phi$ and the truncation
errors exceed the influence of the variation of $h^2_I$.  We have
furthermore used this result for an estimate of the chiral condensate:
\begin{equation}
 \VEV{\ol{\psi}\psi}\equiv
 -\ol{m}_{k_\Phi}^2f_\pi
 Z_{\Phi,k=0}^{-1/2}A\simeq-(195\MeV)^3 
\end{equation} 
where the factor $A\simeq1.7$ accounts for the change of the
normalization scale of $\VEV{\ol{\psi}\psi}$ from $k_\Phi$ to the
commonly used value $1\GeV$. Our value is in reasonable agreement with
results from sum rules~\cite{JM95-1}.  This is non--trivial since not
only $\ol{m}_{k_\Phi}$ and $f_\pi$ enter but also the IR value
$Z_{\Phi,k=0}$.  Integrating \Ref{AAA08} for $\eta_\Phi$ one finds
\begin{equation}
  \label{AAA09}
  Z_{\Phi,k}=Z_{\Phi,k_\Phi}+
  \frac{N_c}{8\pi^2}\ln\frac{k_\Phi}{k}\; .
\end{equation}
Thus $Z_{\Phi,k}$ will indeed be practically independent of its
initial value $Z_{\Phi,k_\Phi}$ already after some running as long as
$Z_{\Phi,k_\Phi}$ is small compared to $0.01$.

The alert reader may have noticed that the beta functions
\Ref{AAA08} correspond exactly to those obtained in the
one--quark--loop approximation or, in other words, to the leading
order in the large--$N_c$ expansion for the Nambu--Jona-Lasinio
model~\cite{BHJ94-1}. The fixed point \Ref{AAA10} is then nothing but
the large--$N_c$ boundary condition on the evolution of $\la_1$ and
$\la_2$ in this model. Yet, we wish to stress that nowhere we have
made the assumption that $N_c$ is a large number.  On the contrary,
the physical value $N_c=3$ suggests that the large--$N_c$ expansion
should a priori only be trusted on a quantitatively rather crude
level. The reason why we expect \Ref{AAA08} to nevertheless give
rather reliable results is based on the fact that for small
$Z_{\Phi,k_\Phi}$ all (renormalized) meson masses are much larger than
the scale $k$ for the initial part of the running.  This implies that
the mesons are effectively decoupled and their contribution to the
beta functions is negligible leading to the one--quark--loop
approximation. Yet, already after some relatively short period of
running the renormalized meson masses approach zero and our
approximation of neglecting mesonic threshold functions breaks down.
Hence, the one--quark--loop approximation is reasonable only for
scales close to $k_\Phi$ but is bound to become inaccurate around
$k_{\chi{\rm SB}}$ and in the SSB regime. What is important in our
context is not the numerical value of the partial fixed points
\Ref{AAA10} but rather their mere existence and the presence of a
large coupling $h^2$ driving the $\tilde{\la}_i$ fast towards them.
This is enough for the IR values of all couplings to become almost
independent of the initial values $\la_{i,I}$. Similar features of IR
stability are expected if the truncation is enlarged, for instance, to
a more general form of the effective potential $U_{k_\Phi}$ as will be
discussed in the next section.

\sect{Flow equation for the scalar potential}
\label{sec8}

In this and the remaining sections we consider the $O(4)$--symmetric
quark meson model without truncating its effective average potential
to a polynomial form~\cite{BJW97-1}. A comparison with the results of
the last sections will give us a feeling for the size of the errors
induced by the quartic truncation used so far.  Furthermore, such an
approach is well suited for a study of the chiral phase transition
close to which the form of the potential deviates substantially from a
polynomial.

It is convenient to work with dimensionless and renormalized variables
therefore eliminating all explicit $k$--dependence. With
($t=\ln(k/k_\Phi)$)
\begin{equation}
  \label{AAA190}
  u(t,\tilde{\rho})\equiv k^{-d}U_k(\rho)\; ,\;\;\;
  \tilde{\rho}\equiv Z_{\Phi,k} k^{2-d}\rho
\end{equation}
and using \Ref{GammaEffective} as a first truncation of the
effective average action $\Gamma_k$ one obtains the flow equation
in arbitrary dimensions $d$
\begin{equation}
  \label{AAA68}
  \begin{array}{rcl}
    \ds{\frac{\prl}{\prl t}u} &=& \ds{
      -d u+\left(d-2+\eta_\Phi\right)
      \tilde{\rho}u^\prime}\\[2mm]
    &+& \ds{
      2v_d\left\{
      3l_0^d(u^\prime;\eta_\Phi)+
      l_0^d(u^\prime+2\tilde{\rho}u^{\prime\prime};\eta_\Phi)-
      2^{\frac{d}{2}+1}N_c
      l_0^{(F)d}(\frac{1}{2}\tilde{\rho}h^2;\eta_\psi)
      \right\} }\; .
  \end{array}
\end{equation}
Here $v_d^{-1}\equiv2^{d+1}\pi^{d/2}\Gamma(d/2)$ and primes denote
derivatives with respect to $\tilde{\rho}$.  We will always use in the
following for the number of quark colors $N_c=3$.  \Ref{AAA68}
is a partial differential equation for the effective potential
$u(t,\tilde{\rho})$ which has to be supplemented by the flow equation
for the Yukawa coupling and expressions for the anomalous dimensions
$\eta_\Phi$, $\eta_\psi$. The definition of the threshold functions
$l_n^d$, $l_n^{(F)d}$ can be found in~\cite{JW95-1}.

The dimensionless renormalized expectation value
$\kappa\equiv2k^{2-d}Z_{\Phi,k}\ol{\si}_{0,k}^2$, with
$\ol{\si}_{0,k}$ the $k$--dependent VEV of $\Phi$, may be computed
for each $k$ directly from the condition
\begin{equation}
  \label{AAA90}
  u^\prime(t,\kappa)=\frac{\jmath}{\sqrt{2\kappa}}
  k^{-\frac{d+2}{2}}Z_{\Phi,k}^{-1/2}\equiv
  \epsilon_g
\end{equation}
where~\cite{BJW97-1}
\begin{equation}
  \label{LAL001}
  \jmath=2\ol{m}_{k_\Phi,k}^2\hat{m}
\end{equation}
and $\hat{m}=(m_u+m_d)/2$ denotes the average light current quark mass
normalized at $k_\Phi$.  Note that $\kappa\equiv0$ in the symmetric
regime for vanishing source term. \Ref{AAA90} allows us to
follow the flow of $\kappa$ according to
\begin{eqnarray}
  \label{AAA91}
  \ds{\frac{d}{d t}\kappa} &=& \ds{
    \frac{\kappa}{\epsilon_g+2\kappa\la}
    \Bigg\{\left[\eta_\Phi-d-2\right]\epsilon_g-
    2\frac{\prl}{\prl t}u^\prime(t,\kappa)\Bigg\} }
\end{eqnarray}
with $\la\equiv u^{\prime\prime}(t,\kappa)$.  We define the Yukawa
coupling for $\tilde{\rho}=\kappa$ and its flow equation
reads~\cite{JW95-1}
\begin{equation}
  \label{AAA70n}
  \begin{array}{rcl}
  \ds{\frac{d}{d t}h^2} &=& \ds{
  \left(d-4+2\eta_\psi+\eta_\Phi\right)h^2-
    2v_d h^4\Bigg\{
    3l_{1,1}^{(F B)d}(\frac{1}{2}h^2\kappa,
    \epsilon_g;\eta_\psi,\eta_\Phi) }\\[2mm]
  &-& \ds{
    l_{1,1}^{(F B)d}(\frac{1}{2}h^2\kappa,
    \epsilon_g+2\la\kappa;
    \eta_\psi,\eta_\Phi)
  \Bigg\} }\; .
  \end{array}
\end{equation}
Similarly, the the scalar and quark anomalous
dimensions are infered from
\begin{equation}
  \label{AAA69n}
  \begin{array}{rcl}
    \ds{\eta_\Phi} &\equiv& \ds{
      -\frac{d}{d t}\ln Z_{\Phi,k}=
      4\frac{v_d}{d}\Bigg\{
      4\kappa\la^2
      m_{2,2}^d(\epsilon_g,\epsilon_g+2\la\kappa;
      \eta_\Phi) }\nnn
    &+& \ds{
      2^{\frac{d}{2}}N_c h^2
      m_4^{(F)d}(\frac{1}{2}h^2\kappa;
      \eta_\psi)
      \Bigg\}\; , }\nnn
    \ds{\eta_\psi} &\equiv& \ds{
      -\frac{d}{d t}\ln Z_{\psi,k}=
      2\frac{v_d}{d}h^2\Bigg\{
      3m_{1,2}^{(F B)d}(\frac{1}{2}h^2\kappa,
      \epsilon_g;\eta_\psi,\eta_\Phi) }\\[2mm]
    &+& \ds{
      m_{1,2}^{(F B)d}(\frac{1}{2}h^2\kappa,
      \epsilon_g+2\la\kappa;
      \eta_\psi,\eta_\Phi)
      \Bigg\}\; , }
  \end{array}
\end{equation}
which is a linear set of equations for the anomalous dimensions.  The
definitions of the threshold functions $l_{n_1,n_2}^{(FB)d}$,
$m_{n_1,n_2}^d$, $m_4^{(F)d}$ and $m_{n_1,n_2}^{(FB)d}$ are again
specified in~\cite{JW95-1}.

The flow equations (\ref{AAA68}), (\ref{AAA91})---(\ref{AAA69n}),
constitute a coupled system of ordinary and partial differential
equations which can be integrated numerically~\cite{ABBFTW95-1,BW97-1}.
Here we take the effective current quark mass dependence of $h$,
$Z_{\Phi,k}$ and $Z_{\psi,k}$ into account by stopping the evolution
according to Eqs.(\ref{AAA70n}), (\ref{AAA69n}), evaluated for the
chiral limit, below the pion mass $m_\pi$.

Similarly to the case of the quartic truncation of the effective
average potential described in the preceding section one finds for
$d=4$ that chiral symmetry breaking indeed occurs for a wide range of
initial values of the parameters. These include the presumably
realistic case of large renormalized Yukawa coupling and a bare mass
$\ol{m}_{k_\Phi}$ of order $100\MeV$.

Most importantly, the approximate partial IR fixed point behavior
encountered for the quartic potential approximation before, carries
over to the truncation of $\Gamma_k$ which maintains the full
effective average potential~\cite{BJW97-1}. To see this explicitly we
study the flow equations (\ref{AAA68}), (\ref{AAA91})---(\ref{AAA69n})
subject to the condition $Z_{\Phi,k_\Phi}\ll1$. For the relevant range
of $\tilde{\rho}$ both $u^\prime(t,\tilde{\rho})$ and
$u^\prime(t,\tilde{\rho})+
2\tilde{\rho}u^{\prime\prime}(t,\tilde{\rho})$ are then much larger
than $\tilde{\rho}h^2(t)$ and we may therefore neglect in the flow
equations all scalar contributions with threshold functions involving
these large masses.  This yields the simplified equations
($d=4,v_4^{-1}=32\pi^2$)
\begin{equation}
  \label{AAA110}
  \begin{array}{rcl}
    \ds{\frac{\prl}{\prl t}u} &=& \ds{
      -4u+\left(2+\eta_\Phi\right)
      \tilde{\rho}u^\prime
      -\frac{N_c}{2\pi^2}
      l_0^{(F)4}(\frac{1}{2}\tilde{\rho}h^2)\; ,
      }\nnn
    \ds{\frac{d}{d t}h^2} &=& \ds{
      \frac{N_c}{8\pi^2}h^4\; ,
      }\nnn
      \ds{\eta_\Phi} &=& \ds{
        \frac{N_c}{8\pi^2}h^2 \; ,\;\;\;
        \eta_\psi=0}\; .
    \end{array}
\end{equation}
Again, this approximation is only valid for the initial range of
running below $k_\Phi$ before the (dimensionless) renormalized scalar
mass squared $u^\prime(t,\tilde{\rho}=0)$ approaches zero near the
chiral symmetry breaking scale.  The system \Ref{AAA110} is exactly
soluble. We find
\begin{equation}
  \label{AAA113}
  \begin{array}{rcl}
    \ds{h^2(t)} &=& \ds{
      Z_\Phi^{-1}(t)=
      \frac{h_I^2}{1-\frac{N_c}{8\pi^2}h_I^2 t}\; ,\;\;\;
      Z_\psi(t)=1\; ,
      }\nnn
    \ds{u(t,\tilde{\rho})} &=& \ds{
      e^{-4t}u_I(e^{2t}\tilde{\rho}\frac{h^2(t)}{h_I^2})-
      \frac{N_c}{2\pi^2}\int_0^t d r e^{-4r}
      l_0^{(F)4}(\frac{1}{2}h^2(t)\tilde{\rho}e^{2r}) }\; .
  \end{array}
\end{equation}
(The integration over $r$ on the right hand side of the solution for
$u$ can be carried out by first exchanging it with the one over
momentum implicit in the definition of the threshold function
$l_0^{(F)4}$.) Here $u_I(\tilde{\rho})\equiv u(0,\tilde{\rho})$
denotes the effective average potential at the compositeness scale and
$h_I^2$ is the initial value of $h^2$ at $k_\Phi$, i.e. for $t=0$.
For simplicity we will use an expansion of the initial value effective
potential $u_I(\tilde{\rho})$ in powers of $\tilde{\rho}$ around
$\tilde{\rho}=0$
\begin{equation}
  \label{AAA140}
  u_I(\tilde{\rho})=
  \sum_{n=0}^\infty
  \frac{u_I^{(n)}(0)}{n!}\tilde{\rho}^n
\end{equation}
even though this is not essential for the forthcoming reasoning.
Expanding also $l_0^{(F)4}$ in \Ref{AAA113} in powers of its
argument one finds for $n>2$
\begin{equation}
  \label{LLL00}
  \ds{\frac{u^{(n)}(t,0)}{h^{2n}(t)}} = \ds{
    e^{2(n-2)t}\frac{u_I^{(n)}(0)}{h_I^{2n}}+
    \frac{N_c}{\pi^2}
    \frac{(-1)^n (n-1)!}{2^{n+2}(n-2)}
    l_n^{(F)4}(0)
    \left[1-e^{2(n-2)t}\right]}\; .
\end{equation}
For decreasing $t\ra-\infty$ the initial values $u_I^{(n)}$ become
rapidly unimportant and $u^{(n)}/h^{2n}$ approaches a fixed point.
For $n=2$, i.e., for the quartic coupling, one finds
\begin{equation}
  \label{LLL01}
  \frac{u^{(2)}(t,0)}{h^2(t)}=
  1-\frac{1-\frac{u_I^{(2)}(0)}{h_I^2}}
  {1-\frac{N_c}{8\pi^2}h_I^2 t}
\end{equation}
leading to the fixed point value $(u^{(2)}/h^2)_*=1$ already
encountered in \Ref{AAA10}. As a consequence of this fixed point
behavior the system looses all its ``memory'' on the initial values
$u_I^{(n\ge2)}$ at the compositeness scale $k_\Phi$! This typically
happens before the approximation
$u^\prime(t,\tilde{\rho}),u^\prime(t,\tilde{\rho})+
2\tilde{\rho}u^{\prime\prime}(t,\tilde{\rho})\gg\tilde{\rho}h^2(t)$
breaks down and the solution \Ref{AAA113} becomes invalid.
Furthermore, the attraction to partial infrared fixed points continues
also for the range of $k$ where the scalar fluctuations cannot be
neglected anymore.  As for the quartic truncation the initial value
for the bare dimensionless mass parameter
\begin{equation}
  \label{AAA142}
  \frac{u_I^\prime(0)}{h_I^2}=
  \frac{\ol{m}^2_{k_\Phi}}{k_\Phi^2}
\end{equation}
is never negligible. (In fact, using the values for
$\ol{m}^2_{k_\Phi}$ and $k_\Phi$ computed previously~\cite{EW94-1} for
a pure quark effective action as described above one obtains
$\ol{m}^2_{k_\Phi}/k_\Phi^2\simeq0.036$.)  For large $h_I$ (and
dropping the constant piece $u_I(0)$) the solution \Ref{AAA113}
therefore behaves with growing $\abs{t}$ as
\begin{equation}
  \label{AAA150}
  \begin{array}{rcl}
    \ds{Z_\Phi(t)} &\simeq& \ds{
      -\frac{N_c}{8\pi^2}t\; ,\;\;\;
      Z_\psi(t)=1\; ,
      }\nnn
    \ds{h^2(t)} &\simeq& \ds{
      -\frac{8\pi^2}{N_c t}\; ,
      }\nnn
    \ds{u(t,\tilde{\rho})} &\simeq& \ds{
      \frac{u_I^\prime(0)}{h_I^2} e^{-2t} h^2(t)\tilde{\rho}
      -\frac{N_c}{2\pi^2}\int_0^t d r e^{-4r}
      l_0^{(F)4}(\frac{1}{2}h^2(t)\tilde{\rho}e^{2r}) }\; .  
  \end{array}
\end{equation}
In other words, for $h_I\ra\infty$ the IR behavior of the linear quark
meson model will depend (in addition to the value of the compositeness
scale $k_\Phi$ and the quark mass $\hat{m}$) only on one parameter,
$\ol{m}^2_{k_\Phi}$.  We have numerically verified this feature by
starting with different values for $u_I^{(2)}(0)$.  Indeed, the
differences in the physical observables were found to be small.  This
IR stability of the flow equations leads to a perhaps surprising
degree of predictive power: Not only the scalar wave function
renormalization but even the full effective potential
$U(\rho)=\lim_{k\ra0}U_k(\rho)$ is (approximately) fixed for
$Z_{\Phi,k_\Phi}\ll1$ once $\ol{m}_{k_\Phi}$ is known!  For
definiteness we will perform our numerical analysis of the full system
of flow equations (\ref{AAA68}), (\ref{AAA91})---(\ref{AAA69n}) with
the idealized initial value
$u_I(\tilde{\rho})=u_I^\prime(0)\tilde{\rho}$ in the limit
$h_I^2\ra\infty$. It should be stressed, though, that deviations from
this idealization will lead only to small numerical deviations in the
IR behavior of the linear quark meson model as long as the condition
$Z_{\Phi,k_\Phi}\ll1$ holds, say for~\cite{JW96-1} $h_I\gta15$.

With this knowledge at hand we may now fix the remaining three
parameters of our model, $k_\Phi$, $\ol{m}^2_{k_\Phi}$ and $\hat{m}$
by using $f_\pi=92.4\MeV$, the pion mass $M_\pi=135\MeV$ and the
constituent quark mass $M_q$ as phenomenological input.  This approach
differs from that of the preceding sections where we took an earlier
determination of $k_\Phi$ as input for the computation of $f_\pi$.  It
will be better suited for precision estimates of the high temperature
behavior in the following sections.  Because of the uncertainty
regarding the precise value of $M_q$ we give in tab.~\ref{tab1} the
results for several values of $M_q$.
\begin{table}
\begin{center}
\begin{tabular}{|c|c||c|c|c||c|c|c|c|} \hline
  $\frac{M_q}{\MeV}$ &
  $\frac{\la_I}{h_I^2}$ &
  $\frac{k_\Phi}{\MeV}$ &
  $\frac{\ol{m}^2_{k_\Phi}}{k_\Phi^2}$ &
  $\frac{\jmath^{1/3}}{\MeV}$ &
  $\frac{\hat{m}(k_\Phi)}{\MeV}$ &
  $\frac{\hat{m}(1\GeV)}{\MeV}$ &
  $\frac{\VEV{\ol{\psi}\psi}(1\GeV)}{\MeV^3}$ &
  $\frac{f_\pi^{(0)}}{\MeV}$
  \\[0.5mm] \hline\hline
  $303$ &
  $1$ &
  $618$ &
  $0.0265$ &
  $66.8$ &
  $14.7$ &
  $11.4$ &
  $-(186)^3$ &
  $80.8$
  \\ \hline
  $300$ &
  $0$ &
  $602$ &
  $0.026$ &
  $66.8$ &
  $15.8$ &
  $12.0$ &
  $-(183)^3$ &
  $80.2$
  \\ \hline
  $310$ &
  $0$ &
  $585$ &
  $0.025$ &
  $66.1$ &
  $16.9$ &
  $12.5$ &
  $-(180)^3$ &
  $80.5$
  \\ \hline
  $339$ &
  $0$ &
  $552$ &
  $0.0225$ &
  $64.4$ &
  $19.5$ &
  $13.7$ &
  $-(174)^3$ &
  $81.4$
  \\ \hline
\end{tabular}
\caption{\footnotesize The table shows the dependence on the
  constituent quark mass $M_q$ of the input parameters $k_\Phi$,
  $\ol{m}^2_{k_\Phi}/k_\Phi^2$ and $\jmath$ as well as some of our
  ``predictions''. The phenomenological input used here besides $M_q$
  is $f_\pi=92.4\MeV$, $m_\pi=135\MeV$.The first line corresponds to
  the values for $M_q$ and $\la_I$ used in the remainder of this
  work. The other three lines demonstrate the insensitivity of our
  results with respect to the precise values of these
  parameters.}
\label{tab1}
\end{center}
\end{table}
The first line of tab.~\ref{tab1} corresponds to the choice of $M_q$
and $\la_I\equiv u_I^{\prime\prime}(\kappa)$ which we will use for the
forthcoming analysis of the model at finite temperature.  As argued
analytically above the dependence on the value of $\la_I$ is weak for
large enough $h_I$ as demonstrated numerically by the second line.
Moreover, we notice that our results, and in particular the value of
$\jmath$, are rather insensitive with respect to the precise value of
$M_q$. It is remarkable that the values for $k_\Phi$ and
$\ol{m}_{k_\Phi}$ are not very different from those
computed~\cite{EW94-1} from four--quark interactions as described
above.  As compared to the analysis of the preceding sections the
present truncation of $\Gamma_k$ is of a higher level of accuracy: We
now consider an arbitrary form of the effective average potential
instead of a polynomial approximation and we have included the pieces
in the threshold functions which are proportional to the anomalous
dimensions. It is encouraging that the results are rather robust with
respect to these improvements.

Once the parameters $k_\Phi$, $\ol{m}^2_{k_\Phi}$ and $\hat{m}$ are
fixed there are a number of ``predictions'' of the linear meson model
which can be compared with the results obtained by other methods or
direct experimental observation. First of all one may compute the
value of $\hat{m}$ at a scale of $1\GeV$ which is suitable for
comparison with results obtained from chiral perturbation
theory~\cite{Leu96-1} and sum rules~\cite{JM95-1}.  For this purpose
one has to account for the running of this quantity with the
normalization scale from $k_\Phi$ as given in tab.~\ref{tab1} to the
commonly used value of $1\GeV$:
$\hat{m}(1\GeV)=A^{-1}\hat{m}(k_\Phi)$. A reasonable estimate of the
factor $A$ is obtained from the three loop running of $\hat{m}$ in the
$\ol{MS}$ scheme~\cite{JM95-1}. For $M_q\simeq300\MeV$ corresponding
to the first two lines in tab.~\ref{tab1} its value is $A\simeq1.3$.
The results for $\hat{m}(1\GeV)$ are in acceptable agreement with
recent results from other methods~\cite{Leu96-1,JM95-1} even though
they tend to be somewhat larger. Closely related to this is the value
of the chiral condensate
\begin{equation}
  \label{LLL04}
  \VEV{\ol{\psi}\psi}(1\GeV)\equiv
  -A\ol{m}^2_{k_\Phi}
  \left[f_\pi Z_{\Phi,k=0}^{-1/2}-
    2\hat{m}\right]
    \; .
\end{equation}
These results are quite non--trivial since not only $f_\pi$ and
$\ol{m}^2_{k_\Phi}$ enter but also the computed IR value
$Z_{\Phi,k=0}$.  We emphasize in this context that there may be
substantial corrections both in the extrapolation from $k_\Phi$ to
$1\GeV$ and because of the neglected influence of the strange quark
which may be important near $k_\Phi$. These uncertainties have only
little effect on the physics at lower scales as relevant for our
analysis of the temperature effects. Only the value of $\jmath$ which
is fixed by $m_\pi$ enters here.

A further more qualitative test concerns the mass of the sigma
resonance or radial mode in the limit $k\ra0$ whose renormalized mass
squared is given by
\begin{equation}
  \label{LAL000}
  m_\si^2=Z_{\Phi,k_\Phi}^{-1/2}
  \frac{\ol{m}_{k_\Phi}^2\hat{m}}
  {\si_0}+4\la\si_0^2\; .
\end{equation}
From our numerical analysis we obtain $\la_{k=0}\simeq20$ which
translates into $m_\si\simeq 430\MeV$.  One should note, though, that
this result is presumably not very accurate as we have employed in
this work the approximation of using the Goldstone boson wave function
renormalization constant also for the radial mode. Furthermore, the
explicit chiral symmetry breaking contribution to $m_\si^2$ is
certainly underestimated as long as the strange quark is neglected.
In any case, we observe that the sigma meson is significantly heavier
than the pions. This is a crucial consistency check for the linear
quark meson model. A low sigma mass would be in conflict with the
numerous successes of chiral perturbation theory~\cite{Leu95-1} which
requires the decoupling of all modes other than the Goldstone bosons
in the IR--limit of QCD. The decoupling of the sigma meson is, of
course, equivalent to the limit $\la\ra\infty$ which formally
describes the transition from the linear to the non--linear sigma
model and which appears to be reasonably well realized by the large
IR--values of $\la$ obtained in our analysis. We also note that the
issue of the sigma mass is closely connected to the value of
$f_\pi^{(0)}$, the value of $f_\pi$ in the chiral limit $\hat{m}=0$
also given in tab.~\ref{tab1}.  To lowest order in
$(f_\pi-f_\pi^{(0)})/f_\pi$ or, equivalently, in $\hat{m}$ one has
\begin{equation}
  \label{ABC00}
  f_\pi-f_\pi^{(0)}=\frac{\jmath}{Z_\Phi^{1/2}m_\si^2}=
  \frac{f_\pi m_\pi^2}{m_\si^2}\; .
\end{equation}
A larger value of $m_\si$ would therefore reduce the difference
between $f_\pi^{(0)}$ and $f_\pi$.

\begin{figure}
\unitlength1.0cm
\begin{center}
\begin{picture}(13.,7.0)
\put(0.0,0.0){
\epsfysize=11.cm
\rotate[r]{\epsffile{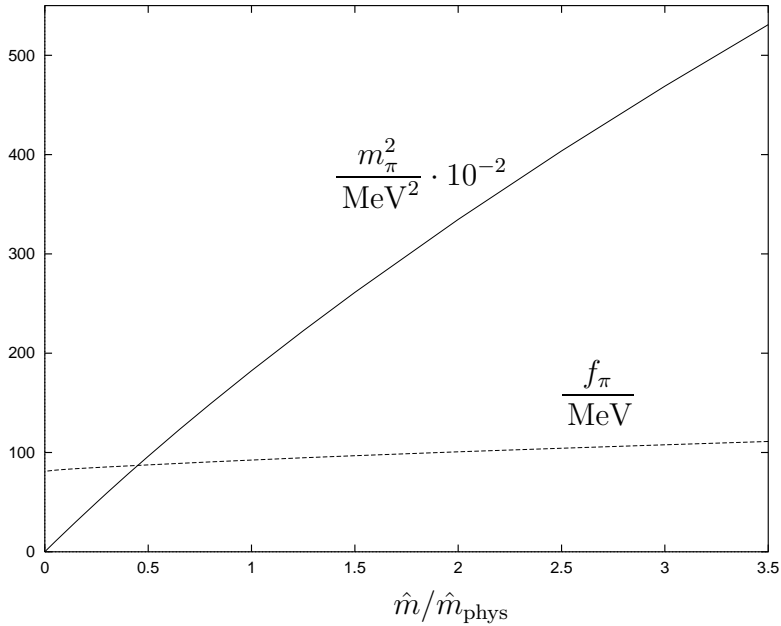}}
}
\put(8.0,2.3){\bf $\ds{\frac{f_\pi}{\MeV}}$}
\put(5.0,5.2){\bf $\ds{\frac{m^2_\pi}{\MeV^2}\cdot10^{-2}}$}
\put(5.8,-0.5){\bf $\ds{\hat{m}/\hat{m}_{\rm phys}}$}
\end{picture}
\end{center}
\caption{\footnotesize The plot shows $m_\pi^2$ (solid line) and
  $f_\pi$ (dashed line) as functions of the current quark mass
  $\hat{m}$ in units of the physical value $\hat{m}_{\rm phys}$.}
\label{mm}
\end{figure}

In fig.~\ref{mm} we show the dependence of the pion mass and decay
constant on the average current quark mass $\hat{m}$.
These curves depend very little on the values of the initial
parameters as demonstrated in tab.~\ref{tab1} by $f_\pi^{(0)}$. We
observe a relatively large difference of $12\MeV$ between the pion
decay constants at $\hat{m}=\hat{m}_{\rm phys}$ and $\hat{m}=0$.
According to \Ref{ABC00} this difference is related to the mass of
the sigma particle and will be modified in the three flavor case. We
will later find that the critical temperature $T_c$ for the second
order phase transition in the chiral limit is almost independent of
the initial conditions. The values of $f_\pi^{(0)}$ and $T_c$
essentially determine the non--universal amplitudes in the critical
scaling region (see below). In summary, we find that the behavior of
our model for small $k$ is quite robust as far as uncertainties in the
initial conditions at the scale $k_\Phi$ are concerned. We will see
that the difference of observables between non--vanishing and
vanishing temperature is entirely determined by the flow of couplings
in the range $0<k\lta3T$.

\sect{Chiral phase transition of two flavor QCD}
\label{ChrialPhaseTransitionInTwoFlavorQCD}

Strong interactions in thermal equilibrium at high temperature $T$ ---
as realized in early stages of the evolution of the Universe ---
differ in important aspects from the well tested vacuum or zero
temperature properties. A phase transition at some critical
temperature $T_c$ or a relatively sharp crossover may separate the
high and low temperature physics~\cite{MO96-1}. Many experimental
activities at heavy ion colliders~\cite{QM96} search for signs of such
a transition.  It was realized early that the transition should be
closely related to a qualitative change in the chiral condensate
according to the general observation that spontaneous symmetry
breaking tends to be absent in a high temperature situation. A series
of stimulating contributions~\cite{PW84-1,RaWi93-1,Raj95-1} pointed
out that for sufficiently small up and down quark masses, $m_u$ and
$m_d$, and for a sufficiently large mass of the strange quark, $m_s$,
the chiral transition is expected to belong to the universality class
of the $O(4)$ Heisenberg model. This means that near the critical
temperature only the pions and the sigma particle play a role for the
behavior of condensates and long distance correlation functions. It
was suggested~\cite{RaWi93-1,Raj95-1} that a large correlation length
may be responsible for important fluctuations or lead to a disoriented
chiral condensate~\cite{Ans88-1}. This was even
related~\cite{RaWi93-1,Raj95-1} to the spectacular ``Centauro
events''~\cite{LFH80-1} observed in cosmic rays. The question how
small $m_u$ and $m_d$ would have to be in order to see a large
correlation length near $T_c$ and if this scenario could be realized
for realistic values of the current quark masses remained, however,
unanswered. The reason was the missing link between the universal
behavior near $T_c$ and zero current quark mass on one hand and the
known physical properties at $T=0$ for realistic quark masses on the
other hand.

It is the purpose of the remaining sections of these lectures to
provide this link~\cite{BJW97-1}. We present here the equation of state
for two flavor QCD within an effective quark meson model. The equation
of state expresses the chiral condensate $\VEV{\ol{\psi}\psi}$ as a
function of temperature and the average current quark mass
$\hat{m}=(m_u+m_d)/2$.  This connects explicitly the universal
critical behavior for $T\ra T_c$ and $\hat{m}\ra0$ with the
temperature dependence for a realistic value $\hat{m}_{\rm phys}$.
Since our discussion covers the whole temperature range $0\le T
\,\ltap\, 1.7\, T_c$ we can fix $\hat{m}_{\rm phys}$ such that the
(zero temperature) pion mass is $m_\pi=135\MeV$. The condensate
$\VEV{\ol{\psi}\psi}$ plays here the role of an order parameter.
fig.~\ref{ccc_T} shows our results for
$\VEV{\ol{\psi}\psi}(T,\hat{m})$:
\begin{figure}
\unitlength1.0cm

\begin{center}
\begin{picture}(13.,7.0)

\put(0.0,0.0){
\epsfysize=11.cm
\rotate[r]{\epsffile{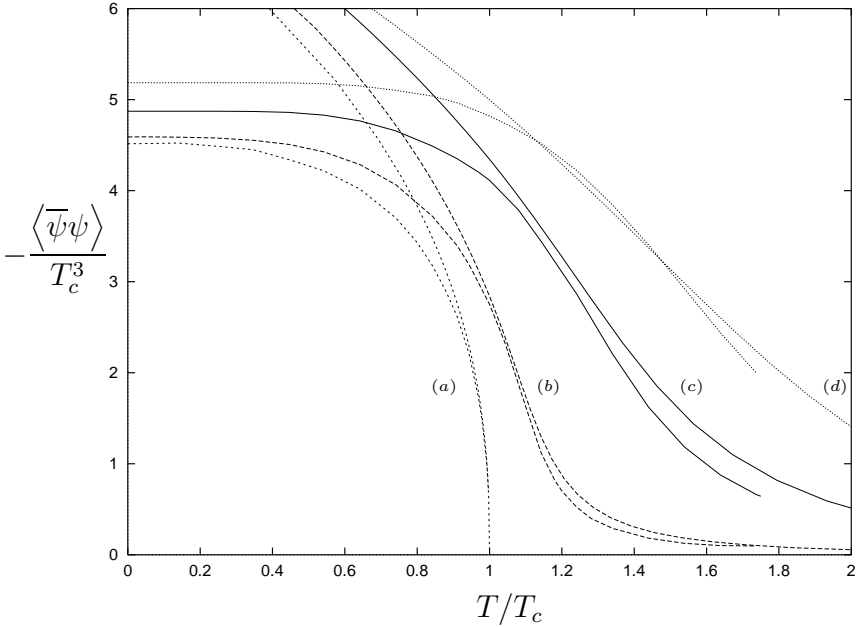}}
}
\put(-0.5,4.2){\bf $\ds{-\frac{\VEV{\ol{\psi}\psi}}{T_{c}^3}}$}
\put(5.8,-0.5){\bf $\ds{T/T_{c}}$}
\put(5.2,2.5){\tiny $(a)$}
\put(6.6,2.5){\tiny $(b)$}
\put(8.5,2.5){\tiny $(c)$}
\put(10.4,2.5){\tiny $(d)$}

\end{picture}
\end{center}
\caption{\footnotesize The plot shows the chiral condensate
  $\VEV{\ol{\psi}\psi}$ as a function of temperature $T$.  Lines
  $(a)$, $(b)$, $(c)$, $(d)$ correspond at zero temperature to
  $m_\pi=0,45\MeV,135\MeV,230\MeV$, respectively. For each pair of
  curves the lower one represents the full $T$--dependence of
  $\VEV{\ol{\psi}\psi}$ whereas the upper one shows for comparison the
  universal scaling form of the equation of state for the $O(4)$
  Heisenberg model. The critical temperature for zero quark mass is
  $T_c=100.7\MeV$. The chiral condensate is normalized at a scale
  $k_{\Phi}\simeq 620\MeV$.}
\label{ccc_T}
\end{figure}
Curve $(a)$ gives the temperature dependence of $\VEV{\ol{\psi}\psi}$
in the chiral limit $\hat{m}=0$. Here the lower curve is the full
result for arbitrary $T$ whereas the upper curve corresponds to the
universal scaling form of the equation of state for the $O(4)$
Heisenberg model.  We see perfect agreement of both curves for $T$
sufficiently close to $T_c=100.7 \MeV$. This demonstrates the
capability of our method to cover the critical behavior and, in
particular, to reproduce the critical exponents of the $O(4)$--model.
We have determined (see below) the universal critical equation of
state as well as the non--universal amplitudes.  This provides the
full functional dependence of $\VEV{\ol{\psi}\psi} (T,\hat{m})$ for
small $T-T_c$ and $\hat{m}$.  The curves $(b)$, $(c)$ and $(d)$ are
for non--vanishing values of the average current quark mass $\hat{m}$.
Curve $(c)$ corresponds to $\hat{m}_{\rm phys}$ or, equivalently,
$m_\pi(T=0)=135\MeV$. One observes a crossover in the range
$T=(1.2-1.5)T_c$. The $O(4)$ universal equation of state (upper curve)
gives a reasonable approximation in this temperature range. The
transition turns out to be much less dramatic than for $\hat{m}=0$. We
have also plotted in curve $(b)$ the results for comparably small
quark masses $\simeq1\MeV$, i.e.~$\hat{m}=\hat{m}_{\rm phys}/10$, for
which the $T=0$ value of $m_\pi$ equals $45\MeV$. The crossover is
considerably sharper but a substantial deviation from the chiral limit
remains even for such small values of $\hat{m}$. In order to
facilitate comparison with lattice simulations which are typically
performed for larger values of $m_\pi$ we also present results for
$m_\pi(T=0)=230\MeV$ in curve $(d)$. One may define a ``pseudocritical
temperature'' $T_{pc}$ associated to the smooth crossover as the
inflection point of $\VEV{\ol{\psi}\psi}(T)$ as often done in lattice
simulations. Our results for this definition of $T_{pc}$ are denoted
by $T_{pc}^{(1)}$ and are presented in tab.~\ref{tab11} for the four
different values of $\hat{m}$ or, equivalently, $m_\pi(T=0)$.
\begin{table}
\begin{center}
\begin{tabular}{|c||c|c|c|c|} \hline
  $\stackrel{ }{\frac{m_\pi}{\MeV}}$ &
  $0$ &
  $45$ &
  $135$ &
  $230$
  \\[1.0mm] \hline
  $\stackrel{ }{\frac{T_{pc}^{(1)}}{\MeV}}$ &
  $100.7$ &
  $\simeq110$ &
  $\simeq130$ &
  $\simeq150$
  \\[1mm] \hline
  $\stackrel{ }{\frac{T_{pc}^{(2)}}{\MeV}}$ &
  $100.7$ &
  $$113 &
  $$128 &
  $$---
  \\[1mm] \hline
\end{tabular}
\caption{\footnotesize The table shows the critical and
  ``pseudocritical'' temperatures for various values of the zero
  temperature pion mass. Here $T_{pc}^{(1)}$ is defined as the
  inflection point of $\VEV{\ol{\psi}\psi}(T)$ whereas $T_{pc}^{(2)}$
  is the location of the maximum of the sigma correlation length.}
\label{tab11}
\end{center}
\end{table}
The value for the pseudocritical temperature for $m_{\pi}=230\MeV$
compares well with the lattice results for two flavor QCD (cf.~the
discussion below). One should mention, though, that a determination of
$T_{pc}$ according to this definition is subject to sizeable numerical
uncertainties for large pion masses as the curve in fig.~\ref{ccc_T}
is almost linear around the inflection point for quite a large
temperature range.  A difficult point in lattice simulations with
large quark masses is the extrapolation to realistic values of $m_\pi$
or even to the chiral limit. Our results may serve here as an analytic
guide. The overall picture shows the approximate validity of the
$O(4)$ scaling behavior over a large temperature interval in the
vicinity of and above $T_c$ once the (non--universal) amplitudes are
properly computed.

\begin{figure}
\unitlength1.0cm
\begin{center}
\begin{picture}(13.,7.0)

\put(0.0,0.0){
\epsfysize=11.cm
\rotate[r]{\epsffile{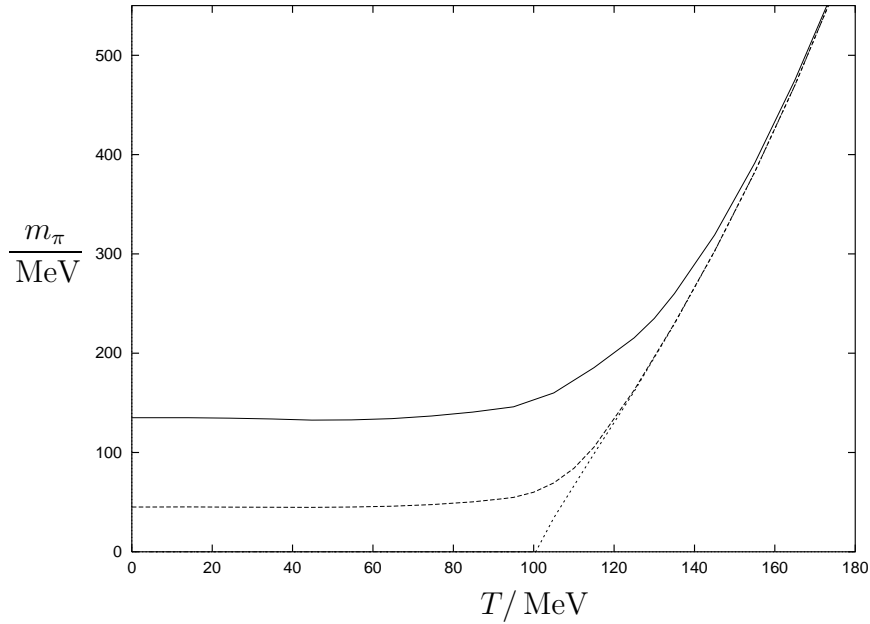}}
}
\put(-0.5,4.2){\bf $\ds{\frac{m_\pi}{\MeV}}$}
\put(5.8,-0.5){\bf $\ds{T/\MeV}$}

\end{picture}
\end{center}
\caption{\footnotesize The plot shows $m_\pi$ as a function of
  temperature $T$ for three different values of the average light
  current quark mass $\hat{m}$. The solid line corresponds to the
  realistic value $\hat{m}=\hat{m}_{\rm phys}$ whereas the dotted line
  represents the situation without explicit chiral symmetry breaking,
  i.e., $\hat{m}=0$. The intermediate, dashed line assumes 
  $\hat{m}=\hat{m}_{\rm phys}/10$.}
\label{mpi_T}
\end{figure}

A second important result of our investigations is the temperature
dependence of the space--like pion correlation length
$m_\pi^{-1}(T)$. (We will often call $m_\pi(T)$ the temperature
dependent pion mass since it coincides with the physical pion mass for
$T=0$.) The plot for $m_\pi(T)$ in fig.~\ref{mpi_T} again shows the
second order phase transition in the chiral limit $\hat{m}=0$. 
For $T<T_c$ the pions are massless Goldstone bosons whereas for
$T>T_c$ they form together with the sigma particle a degenerate vector
of $O(4)$ whose mass increases as a function of temperature.  For
$\hat{m}=0$ the behavior for small positive $T-T_c$ is characterized
by the critical exponent $\nu$, i.e.
$m_\pi(T)=\left(\xi^+\right)^{-1}T_c \left( (T-T_c)/T_c\right)^\nu$
and we obtain $\nu=0.787$, $\xi^+=0.270$. For $\hat{m}>0$ we find that
$m_\pi(T)$ remains almost constant for $T\lta T_c$ with only a very
slight dip for $T$ near $T_c/2$. For $T>T_c$ the correlation length
decreases rapidly and for $T\gg T_c$ the precise value of $\hat{m}$
becomes irrelevant. We see that the universal critical behavior near
$T_c$ is quite smoothly connected to $T=0$.  The full functional
dependence of $m_\pi(T,\hat{m})$ allows us to compute the overall size
of the pion correlation length near the critical temperature and we
find $ m_\pi(T_{pc})\simeq 1.7 m_\pi(0)$ for the realistic value
$\hat{m}_{\rm phys}$. This correlation length is even smaller than the
vacuum ($T=0$) one and shows no indication for strong fluctuations of
pions with long wavelength. It would be interesting to see if a
decrease of the pion correlation length at and above $T_c$ is
experimentally observable.  It should be emphasized, however, that a
tricritical behavior with a massless excitation remains possible for
three flavors. This would not be characterized by the universal
behavior of the $O(4)$--model. We also point out that the present
investigation for the two flavor case does not take into account a
speculative ``effective restoration'' of the axial $U_A(1)$ symmetry
at high temperature~\cite{PW84-1,Shu94-1}. We will comment on these
issues in the last section. In the next sections we will describe the
formalism which leads to these results.

\sect{Thermal equilibrium and dimensional reduction}
\label{FiniteTemperatureFormalism}

The extension of flow equations to thermal equilibrium situations at
non--vanishing temperature $T$ is straightforward~\cite{TW93-1}. In
the Euclidean formalism non--zero temperature results in
(anti--)periodic boundary conditions for (fermionic) bosonic fields in
the Euclidean time direction with periodicity~\cite{Kap89-1} $1/T$.
This leads to the replacement
\begin{equation}
  \label{AAA120}
  \int\frac{d^d q}{(2\pi)^d}f(q^2)\ra
  T\sum_{l\in\ZZZ}\int\frac{d^{d-1}\vec{q}}{(2\pi)^{d-1}}
  f(q_0^2(l)+\vec{q}^{\,2})
\end{equation}
in the trace in \Ref{ERGE} when represented as a momentum
integration, with a discrete spectrum for the zero component
\begin{equation}
  \label{AAA121}
  q_0(l)=\left\{
  \begin{array}{lll}
    2l\pi T &{\rm for}& {\rm bosons}\\
    (2l+1)\pi T &{\rm for}& {\rm fermions}\; .
  \end{array}\right.
\end{equation}
Hence, for $T>0$ a four--dimensional QFT can be interpreted as a
three--dimensional model with each bosonic or fermionic degree of
freedom now coming in an infinite number of copies labeled by
$l\in\ZZ$ (Matsubara modes). Each mode acquires an additional
temperature dependent effective mass term $q_0^2(l)$. In a high
temperature situation where all massive Matsubara modes decouple from
the dynamics of the system one therefore expects to observe an
effective three--dimensional theory with the bosonic zero modes as the
only relevant degrees of freedom. In other words, if the
characteristic length scale associated with the physical system is
much larger than the inverse temperature the compactified Euclidean
``time'' dimension cannot be resolved anymore. This phenomenon is
known as ``dimensional reduction''~\cite{Gin80-1}.

\begin{figure}
\unitlength1.0cm
\begin{center}
\begin{picture}(13.,17.0)

\put(0.0,9.5){
\epsfysize=11.cm
\rotate[r]{\epsffile{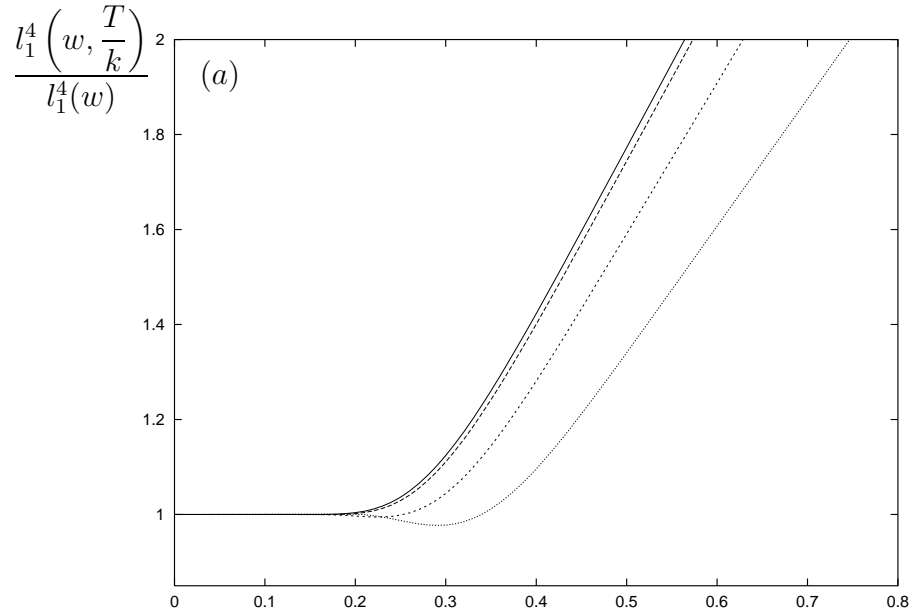}}
}
\put(-1.0,16.5){\bf $\ds{\frac{l_1^4\left(w,\ds{\frac{T}{k}}\right)}
    {l_1^4(w)}}$}
\put(1.5,16.5){\bf $\ds{(a)}$}

\put(0.0,0.5){
\epsfysize=11.cm
\rotate[r]{\epsffile{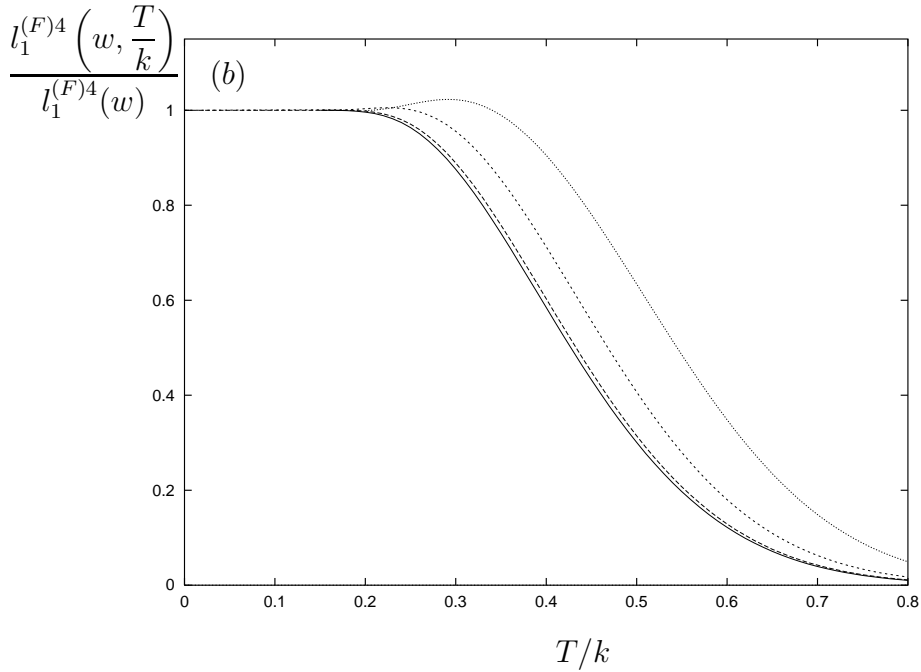}}
}
\put(-1.2,7.5){\bf
  $\ds{\frac{l_1^{(F)4}\left(w,\ds{\frac{T}{k}}\right)}
    {l_1^{(F)4}(w)}}$}
\put(6.1,-0.2){\bf $\ds{T/k}$}
\put(1.5,7.5){\bf $\ds{(b)}$}
\end{picture}
\end{center}
\caption{\footnotesize The plot shows the temperature 
  dependence of the bosonic (a) and the fermionic (b) threshold
  functions $l_1^4(w,T/k)$ and $l_1^{(F)4}(w,T/k)$, respectively, for
  different values of the dimensionless mass term $w$.  The solid line
  corresponds to $w=0$ whereas the dotted ones correspond to $w=0.1$,
  $w=1$ and $w=10$ with decreasing size of the dots.  For $T \gg k$
  the bosonic threshold function becomes proportional to $T/k$ whereas
  the fermionic one tends to zero.  In this range the theory with
  properly rescaled variables behaves as a classical
  three--dimensional theory.}
\label{Thresh}
\end{figure}

The formalism of the effective average action automatically provides
the tools for a smooth decoupling of the massive Matsubara modes as
the scale $k$ is lowered from $k\gg T$ to $k\ll T$.  It therefore
allows us to directly link the low--$T$, four--dimensional QFT to the
effective three--dimensional high--$T$ theory. The replacement
\Ref{AAA120} in \Ref{ERGE} manifests itself in the flow equations
(\ref{AAA68}), (\ref{AAA91})---(\ref{AAA69n}) only through a change to
$T$--dependent threshold functions.  For instance, the dimensionless
functions $l_n^d(w;\eta_\Phi)$ defined in \Ref{AAA85n} are
replaced by
\begin{equation}
  \label{AAA200}
  l_n^d(w,\frac{T}{k};\eta_\Phi)\equiv
  \frac{n+\delta_{n,0}}{4}v_d^{-1}k^{2n-d}
  T\sum_{l\in\ZZZ}\int
  \frac{d^{d-1}\vec{q}}{(2\pi)^{d-1}}
  \left(\frac{1}{Z_{\Phi,k}}\frac{\prl R_k(q^2)}{\prl t}\right)
  \frac{1}{\left[P(q^2)+k^2 w\right]^{n+1}}
\end{equation}
where $q^2=q_0^2+\vec{q}^{\,2}$ and $q_0=2\pi l T$. A list of the
various temperature dependent threshold functions appearing in the
flow equations can be found~\cite{BJW97-1}.
There we also discuss some subtleties regarding the definition of the
Yukawa coupling and the anomalous dimensions for $T\neq0$. In the
limit $k\gg T$ the sum over Matsubara modes approaches the integration
over a continuous range of $q_0$ and we recover the zero temperature
threshold function $l_n^d(w;\eta_\Phi)$.  In the opposite limit $k\ll
T$ the massive Matsubara modes ($l\neq0$) are suppressed and we expect
to find a $d-1$ dimensional behavior of $l_n^d$. In fact, one obtains
from \Ref{AAA200}
\begin{equation}
  \label{AAA201}
  \begin{array}{rclcrcl}
    \ds{l_n^d(w,T/k;\eta_\Phi)} &\simeq& \ds{
      l_n^{d}(w;\eta_\Phi)}
    &{\rm for}& \ds{T\ll k}\; ,\nnn
    \ds{l_n^d(w,T/k;\eta_\Phi)} &\simeq& \ds{
      \frac{T}{k}\frac{v_{d-1}}{v_d}
      l_n^{d-1}(w;\eta_\Phi)}
    &{\rm for}& \ds{T\gg k}\; .
  \end{array}
\end{equation}
For our choice of the infrared cutoff function $R_k$, \Ref{AAA61}, the
temperature dependent Matsubara modes in $l_n^d(w,T/k;\eta_\Phi)$ are
exponentially suppressed for $T\ll k$ whereas the behavior is more
complicated for other threshold functions appearing in the flow
equations (\ref{AAA68}), (\ref{AAA91})---(\ref{AAA69n}).
Nevertheless, all bosonic threshold functions are proportional to
$T/k$ for $T\gg k$ whereas those with fermionic contributions vanish
in this limit\footnote{For the present choice of $R_k$ the temperature
  dependence of the threshold functions is considerably smoother than
  in that of previous investigations~\cite{TW93-1}.}.  This behavior
is demonstrated in fig.~\ref{Thresh} where we have plotted the
quotients $l_1^4(w,T/k)/l_1^4(w)$ and
$l_1^{(F)4}(w,T/k)/l_1^{(F)4}(w)$ of bosonic and fermionic threshold
functions, respectively.
One observes that for $k\gg T$ both threshold functions essentially
behave as for zero temperature. For growing $T$ or decreasing $k$ this
changes as more and more Matsubara modes decouple until finally all
massive modes are suppressed. The bosonic threshold function $l^4_1$
shows for $k \ll T$ the linear dependence on $T/k$ derived in
\Ref{AAA201}.  In particular, for the bosonic excitations the
threshold function for $w\ll1$ can be approximated with reasonable
accuracy by $l_n^4(w;\eta_\Phi)$ for $T/k<0.25$ and by
$(4T/k)l_n^3(w;\eta_\Phi)$ for $T/k>0.25$. The fermionic threshold
function $l_1^{(F)4}$ tends to zero for $k\ll T$ since there is no
massless fermionic zero mode, i.e.~in this limit all fermionic
contributions to the flow equations are suppressed.  On the other
hand, the fermions remain quantitatively relevant up to $T/k\simeq0.6$
because of the relatively long tail in fig.~\ref{Thresh}b.  The
transition from four to three--dimensional threshold functions leads
to a {\em smooth dimensional reduction} as $k$ is lowered from $k\gg
T$ to $k\ll T$!  Whereas for $k\gg T$ the model is most efficiently
described in terms of standard four--dimensional fields $\Phi$ a
choice of rescaled three--dimensional variables
$\Phi_{3}=\Phi/\sqrt{T}$ becomes better adapted for $k\ll T$.
Accordingly, for high temperatures one will use a potential
\begin{equation}
  \label{CCC01}
  u_{3}(t,\tilde{\rho}_{3})=\frac{k}{T}
  u(t,\tilde{\rho})\; ;\;\;\;
  \tilde{\rho}_{3}=\frac{k}{T}\tilde{\rho}\; .
\end{equation}
In this regime $\Gamma_{k\ra0}$ corresponds to the free energy of
classical statistics and $\Gamma_{k>0}$ is a classical coarse grained
free energy.

For our numerical calculations at non--vanishing temperature we
exploit the discussed behavior of the threshold functions by using the
zero temperature flow equations in the range $k\ge10T$. For smaller
values of $k$ we approximate the infinite Matsubara sums
(cf.~\Ref{AAA200}) by a finite series such that the numerical
uncertainty at $k=10T$ is better than $10^{-4}$. This approximation
becomes exact in the limit $k\ll10T$.

\sect{The quark meson model at $T\neq0$}
\label{TheQuarkMesonModelAtTNeq0}

So far we have considered the relevant fluctuations that contribute to
the flow of $\Gamma_k$ in dependence on the scale $k$. In a thermal
equilibrium situation $\Gamma_k$ also depends on the temperature $T$
and one may ask for the relevance of thermal fluctuations at a given
scale $k$.  In particular, for not too high values of $T$ the
``initial condition'' $\Gamma_{k_\Phi}$ for the solution of the flow
equations should essentially be independent of temperature.  This will
allow us to fix $\Gamma_{k_\Phi}$ from phenomenological input at $T=0$
and to compute the temperature dependent quantities in the infrared
($k \to 0$).  We note that the thermal fluctuations which contribute
to the r.h.s.\ of the flow equation for the meson potential
\Ref{AAA68} are effectively suppressed for $T \lta k/4$ as discussed
in detail in the last section.  Clearly for $T \gta k_{\Phi}/3$
temperature effects become important at the compositeness scale. We
expect the linear quark meson model with a compositeness scale
$k_{\Phi} \simeq 600 \MeV$ to be a valid description for two flavor
QCD below a temperature of about $170 \MeV$. We note that there will
be an effective temperature dependence of $\Gamma_{k_{\Phi}}$ induced
by the fluctuations of other degrees of freedom besides the quarks,
the pions and the sigma which are taken into account here.  We will
comment on this issue in the last section.  For realistic three flavor
QCD the thermal kaon fluctuations will become important for
$T\gta170\MeV$.

We compute the quantities of interest for temperatures $T\lta170\MeV$
by numerically solving the $T$--dependent version of the flow
equations~\cite{BJW97-1} (\ref{AAA68}), (\ref{AAA91})---(\ref{AAA69n})
by lowering $k$ from $k_\Phi$ to zero. For this range of temperatures
we use the initial values as given in the first line of
tab.~\ref{tab1}.  This corresponds to choosing the zero temperature
pion mass and the pion decay constant ($f_{\pi}=92.4 \MeV$ for
$m_{\pi}=135 \MeV$) as phenomenological input. The only further input
is the constituent quark mass $M_q$ which we vary in the range $M_q
\simeq 300 - 350 \MeV$. We observe only a minor dependence of our
results on $M_q$ for the considered range of values. In particular,
the value for the critical temperature $T_c$ of the model remains
almost unaffected by this variation.

\begin{figure}
\unitlength1.0cm

\begin{center}
\begin{picture}(13.,7.0)

\put(0.0,0.0){
\epsfysize=11.cm
\rotate[r]{\epsffile{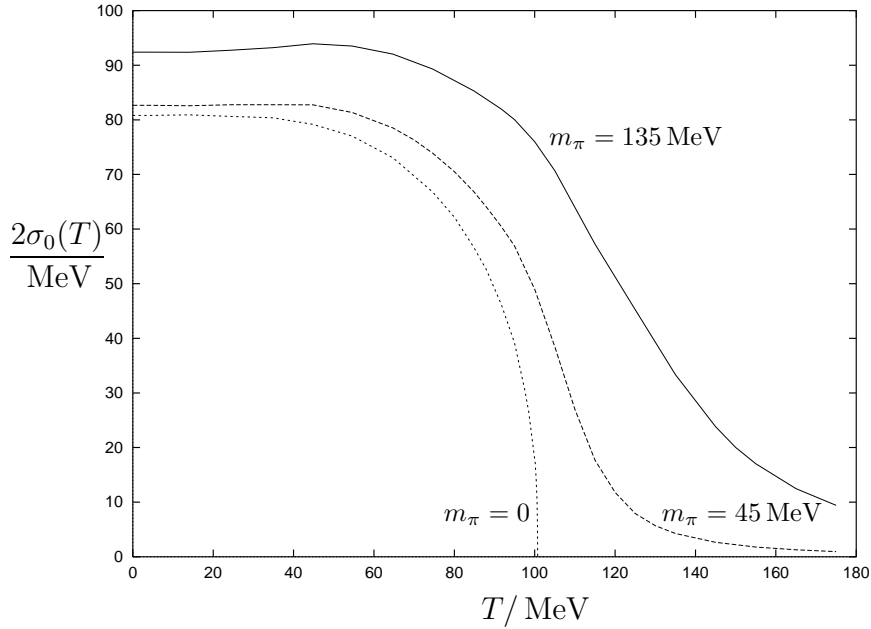}}
}
\put(-0.5,4.2){\bf $\ds{\frac{2\si_0(T)}{\MeV}}$}
\put(5.8,-0.5){\bf $\ds{T/\MeV}$}
\put(5.3,0.8){\footnotesize\bf $m_\pi=0$}
\put(8.2,0.8){\footnotesize\bf $m_\pi=45\MeV$}
\put(6.7,5.8){\footnotesize\bf $m_\pi=135\MeV$}

\end{picture}
\end{center}
\caption{\footnotesize The expectation value $2\si_0$ is shown as a
  function of temperature $T$ for three different values of the
  zero temperature pion mass.}
\label{fpi_T}
\end{figure}

We have plotted in fig.~\ref{fpi_T} the renormalized expectation
value $2\si_0$ of the scalar field as a function of temperature for
three different values of the average light current quark mass
$\hat{m}$. (We remind that $2\si_0(T=0)=f_{\pi}$.)
For $\hat{m}=0$ the order parameter $\si_0$ of chiral symmetry
breaking continuously goes to zero for $T\ra T_c = 100.7\MeV$
characterizing the phase transition to be of second order.  The
universal behavior of the model for small $T-T_c$ and small $\hat{m}$
is discussed in more detail in the following section.  We point out
that the value of $T_c$ corresponds to $f_\pi^{(0)}=80.8\MeV$,
i.e.~the value of the pion decay constant for $\hat{m}=0$, which is
significantly lower than $f_\pi=92.4\MeV$ obtained for the realistic
value $\hat{m}_{\rm phys}$.  If we would fix the value of the pion
decay constant to be $92.4\MeV$ also in the chiral limit
($\hat{m}=0$), the value for the critical temperature would raise to
$115\MeV$.  The nature of the transition changes qualitatively for
$\hat{m}\neq0$ where the second order transition is replaced by a
smooth crossover.  The crossover for a realistic $\hat{m}_{\rm phys}$
or $m_{\pi}(T=0)=135 \MeV$ takes place in a temperature range $T
\simeq(120-150)\MeV$.  The middle curve in fig.~\ref{fpi_T}
corresponds to a value of $\hat{m}$ which is only a tenth of the
physical value, leading to a zero temperature pion mass
$m_\pi=45\MeV$. Here the crossover becomes considerably sharper but
there remain substantial deviations from the chiral limit even for
such small quark masses $\hat{m}\simeq 1 \MeV$. The temperature
dependence of $m_\pi$ has already been mentioned (see
fig.~\ref{mpi_T}) for the same three values of $\hat{m}$.  As
expected, the pions behave like true Goldstone bosons for $\hat{m}=0$,
i.e.~their mass vanishes for $T\le T_c$. Interestingly, $m_\pi$
remains almost constant as a function of $T$ for $T<T_c$ before it
starts to increase monotonically. We therefore find for two flavors no
indication for a substantial decrease of $m_\pi$ around the critical
temperature.

\begin{figure}
\unitlength1.0cm
\begin{center}
\begin{picture}(13.,7.0)

\put(0.0,0.0){
\epsfysize=11.cm
\rotate[r]{\epsffile{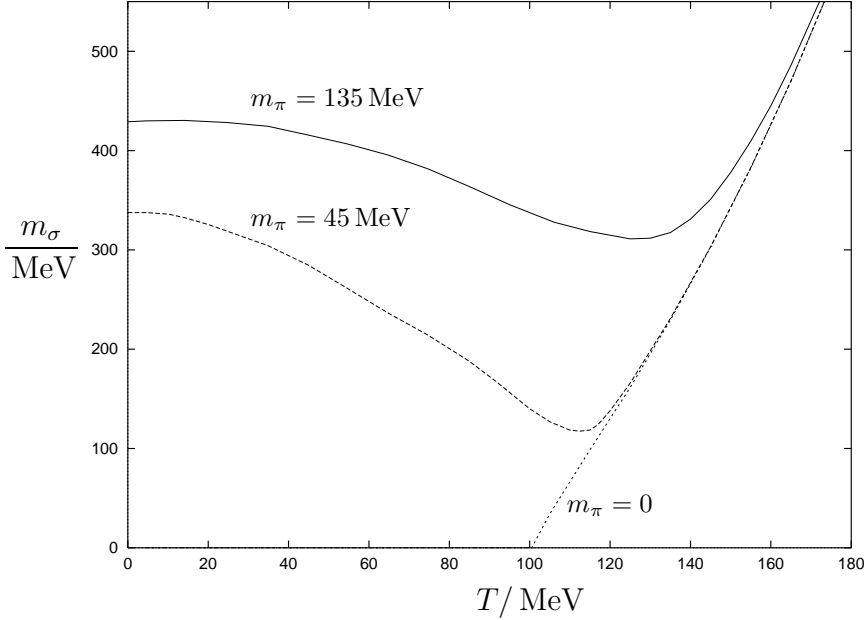}}
}
\put(-0.5,4.2){\bf $\ds{\frac{m_\si}{\MeV}}$}
\put(5.8,-0.5){\bf $\ds{T/\MeV}$}
\put(7.0,0.8){\footnotesize\bf $m_\pi=0$}
\put(2.8,4.6){\footnotesize\bf $m_\pi=45\MeV$}
\put(2.8,6.2){\footnotesize\bf $m_\pi=135\MeV$}

\end{picture}
\end{center}
\caption{\footnotesize The plot shows the $m_\si$ as a function of
  temperature $T$ for three different values of the 
  zero temperature pion mass.}
\label{ms_T}
\end{figure}

The dependence of the mass of the sigma resonance $m_\si$ on the
temperature is displayed in fig.~\ref{ms_T} for the above three
values of $\hat{m}$.
In the absence of explicit chiral symmetry breaking, $\hat{m}=0$, the
sigma mass vanishes for $T\le T_c$. For $T<T_c$ this is a consequence
of the presence of massless Goldstone bosons in the Higgs phase which
drive the renormalized quartic coupling $\la$ to zero.  In fact,
$\la$ runs linearly with $k$ for $T \gta k/4$ and only evolves
logarithmically for $T \lta k/4$.  Once $\hat{m}\neq0$ the pions
acquire a mass even in the spontaneously broken phase and the
evolution of $\la$ with $k$ is effectively stopped at $k\sim
m_\pi$. Because of the temperature dependence of $\si_{0,k=0}$
(cf.~fig.~\ref{fpi_T}) this leads to a monotonically decreasing
behavior of $m_\si$ with $T$ for $T\lta T_c$. This changes into the
expected monotonic growth once the system reaches the symmetric phase
for $T>T_c$. For low enough $\hat{m}$ one may use the minimum of
$m_{\si}(T)$ for an alternative definition of the (pseudo-)critical
temperature denoted as $T_{pc}^{(2)}$. Tab.~\ref{tab11} in the
introduction shows our results for the pseudocritical temperature for
different values of $\hat{m}$ or, equivalently, $m_{\pi}(T=0)$. For a
zero temperature pion mass $m_{\pi}=135 \MeV$ we find
$T_{pc}^{(2)}=128 \MeV$. At larger pion masses of about $230 \MeV$ we
observe no longer a characteristic minimum for $m_{\si}$ apart from
a very broad, slight dip at $T \simeq 90 \MeV$.  A comparison of our
results with lattice data is given below in the next section.

Our results for the chiral condensate $\VEV{\ol{\psi}\psi}$ as a
function of temperature for different values of the average current
quark mass are presented in fig.~\ref{ccc_T}. We will compare
$\VEV{\ol{\psi}\psi}(T,\hat{m})$ with its universal scaling form for
the $O(4)$ Heisenberg model in the following section.

\begin{figure}
\unitlength1.0cm
\begin{center}
\begin{picture}(13.,7.0)
\put(0.0,0.0){
\epsfysize=11.cm
\rotate[r]{\epsffile{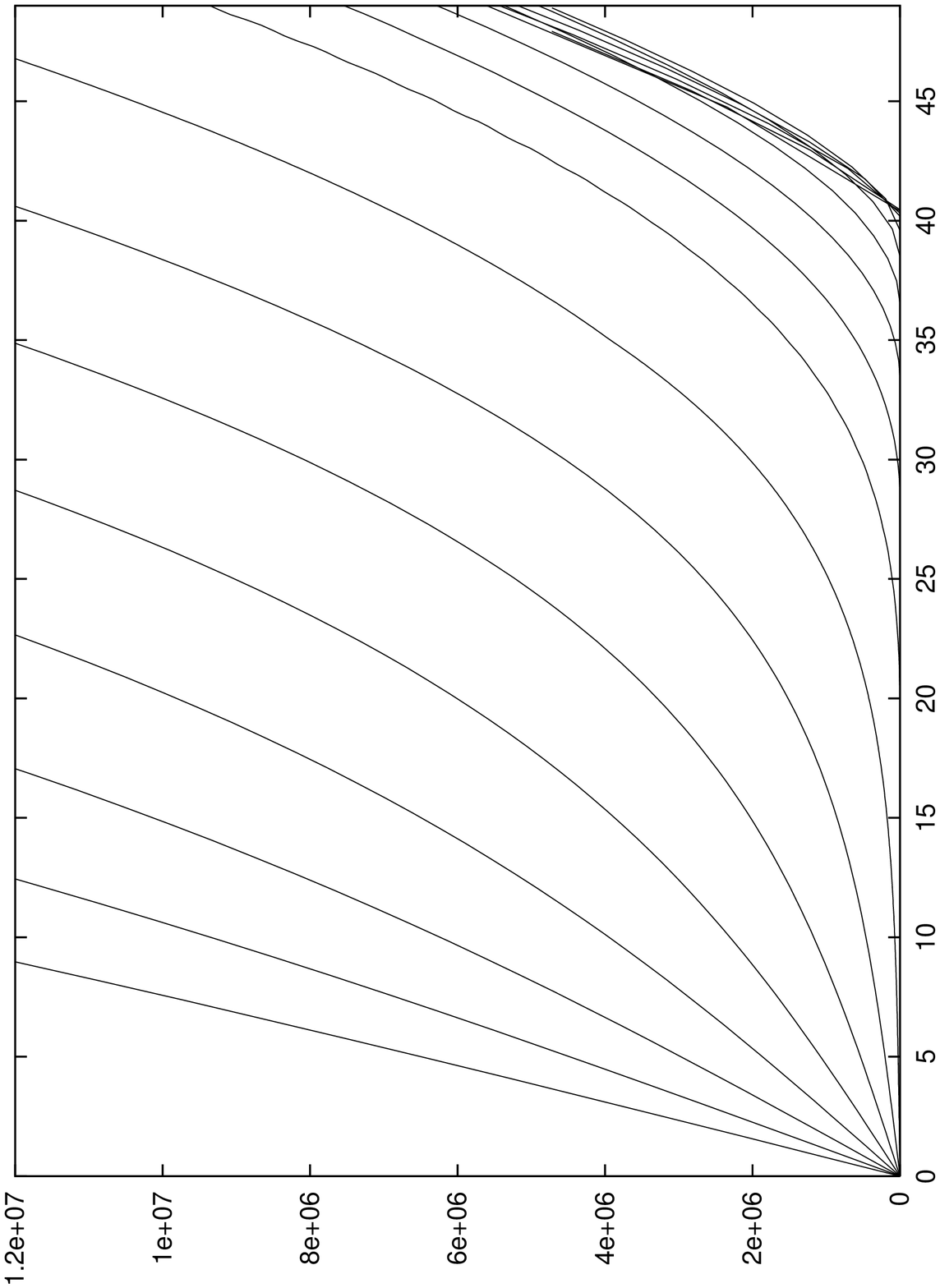}}
}
\put(-1.3,4.2){\bf $\ds{\frac{\partial U(T)/\partial \phi_R}
    {\MeV^{3}}}$}
\put(5.8,-0.5){\bf $\ds{\phi_R/\MeV}$}
\end{picture}
\end{center}
\caption{\footnotesize The plot shows the derivative of the
  meson potential $U(T)$ with respect to the renormalized field
  $\phi_R=(Z_\Phi\rho/2)^{1/2}$ for different values of $T$.  The
  first curve on the left corresponds to $T=175 \MeV$. The successive
  curves to the right differ in temperature by $\Delta T=10 \MeV$ down
  to $T=5 \MeV$. }
\label{Usig}
\end{figure}

Our ability to compute the complete temperature dependent effective
meson potential $U$ is demonstrated in fig.~\ref{Usig} where we
display the derivative of the potential with respect to the
renormalized field $\phi_R=(Z_\Phi\rho/2)^{1/2}$, for different values
of $T$.
The curves cover a temperature range $T = (5 - 175) \MeV$.  The first
one to the left corresponds to $T=175 \MeV$ and neighboring curves
differ in temperature by $\Delta T = 10 \MeV$. One observes only a
weak dependence of $\partial U(T)/\partial\phi_R$ on the temperature
for $T\lta60\MeV$.  Evaluated for $\phi_R=\si_{0}$ this
function connects the renormalized field expectation value with
$m_{\pi}(T)$, the source $\jmath$ and the mesonic wave function
renormalization $Z_{\Phi}(T)$ according to
\begin{equation}
  \label{Usigeq}
  \ds{\frac{\partial U(T)}{\partial\phi_R}}
  (\phi_R=\si_{0})=
  \ds{\frac{2\jmath}{Z_{\Phi}^{1/2}(T)}}=4 \si_{0}(T) 
  m_{\pi}^2(T) \; .
\end{equation} 

We close this section with a short assessment of the validity of our
effective quark meson model as an effective description of two flavor
QCD at non--vanishing temperature.  The identification of
qualitatively different scale intervals which appear in the context of
chiral symmetry breaking, as presented in the preceding sections for
the zero temperature case, can be generalized to $T \neq 0$: For
scales below $k_{\Phi}$ there exists a hybrid description in terms of
quarks and mesons. For $k_{\chi SB} \leq k \lta 600 \MeV$ chiral
symmetry remains unbroken where the symmetry breaking scale $k_{\chi
  SB}(T)$ decreases with increasing temperature. Also the constituent
quark mass decreases with $T$. The running Yukawa coupling depends
only mildly on temperature for $T\lta120\MeV$.  (Only near the
critical temperature and for $\hat{m}=0$ the running is extended
because of massless pion fluctuations.) On the other hand, for
$k\lta4T$ the effective three--dimensional gauge coupling increases
faster than at $T=0$ leading~\cite{RW93-1} to an increase of
$\Lambda_{\rm QCD}(T)$ with $T$. As $k$ gets closer to the scale
$\Lambda_{\rm QCD}(T)$ it is no longer justified to neglect in the
quark sector confinement effects which go beyond the dynamics of our
present quark meson model.  Here it is important to note that the
quarks remain quantitatively relevant for the evolution of the meson
degrees of freedom only for scales $k \gta T/0.6$
(cf.~fig.~\ref{Thresh}).  In the limit $k \ll T/0.6$ all fermionic
Matsubara modes decouple from the evolution of the meson potential
according to the temperature dependent version of \Ref{AAA68}.
Possible sizeable confinement corrections to the meson physics may
occur if $\Lambda_{\rm QCD}(T)$ becomes larger than the maximum of
$M_q(T)$ and $T/0.6$.  This is particularly dangerous for small
$\hat{m}$ in a temperature interval around $T_c$. Nevertheless, the
situation is not dramatically different from the zero temperature case
since only a relatively small range of $k$ is concerned. We do not
expect that the neglected QCD non--localities lead to qualitative
changes.  Quantitative modifications, especially for small $\hat{m}$
and $\abs{T-T_c}$ remain possible. This would only effect the
non--universal amplitudes which will be discussed in the next section.
The size of these corrections depends on the strength of (non--local)
deviations of the quark propagator and the Yukawa coupling from the
values computed in the quark meson model.

\sect{Critical behavior near the chiral phase transition}
\label{CriticalBehavior}

In this section we study the linear quark meson model in the vicinity
of the critical temperature $T_c$ close to the chiral limit
$\hat{m}=0$. In this region we find that the sigma mass
$m_\si^{-1}$ is much larger than the inverse temperature $T^{-1}$,
and one observes an effectively three--dimensional behavior of the
high temperature quantum field theory.  We also note that the fermions
are no longer present in the dimensionally reduced system as has been
discussed above. We therefore have to deal with a purely bosonic
$O(4)$--symmetric linear sigma model.  At the phase transition the
correlation length becomes infinite and the effective
three--dimensional theory is dominated by classical statistical
fluctuations. In particular, the critical exponents which describe the
singular behavior of various quantities near the second order phase
transition are those of the corresponding classical system.

Many properties of this system are universal, i.e.~they only depend
on its symmetry ($O(4)$), the dimensionality of space (three) and its
degrees of freedom (four real scalar components). Universality means
that the long--range properties of the system do not depend on the
details of the specific model like its short distance
interactions. Nevertheless, important properties as the value of the
critical temperature are non--universal. We emphasize that although we
have to deal with an effectively three--dimensional bosonic theory,
the non--universal properties of the system crucially depend on the
details of the four--dimensional theory and, in particular, on the
fermions. 

Our aim is a computation of the critical equation of state which
relates for arbitrary $T$ near $T_c$ 
the derivative of the free energy or effective potential $U$
to the average current quark mass $\hat{m}$. The equation of state
then permits to study the temperature and quark mass dependence of
properties of the chiral phase transition.

At the critical temperature and in the chiral limit there is no scale
present in the theory. In the vicinity of $T_c$ and for small enough
$\hat{m}$ one therefore expects a scaling behavior of the effective
average potential $u_k$ and accordingly a universal scaling form of
the equation of state~\cite{TW94-1}. There are only two independent
scales close to the transition point which can be related to the
deviation from the critical temperature, $T-T_c$, and to the explicit
symmetry breaking through the quark mass $\hat{m}$.  As a consequence,
the properly rescaled potential can only depend on one scaling
variable.  A possible choice for the parameterization of the rescaled
``unrenormalized'' potential is the use of the Widom scaling
variable~\cite{Wid65-1}
\begin{equation}
  \label{XXX20}
  x=\frac{\left( T-T_c\right)/T_c}
  {\left(2\ol{\si}_0/T_c\right)^{1/\beta}}\; .
\end{equation}
Here $\beta$ is the critical exponent of the order parameter
$\ol{\si}_0$ in the chiral limit $\hat{m}=0$ (see \Ref{NNN21}).  With
$U^\prime(\rho=2\ol{\si}_0^2)=\jmath/(2\ol{\si}_0)$ the Widom scaling
form of the equation of state reads~\cite{Wid65-1}
\begin{equation}
  \label{XXX21}
  \frac{\jmath}{T_c^3}=
  \left(\frac{2\ol{\si}_0}{T_c}\right)^\delta f(x)
\end{equation}
where the exponent $\delta$ is related to the behavior of the order
parameter according to \Ref{NNN21b}.  The equation of state
\Ref{XXX21} is written for convenience directly in terms of
four--dimensional quantities.  They are related to the corresponding
effective variables of the three--dimensional theory by appropriate
powers of $T_c$.  The source $\jmath$ is determined by the average
current quark mass $\hat{m}$ as $\jmath=2\ol{m}^2_{k_\Phi}\hat{m}$.
The mass term at the compositeness scale, $\ol{m}^2_{k_\Phi}$, also
relates the chiral condensate to the order parameter according to
$\VEV{\ol{\psi}\psi}=-2\ol{m}^2_{k_\Phi}(\ol{\si}_0-\hat{m})$.  The
critical temperature of the linear quark meson model was found above
to be $T_c=100.7\MeV$.

\begin{figure}
\unitlength1.0cm
\begin{center}
\begin{picture}(17.,12.)
\put(0.3,5.5){$\ds{\frac{2\ol{\si}_0/T_c}
{(\jmath/T_c^3 D)^{1/\delta}}}$}
\put(8.5,-0.2){$\ds{\frac{(T-T_c)/T_c}
{(\jmath/T_c^3 B^{\delta} D)^{1/\beta \delta}}}$}
\put(8.,2.5){\footnotesize $\mbox{average action}$}
\put(4.19,11.2){\footnotesize $\mbox{average action}$}
\put(13.8,2.2){\footnotesize $\epsilon$}
\put(3.53,11.19){\footnotesize $\epsilon$}
\put(11.3,2.05){\footnotesize $\mbox{MC}$}
\put(3.9,10.){\footnotesize $\mbox{MC}$}
\put(13.2,1.8){\footnotesize $\mbox{mf}$}
\put(6.5,9.2){\footnotesize $\mbox{mf}$}
\put(-1.,-5.9){
\epsfysize=21.cm
\epsfxsize=18.cm
\epsffile{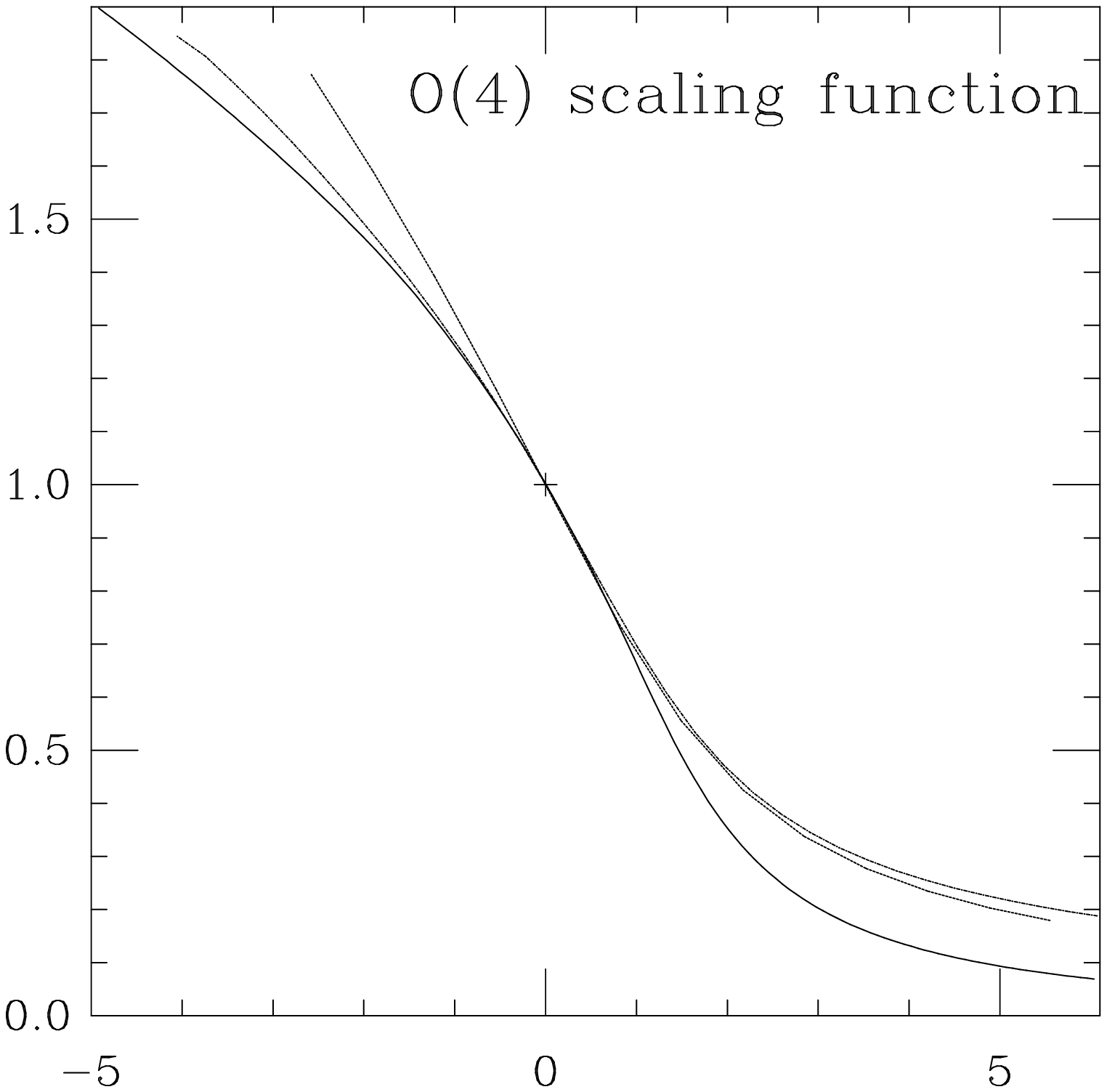}}
\put(1.25,0.483){
\epsfysize=13.45cm
\epsfxsize=11.22cm
\rotate[r]{\epsffile{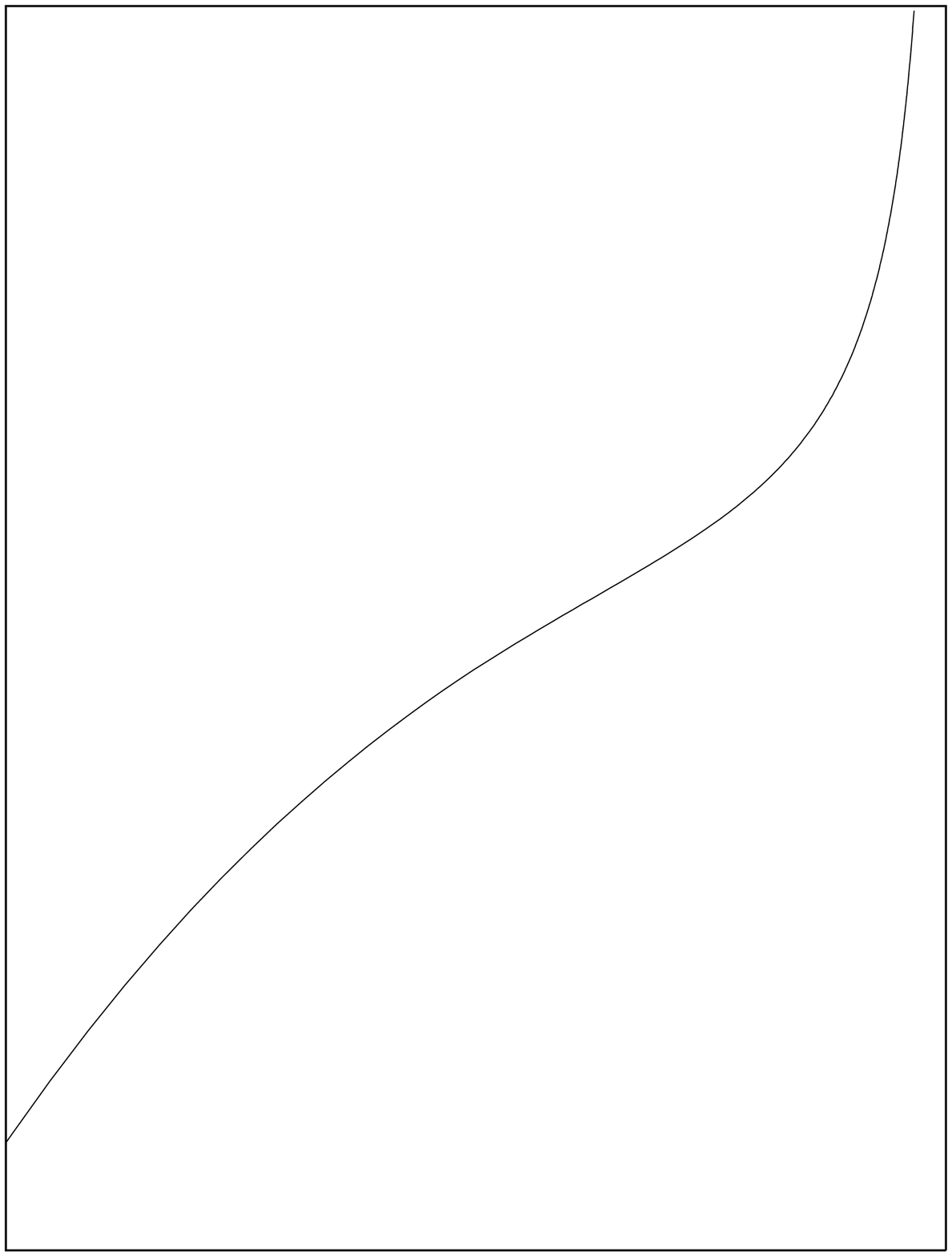}}}
\end{picture}
\end{center}
\caption[]{\footnotesize
  The figure shows a comparison of our results, denoted by ``average
  action'', with results of other methods for the scaling function of
  the three--dimensional $O(4)$ Heisenberg model. We have labeled the
  axes for convenience in terms of the expectation value $\ol{\si}_0$
  and the source $\jmath$ of the corresponding four--dimensional
  theory.  The constants $B$ and $D$ specify the non--universal
  amplitudes of the model (cf.~\Ref{norm}.  The curve labeled by
  ``MC'' represents a fit to lattice Monte Carlo data. The second
  order epsilon expansion~\cite{BWW73-1} and mean field results are
  denoted by ``$\epsilon$'' and ``mf'', respectively.  Apart from our
  results the curves are taken from~\cite{Tou97-1}.
  \label{scalfunc}
  }
\end{figure}

The scaling function $f$ is universal up to the model specific
normalization of $x$ and itself. Accordingly, all models in the same
universality class can be related by a rescaling of $\ol{\si}_0$
and $T-T_c$. The non--universal normalizations for the quark meson
model discussed here are defined according to
\begin{equation}
  \label{norm}
  f(0)=D\quad, \qquad f(-B^{-1/\beta})=0\; .
\end{equation}
We find $D=1.82\cdot10^{-4}$, $B=7.41$ and our result for $\beta$ is
given in tab.~\ref{tab2}. Apart from the immediate vicinity of the
zero of $f(x)$ we find the following two parameter fit for the scaling
function~\cite{BTW96-1},
\begin{equation}
  \label{ffit}
  \begin{array}{rcl}
    \ds{f_{\rm fit}(x)}&=&\ds{1.816 \cdot 10^{-4} (1+136.1\, x)^2 \,
      (1+160.9\, \theta\,
      x)^{\Delta}}\nnn
    &&
    \ds{(1+160.9\, (0.9446\, \theta^{\Delta})^{-1/(\gamma-2-\Delta)} 
      \, x)^{\gamma-2-\Delta}}
  \end{array}
\end{equation}
to reproduce the numerical results for $f$ and $df/dx$ at the $1-2\%$
level with $\theta=0.625$ $(0.656)$, $\Delta=-0.490$ $(-0.550)$ for $x
> 0$ $(x < 0)$ and $\gamma$ as given in tab.~\ref{tab2}.  The
universal properties of the scaling function can be compared with
results obtained by other methods for the three--dimensional $O(4)$
Heisenberg model.  In fig.~\ref{scalfunc} we display our results
along with those obtained from lattice Monte Carlo
simulation~\cite{Tou97-1}, second order epsilon expansion~\cite{BWW73-1}
and mean field theory.
We observe a good agreement of average action, lattice and epsilon
expansion results within a few per cent for $T < T_c$. Above $T_c$ the
average action and the lattice curve go quite close to each other with
a substantial deviation from the epsilon expansion and mean field
scaling function. (We note that the question of a better agreement of
the curves for $T<T_c$ or $T>T_c$ depends on the chosen non--universal
normalization conditions for $x$ and $f$ (cf.~\Ref{norm}).)

Before we use the scaling function $f(x)$ to discuss the general
temperature and quark mass dependent case, we consider the limits
$T=T_c$ and $\hat{m}=0$, respectively.  In these limits the behavior
of the various quantities is determined solely by critical amplitudes
and exponents. In the spontaneously broken phase ($T<T_c$) and in the
chiral limit we observe that the renormalized and unrenormalized order
parameters scale according to
\begin{equation}
  \label{NNN21}
  \begin{array}{rcl}
    \ds{\frac{2\si_0(T)}{T_c}} &=& \ds{
      \left(2E\right)^{1/2}
        \left(\frac{T_c-T}{T_c}\right)^{\nu/2}
      }\; ,\nnn
    \ds{\frac{2\ol{\si}_0(T)}{T_c}} &=& \ds{
      B \left(\frac{T_c-T}{T_c}\right)^{\beta}
      }\; ,
  \end{array}
\end{equation}
respectively, with $E=0.814$ and the value of $B$ given above.  In the
symmetric phase the renormalized mass $m=m_\pi=m_\si$ and the
unrenormalized mass $\ol{m}=Z_\Phi^{1/2}m$ behave as
\begin{equation}
  \label{NNN21a}
  \begin{array}{rcl}
    \ds{\frac{m(T)}{T_c}} &=& \ds{
      \left(\xi^+\right)^{-1}
      \left(\frac{T-T_c}{T_c}\right)^\nu
      }\; ,\nnn
    \ds{\frac{\ol{m}(T)}{T_c}} &=& \ds{
      \left( C^+\right)^{-1/2}
      \left(\frac{T-T_c}{T_c}\right)^{\gamma/2}
       \; , }
  \end{array}
\end{equation}
where $\xi^+=0.270$, $C^+=2.79$. For $T=T_c$ and
non--vanishing current quark mass we have
\begin{equation}
  \label{NNN21b}
  \begin{array}{rcl}
    \ds{\frac{2\ol{\si}_0}{T_c}} &=& \ds{
      D^{-1/\delta}
        \left(\frac{\jmath}{T_c^3}\right)^{1/\delta}
      }
  \end{array}
\end{equation}
with the value of $D$ given above. 

Though the five amplitudes $E$, $B$, $\xi^+$, $C^+$ and $D$ are not
universal there are ratios of amplitudes which are invariant under a
rescaling of $\ol{\si}_0$ and $T-T_c$. Our results for the
universal amplitude ratios are
\begin{equation}
  \label{ABC01}
  \begin{array}{rcl}
    \ds{R_\chi} &=& \ds{C^+ D B^{\delta-1}=1.02}\; ,\nnn
    \ds{\tilde{R}_\xi} &=& \ds{
      (\xi^+)^{\beta/\nu}D^{1/(\delta+1)}B=0.852}\; ,\nnn
    \ds{\xi^+ E} &=& \ds{0.220}\; .
  \end{array}
\end{equation}
Those for the critical exponents are given in tab.~\ref{tab2}.
\begin{table}
\begin{center}
\begin{tabular}{|c||l|l|l|l|l|} \hline
   &
  $\nu$ &
  $\gamma$ &
  $\delta$ &
  $\beta$ &
  $\eta$
  \\[0.5mm] \hline\hline
  average action &
  $0.787$ &
  $1.548$ &
  $4.80$ &
  $0.407$ &
  $0.0344$
  \\ \hline
  $3d$ PT &
  $0.73(2)$ &
  $1.44(4)$ &
  $4.82(5)$ &
  $0.38(1)$ &
  $0.03(1)$
  \\ \hline
  MC &
  $0.7479(90)$ &
  $1.477(18)$ &
  $4.851(22)$ &
  $0.3836(46)$ &
  $0.0254(38)$
  \\ \hline
\end{tabular}
\caption[]{\footnotesize The table shows the critical exponents 
  corresponding to the three--dimensional $O(4)$--Heisenberg model.
  Our results are denoted by ``average action'' whereas ``$3d$ PT''
  labels the exponents obtained from a $3d$ perturbative expansion
  using Pade techniques~\cite{BMN78-1,Zin93-1}. The bottom line
  contains lattice Monte Carlo results.~\cite{KK95-1}
  \label{tab2}}
\end{center}
\end{table}
Here the value of $\eta$ is obtained from the temperature dependent
version of \Ref{AAA69n} at the critical temperature~\cite{BJW97-1}. For
comparison tab.~\ref{tab2} also gives the
results~\cite{BMN78-1,Zin93-1} of a $3d$ perturbative computation as
well as lattice Monte Carlo results~\cite{KK95-1} which have been used
for the lattice form of the scaling function in
fig.~\ref{scalfunc}.~\footnote{See also ref~\cite{MT97-1} and
  references therein for a recent calculation of critical exponents
  using similar methods as in this work. For high precision estimates
  of the critical exponents see also refs~\cite{BC95-1,Rei95-1}.} There
are only two independent amplitudes and critical exponents,
respectively.  They are related by the usual scaling relations of the
three--dimensional scalar $O(N)$--model~\cite{Zin93-1} which we have
explicitly verified by the independent calculation of our exponents.

We turn to the discussion of the scaling behavior of the chiral
condensate $\VEV{\ol{\psi}\psi}$ for the general case of a temperature
and quark mass dependence.  In fig.~\ref{ccc_T} we have displayed
our results for the scaling equation of state in terms of the chiral
condensate
\begin{equation}
  \label{XXX30}
  \VEV{\ol{\psi}\psi}=
  -\ol{m}^2_{k_\Phi}T_c
  \left(\frac{\jmath/T_c^3}{f(x)}\right)^{1/\delta}+
    \jmath
\end{equation}
as a function of $T/T_c=1+x(\jmath/T_c^3 f(x))^{1/\beta\delta}$ for
different quark masses or, equivalently, different values of $\jmath$.
The curves shown in fig.~\ref{ccc_T} correspond to quark masses
$\hat{m}=0$, $\hat{m}=\hat{m}_{\rm phys}/10$, $\hat{m}=\hat{m}_{\rm
  phys}$ and $\hat{m}=3.5\hat{m}_{\rm phys}$ or, equivalently, to zero
temperature pion masses $m_\pi=0$, $m_\pi=45\MeV$, $m_\pi=135\MeV$ and
$m_\pi=230\MeV$, respectively (cf.~fig.~\ref{mm}). One observes that
the second order phase transition with a vanishing order parameter at
$T_c$ for $\hat{m}=0$ is turned into a smooth crossover in the
presence of non--zero quark masses.

The scaling form \Ref{XXX30} for the chiral condensate is exact only
in the limit $T\to T_c$, $\jmath\ra0$.  It is interesting to find the
range of temperatures and quark masses for which $\VEV{\ol{\psi}\psi}$
approximately shows the scaling behavior \Ref{XXX30}.  This can be
infered from a comparison (see fig.~\ref{ccc_T}) with our full
non--universal solution for the $T$ and $\jmath$ dependence of
$\VEV{\ol{\psi}\psi}$. For $m_\pi=0$ one observes approximate scaling
behavior for temperatures $T\gta90\MeV$. This situation persists up to
a pion mass of $m_\pi=45\MeV$. Even for the realistic case,
$m_\pi=135\MeV$, and to a somewhat lesser extent for $m_\pi=230\MeV$
the scaling curve reasonably reflects the physical behavior for $T\gta
T_c$. For temperatures below $T_c$, however, the zero temperature mass
scales become important and the scaling arguments leading to
universality break down.

The above comparison may help to shed some light on the use of
universality arguments away from the critical temperature and the
chiral limit. One observes that for temperatures above $T_c$ the
scaling assumption leads to quantitatively reasonable results even for
a pion mass almost twice as large as the physical value. This in turn
has been used for two flavor lattice QCD as theoretical input to guide
extrapolation of results to light current quark masses.  From
simulations based on a range of pion masses $0.3\lta
m_\pi/m_\rho\lta0.7$ and temperatures $0<T\lta250\MeV$ a
``pseudocritical temperature'' of approximately $140\MeV$ with a weak
quark mass dependence is reported~\cite{MILC97-1}. Here the
``pseudocritical temperature'' $T_{pc}$ is defined as the inflection
point of $\VEV{\ol{\psi}\psi}$ as a function of temperature.  The
values of the lattice action parameters used in~\cite{MILC97-1} with
$N_t=6$ were $a\hat{m}=0.0125$, $6/g^2=5.415$ and $a\hat{m}=0.025$,
$6/g^2=5.445$. For comparison with lattice data we have displayed in
fig.~\ref{ccc_T} the temperature dependence of the chiral condensate
for a pion mass $m_\pi=230\MeV$.  From the free energy of the linear
quark meson model we obtain in this case a pseudocritical temperature
of about $150\MeV$ in reasonable agreement with the results of
ref~\cite{MILC97-1}.  In contrast, for the critical temperature in
the chiral limit we obtain $T_c=100.7\MeV$.  This value is
considerably smaller than the lattice results of about $(140 - 150)
\MeV$ obtained by extrapolating to zero quark mass in
ref~\cite{MILC97-1}.  We point out that for pion masses as large as
$230\MeV$ the condensate $\VEV{\ol{\psi}\psi}(T)$ is almost linear
around the inflection point for quite a large range of temperature.
This makes a precise determination of $T_c$ somewhat difficult.
Furthermore, fig.~\ref{ccc_T} shows that the scaling form of
$\VEV{\ol{\psi}\psi}(T)$ underestimates the slope of the physical
curve. Used as a fit with $T_c$ as a parameter this can lead to an
overestimate of the pseudocritical temperature in the chiral limit.
We also mention here the results of ref~\cite{Got97-1}.  There two
values of the pseudocritical temperature, $T_{pc}=150(9)\MeV$ and
$T_{pc}=140(8)$, corresponding to $a\hat{m}=0.0125$, $6/g^2=5.54(2)$
and $a\hat{m}=0.00625$, $6/g^2=5.49(2)$, respectively, (both for
$N_t=8$) were computed.  These values show a somewhat stronger quark
mass dependence of $T_{pc}$ and were used for a linear extrapolation
to the chiral limit yielding $T_c=128(9)\MeV$.

The linear quark meson model exhibits a second order phase transition
for two quark flavors in the chiral limit. As a consequence the model
predicts a scaling behavior near the critical temperature and the
chiral limit which can, in principle, be tested in lattice
simulations. For the quark masses used in the present lattice studies
the order and universality class of the transition in two flavor QCD
remain a partially open question. Though there are results from the
lattice giving support for critical scaling~\cite{Kar94-1,IKKY97-1}
there are also recent simulations with two flavors that reveal
significant finite size effects and problems with
$O(4)$ scaling~\cite{BKLO96-1,Uka97-1}.

\sect{Additional degrees of freedom}
\label{AdditionalDegreesOfFreedom}

So far we have investigated the chiral phase transition of QCD as
described by the linear $O(4)$--model containing the three pions and
the sigma resonance as well as the up and down quarks as degrees of
freedom. Of course, it is clear that the spectrum of QCD is much
richer than the states incorporated in our model. It is therefore
important to ask to what extent the neglected degrees of freedom like
the strange quark, strange (pseudo)scalar mesons, (axial)vector
mesons, baryons, etc., might be important for the chiral dynamics of
QCD.  Before doing so it is perhaps instructive to first look into the
opposite direction and investigate the difference between the linear
quark meson model described here and chiral perturbation theory based
on the non--linear sigma model~\cite{Leu95-1}.  In some sense, chiral
perturbation theory is the minimal model of chiral symmetry breaking
containing only the Goldstone degrees of freedom. By construction it
is therefore only valid in the spontaneously broken phase and can not
be expected to yield realistic results for temperatures close to $T_c$
or for the symmetric phase.  However, for small temperatures (and
momentum scales) the non--linear model is expected to describe the
low--energy and low--temperature limit of QCD reliably as an expansion
in powers of the light quark masses. For vanishing temperature it has
been demonstrated recently~\cite{JW96-1,JW96-3,JW97-1} that the results
of chiral perturbation theory can be reproduced within the linear
meson model once certain higher dimensional operators in its effective
action are taken into account for the three flavor case.  Moreover,
some of the parameters of chiral perturbation theory
($L_4,\ldots,L_8$) can be expressed and therefore also numerically
computed in terms of those of the linear model. For non--vanishing
temperature one expects agreement only for low $T$ whereas deviations
from chiral perturbation theory should become large close to $T_c$.
Yet, even for $T\ll T_c$ small quantitative deviations should exist
because of the contributions of (constituent) quark and sigma meson
fluctuations in the linear model which are not taken into account in
chiral perturbation theory.

\begin{figure}
\unitlength1.0cm
\begin{center}
\begin{picture}(13.,7.0)

\put(0.0,0.0){
\epsfysize=11.cm
\rotate[r]{\epsffile{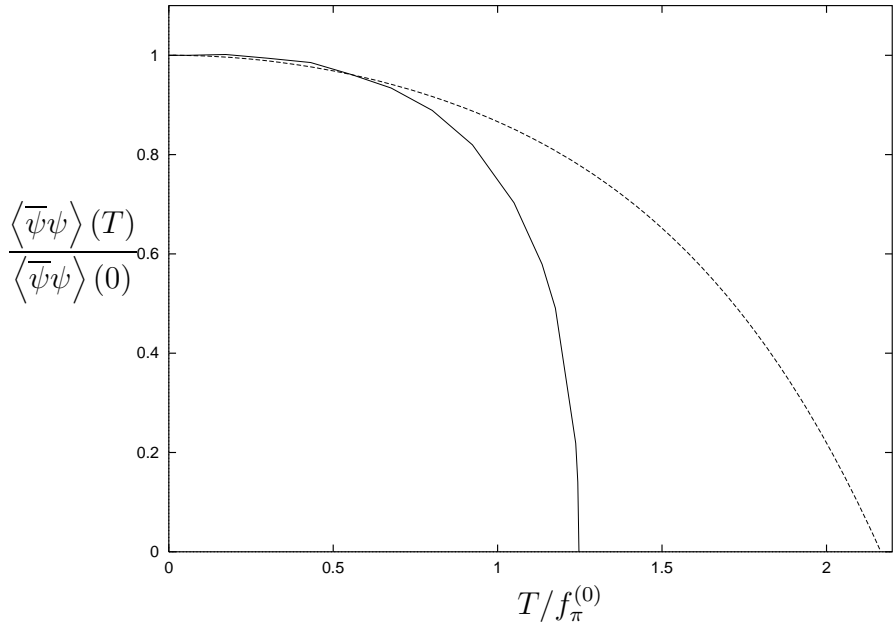}}
}
\put(-1.0,4.2){\bf 
  $\ds{\frac{\VEV{\ol{\psi}\psi}(T)}{\VEV{\ol{\psi}\psi}(0)} }$}
\put(5.8,-0.5){\bf $\ds{T/f_\pi^{(0)}}$}

\end{picture}
\end{center}
\caption{\footnotesize The plot displays the chiral condensate
  $\VEV{\ol{\psi}\psi}$ as a function of $T/f_\pi^{(0)}$. The solid
  line corresponds to our results for vanishing average current quark
  mass $\hat{m}=0$ whereas the dashed line shows the corresponding
  three--loop chiral perturbation theory result for
  $\Gamma_1=470\MeV$.}
\label{cc_T}
\end{figure}

From~\cite{GL87-1} we infer the three--loop result for the
temperature dependence of the chiral condensate in the chiral limit
for $N$ light flavors
\begin{equation}
  \label{BBB100}
  \begin{array}{rcl}
    \ds{\VEV{\ol{\psi}\psi}(T)_{\chi PT}} &=& \ds{
      \VEV{\ol{\psi}\psi}_{\chi PT}(0)
      \Bigg\{1-\frac{N^2-1}{N}\frac{T^2}{12F_0^2}-
        \frac{N^2-1}{2N^2}
        \left(\frac{T^2}{12F_0^2}\right)^2}\nnn
    &+& \ds{
      N(N^2-1)\left(\frac{T^2}{12F_0^2}\right)^3
      \ln\frac{T}{\Gamma_1}
      \Bigg\} +\Oc(T^8)}\; .
  \end{array}
\end{equation}
The scale $\Gamma_1$ can be determined from the $D$--wave isospin zero
$\pi\pi$ scattering length and is given by $\Gamma_1=(470\pm100)\MeV$.
The constant $F_0$ is (in the chiral limit) identical to the pion
decay constant $F_0=f_\pi^{(0)}=80.8\MeV$. In fig.~\ref{cc_T} we
have plotted the chiral condensate as a function of $T/F_0$ for both,
chiral perturbation theory according to \Ref{BBB100} and for the
linear quark meson model.
As expected the agreement for small $T$ is very good. Nevertheless,
the anticipated small numerical deviations present even for $T\ll T_c$
due to quark and sigma meson loop contributions are manifest.  For
larger values of $T$, say for $T\gta0.8f_\pi^{(0)}$ the deviations
become significant because of the intrinsic inability of chiral
perturbation theory to correctly reproduce the critical behavior of
the system near its second order phase transition.

Within the language of chiral perturbation theory the neglected
effects of thermal quark fluctuations may be described by an effective
temperature dependence of the parameter $F_0(T)$. We notice that the
temperature at which these corrections become important equals
approximately one third of the constituent quark mass $M_q(T)$ or the
sigma mass $m_\si(T)$, respectively, in perfect agreement with
fig.~\ref{Thresh}. As suggested by this figure the onset of the
effects from thermal fluctuations of heavy particles with a
$T$--dependent mass $m_H(T)$ is rather sudden for $T\gta m_H(T)/3$.
These considerations also apply to our two flavor quark meson model.
Within full QCD we expect temperature dependent initial values at
$k_\Phi$.

The dominant contribution to the temperature dependence of the initial
values presumably arises from the influence of the mesons containing
strange quarks as well as the strange quark itself.  Here the quantity
$\ol{m}^2_{k_\Phi}$ seems to be the most important one.  (The
temperature dependence of higher couplings like $\la(T)$ is not
very relevant if the IR attractive behavior remains valid, i.e.~if
$Z_{\Phi,k_\Phi}$ remains small for the range of temperatures
considered. We neglect a possible $T$--dependence of the current quark
mass $\hat{m}$.) In particular, for three flavors the potential
$U_{k_\Phi}$ contains a term
\begin{equation}
  \label{LLL12}
  -\frac{1}{2}\ol{\nu}_{k_\Phi}
  \left(\det\Phi+\det\Phi^\dagger\right)=
  -\ol{\nu}_{k_\Phi}\vph_s\Phi_{uu}\Phi_{dd}+\ldots
\end{equation}
which reflects the axial $U_A(1)$ anomaly. It yields a contribution to
the effective mass term proportional to the expectation value
$\VEV{\Phi_{ss}}\equiv\vph_s$, i.e.
\begin{equation}
  \label{LLL13}
  \Delta\ol{m}^2_{k_\Phi}=
  -\frac{1}{2}\ol{\nu}_{k_\Phi}\vph_s\; .
\end{equation}
Both, $\ol{\nu}_{k_\Phi}$ and $\vph_s$, depend on $T$.  We expect
these corrections to become relevant only for temperatures exceeding
$m_K(T)/3$ or $M_s(T)/3$. We note that the temperature dependent kaon
and strange quark masses, $m_K(T)$ and $M_s(T)$, respectively, may be
somewhat different from their zero temperature values but we do not
expect them to be much smaller. A typical value for these scales is
around $500\MeV$. Correspondingly, the thermal fluctuations neglected
in our model should become important for $T\gta170\MeV$. It is even
conceivable that a discontinuity appears in $\vph_s(T)$ for
sufficiently high $T$ (say $T\simeq170\MeV$). This would be reflected
by a discontinuity in the initial values of the $O(4)$--model leading
to a first order transition within this model.  Obviously, these
questions should be addressed in the framework of the three flavor
$SU_L(3)\times SU_R(3)$ quark meson model. Work in this direction is
in progress.

We note that the temperature dependence of $\ol{\nu}(T)\vph_s(T)$ is
closely related to the question of an effective high temperature
restoration of the axial $U_A(1)$ symmetry~\cite{PW84-1,Shu94-1}.  The
$\eta^\prime$ mass term is directly proportional to this
combination~\cite{JW96-1},
$m_{\eta^\prime}^2(T)-m_\pi^2(T)\simeq\frac{3}{2}\ol{\nu}(T)
\vph_s(T)$. Approximate $U_A(1)$ restoration would occur if
$\vph_s(T)$ or $\ol{\nu}(T)$ would decrease sizeable for large $T$.
For realistic QCD this question should be addressed by a three flavor
study. Within two flavor QCD the combination $\ol{\nu}_k\vph_s$ is
replaced by an effective anomalous mass term $\ol{\nu}_k^{(2)}$. The
temperature dependence of $\ol{\nu}^{(2)}(T)$ could be studied by
introducing quarks and the axial anomaly in the two flavor matrix
model of ref~\cite{BW97-1}.  We add that this question has also been
studied within full two flavor QCD in lattice
simulations~\cite{BKLO96-1,MILC97-2,KLS97-1}. So far there does not
seem to be much evidence for a restoration of the $U_A(1)$ symmetry
near $T_c$ but no final conclusion can be drawn yet.

To summarize, we have found that the effective two flavor quark meson
model presumably gives a good description of the temperature effects
in two flavor QCD for a temperature range $T\lta170\MeV$. Its
reliability should be best for low temperature where our results agree
with chiral perturbation theory.  However, the range of validity is
considerably extended as compared to chiral perturbation theory and
includes, in particular, the critical temperature of the second order
phase transition in the chiral limit.  We have explicitly connected
the universal critical behavior for small $\abs{T-T_c}$ and small
current quark masses with the renormalized couplings at $T=0$ and
realistic quark masses. The main quantitative uncertainties from
neglected fluctuations presumably concern the values of $f_\pi^{(0)}$
and $T_c$ which, in turn, influence the non--universal amplitudes $B$
and $D$ in the critical region. We believe that our overall picture is
rather solid. Where applicable our results compare well with numerical
simulations of full two flavor QCD.

\sect{Conclusions}

Our conclusions may be summarized in the following points:
\begin{enumerate}
\item The connection between short distance perturbative QCD and long
  distance meson physics by analytical methods seems to emerge step by
  step. The essential ingredients are nonperturbative flow equations
  as approximations of exact renormalization group equations and a
  formalism which allows a change of variables by introducing
  meson--like composite fields.
\item The relevance of meson--like $\ol{\psi}\psi$ composite objects
  is established in this framework. A typical scale where the mesonic
  bound states appear is the compositeness scale
  $k_\Phi\simeq(600-700)\MeV$. The occurrence of chiral symmetry
  breaking depends on the effective meson mass $\ol{m}_{k_\Phi}$ at
  the compositeness scale. For a certain range of values of the ratio
  $\ol{m}_{k_\Phi}^2/k_\Phi^2$ spontaneous chiral symmetry breaking is
  induced by quark fluctuations. A definite analytical establishment
  of spontaneous chiral symmetry breaking from ``first principles''
  (i.e., short distance QCD) still awaits a reliable calculation of
  this ratio.
\item Phenomenologically, the ratio $\ol{m}_{k_\Phi}^2/k_\Phi^2$ may
  be determined from the value of the constituent quark mass in units
  of the pion decay constant $f_\pi$. For this value the ratios
  $f_\pi/k_\Phi$ and $\VEV{\ol{\psi}\psi}/k_\Phi^3$ can be computed.
  Together with an earlier estimate of $k_\Phi$ this yields rather
  encouraging results: $f_\pi\simeq100\MeV$ and
  $\VEV{\ol{\psi}\psi}\simeq-(190\MeV)^3$ for two flavor QCD. It will
  be very interesting to generalize these results to the realistic
  three flavor case.
\item The formalism presented here naturally leads to an effective
  linear quark meson model for the description of mesons below the
  compositeness scale $k_\Phi$. For this model the standard
  non--linear sigma model of chiral perturbation theory emerges as a
  low energy approximation due to the large sigma mass.
\item The old puzzle about the precise connection between the current
  and the constituent quark mass reveals new interesting aspects in
  this formalism. In the context of the effective average action one
  may define these masses by the quark propagator at zero or very
  small momentum, either in the quark gluon picture (current quark
  mass) or the effective quark meson model (constituent quark mass).
  There is no conceptual difference between the two situations. As a
  function of $k$ the running quark mass smoothly interpolates between
  the standard current quark mass $m_q$ for high $k$ and the standard
  constituent quark mass $M_q$ for low $k$. (This holds at least as
  long as the minimum of the effective scalar potential is unique.) A
  rapid quantitative change occurs for $k\simeq(400-500)\MeV$ because
  of the onset of chiral symmetry breaking. For $k=0$ one expects this
  behavior to carry over to the momentum dependence of the quark
  propagator: For small $q^2$ the inverse quark propagator is
  dominated by $M_q$. In contrast, for high $q^2$ the constant term in
  the inverse propagator is reduced to $m_q$ since $q^2$ replaces
  $k^2$ as an effective infrared cutoff.  One expects a smooth
  interpolation between the two limits and it would be interesting to
  know the form of the propagator for momenta in the transition
  region. Furthermore, the symmetry breaking source term which
  determines the pion mass in the linear or non--linear meson model
  can be related to the current quark mass at the compositeness scale.
  This constitutes a bridge between the low energy meson properties
  and the running quark mass at a scale which is not too far from the
  validity of perturbation theory.
\item A particular version of the Nambu--Jona-Lasinio model appears in
  our formalism as a limiting case (infinitely strong renormalized
  Yukawa coupling at the compositeness scale $k_\Phi$). Here the
  ultraviolet cutoff which is implicit in the NJL model is dictated by
  the momentum dependence of the infrared cutoff function $R_k$ in the
  effective average action. The characteristic cutoff scale is
  $k_\Phi$. In this context our results can be interpreted as an
  approximative solution of the NJL model. Our method includes many
  contributions beyond the leading order contribution of the $1/N_c$
  expansion.
\item Based on a satisfactory understanding of the meson properties in
  the vacuum we have described their behavior in a thermal equilibrium
  situation. Our method should remain valid for temperature below
  $\sim170\MeV$. This extends well beyond the validity of chiral
  perturbation theory. In particular, in the chiral limit of vanishing
  quark masses we can describe the universal critical behavior near a
  second order phase transition of two flavor QCD. The universal
  behavior is quantitatively connected to observed quantities at zero
  temperature and realistic quark masses.
\end{enumerate}

\nopagebreak\vspace{1cm}\noindent {\bf Acknowledgement:} We thank
J.~Berges and B.~Bergerhoff for collaboration on many subjects covered
by these lectures. We also wish to express our gratitude to the
organizers of the {\em NATO Advanced Study Institute: Confinement,
  Duality and Non--perturbative Aspects of QCD} for their excellent
organization and for providing a most stimulating environment.

\end{document}